\documentclass[final,3p,12pt]{elsarticle}

\usepackage{amssymb}
\usepackage{mathtools}
\usepackage{bm}
\usepackage{amsmath}
\usepackage{graphicx}
\usepackage[unicode=true,
 bookmarks=true,bookmarksnumbered=false,bookmarksopen=false,
 breaklinks=false,pdfborder={0 0 1},backref=false,colorlinks=false]
 {hyperref}

\usepackage{booktabs}
\usepackage{multirow}
\usepackage{siunitx}
\usepackage{xargs}                      
\usepackage[pdftex,dvipsnames]{xcolor}  
\usepackage[colorinlistoftodos,prependcaption]{todonotes}
\newcommandx{\toby}[2][1=]{\todo[inline,linecolor=red,backgroundcolor=red!25,bordercolor=red,#1]{Toby: #2}}
\newcommandx{\adam}[2][1=]{\todo[inline,linecolor=orange,backgroundcolor=orange!25,bordercolor=orange,#1]{Adam: #2}}
\newcommandx{\jon}[2][1=]{\todo[inline,linecolor=blue,backgroundcolor=blue!25,bordercolor=blue,#1]{Jon: #2}}
\newcommandx{\colin}[2][1=]{\todo[inline,linecolor=OliveGreen,backgroundcolor=OliveGreen!25,bordercolor=OliveGreen,#1]{Colin: #2}}
\newcommandx{\david}[2][1=]{\todo[inline,linecolor=Plum,backgroundcolor=Plum!25,bordercolor=Plum,#1]{David: #2}}

\journal{Journal of the Electrochemical Society}

\begin{document}

\begin{frontmatter}



\title{
Nonlinear electrochemical impedance spectroscopy for lithium-ion battery model parameterization}



\author[inst1,inst2]{Toby L. Kirk}

\affiliation[inst1]{organization={Mathematical Institute, University of Oxford},
            addressline={Andrew Wiles Building, Woodstock Road}, 
            city={Oxford},
            postcode={OX2 6GG},
            country={UK}}

\author[inst2,inst3]{Adam Lewis-Douglas}
\author[inst2,inst3]{David Howey}
\author[inst1,inst2]{Colin P. Please}
\author[inst1,inst2]{S. Jon Chapman}

\affiliation[inst2]{organization={The Faraday Institution},
            addressline={Quad One, Becquerel Avenue, Harwell Campus}, 
            city={Didcot},
            postcode={OX11 0RA}, 
            country={UK}}
\affiliation[inst3]{organization={Battery Intelligence Lab, Department of Engineering Science, University of Oxford},
            city={Oxford},
            postcode={OX1 3PJ}, 
            country={UK}}

\begin{abstract}
In this work we analyse the local nonlinear electrochemical impedance spectroscopy (NLEIS) response of a lithium-ion battery and estimate model parameters from measured NLEIS data. The analysis assumes a single-particle model including nonlinear diffusion of lithium within the electrode particles and asymmetric charge transfer kinetics at their surface. Based on this model and assuming a moderately-small excitation amplitude, we systematically derive analytical formulae for the impedances up to the second harmonic response, allowing the meaningful interpretation of each contribution in terms of physical processes and nonlinearities in the model. The implications of this for parameterization are explored, including structural identifiability analysis and parameter estimation using maximum likelihood, with both synthetic and experimentally measured impedance data. Accurate fits to impedance data are possible, however inconsistencies in the fitted diffusion timescales suggest that a nonlinear diffusion model may not be appropriate for the cells considered. Model validation is also demonstrated by predicting time-domain voltage response using the parameterized model and this is shown to have excellent agreement with measured voltage time-series data (\SI{11.1}{mV} RMSE).
\end{abstract}




\end{frontmatter}


\section{Introduction}

Lithium-ion batteries have emerged as the dominant energy storage solution for portable electronics and electric vehicles, enabled by their high energy and power density and decreasing costs \cite{Iclodean2017}. They are also increasingly of interest for renewable energy integration within the power grid. In all of these applications, battery models are used to improve understanding of performance trade-offs at design stage and for operational monitoring of key factors such as state of charge, temperature, and lifetime. Equivalent circuit models are widely used in battery management systems, but electrochemical models, such as those derived from porous electrode theory within the Doyle-Fuller-Newman (DFN) framework \cite{DoyleFullerNewman1993}, are also increasingly of interest for high-fidelity predictions in academia and industry.


The effectiveness of electrochemical battery models as a tool for performance understanding and improvement depends on the realism of their internal structure and parameters for the particular questions being asked. Even if measured input-output quantities such as voltage can be predicted accurately, the internal states may differ significantly from reality. This is compounded by the large number of parameters required in these models (e.g.\, the DFN model has more than 30 parameters) and hence values are often taken from literature, but may not apply to the specific cell at hand. Fast and noninvasive characterisation techniques that can completely parameterize a physical model for a specific cell are therefore desirable.


Electrochemical impedance spectroscopy (EIS) is a widely used noninvasive frequency-domain characterization technique where a small amplitude sinusoidal current (or voltage) is applied to a device and the corresponding  voltage (or current) response is measured \cite{Barsukov2012}. 
The system response is assumed to be linear, and is traditionally fitted using `local' equivalent circuit models containing resistors, capacitors, and sometimes other \emph{ad hoc} components (e.g.\ constant phase elements), although this may lead to overfitting and difficulties in interpretation of the underlying processes\textemdash see Ciucci et al.\ \cite{Ciucci2019} for a review of the application of EIS to lithium-ion batteries. The assumption of a linear response in standard EIS necessarily discards  information about nonlinearities that may be useful for model selection and parameterization. This information may be measured by applying a larger amplitude sinusoidal excitation to generate harmonics in the response of the device under test. 

\begin{figure}
    \centering
    \includegraphics[width=0.75\textwidth]{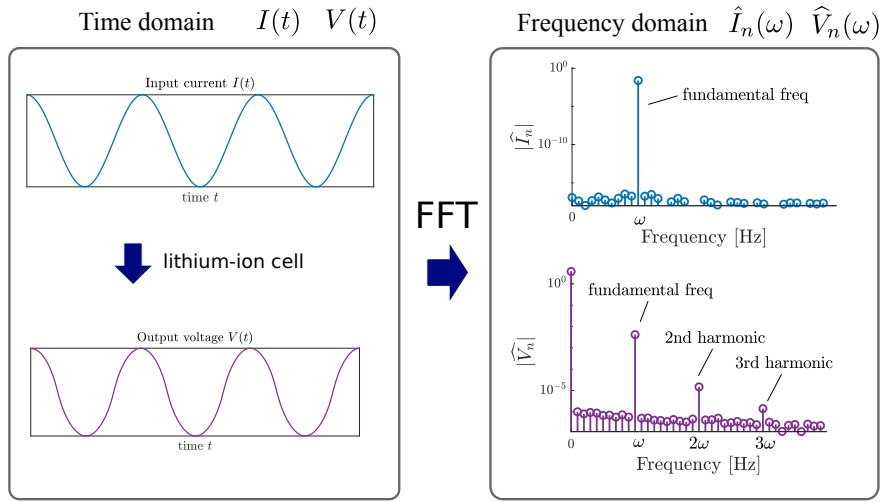}
    \caption{Diagram of the NLEIS process, where a sinusoidal current input and voltage output are transformed to the frequency domain using Fourier transforms. In practice experimental measurements are pre-processed to remove transients and windowed (e.g.\ with a Hann window) before transforming.}
    \label{fig:FFT_schematic}
\end{figure}

The injection of a sufficiently large amplitude sinusoidal current
into a battery will result in a nonlinear voltage response, with
harmonic components not only at the fundamental excitation frequency
$\omega$, but at all integer multiples of this i.e.\
$0,\omega,2\omega,3\omega,\ldots$; see Fig.\
\ref{fig:FFT_schematic}. The components $n\omega$ with $n\geq 2$ are
referred to as higher harmonics, and are due entirely to
nonlinearities. Several ways of analysing these harmonics have been
considered \cite{Murbach2017,Harting2017}, but one approach is to assume the input amplitude is
small, but larger than for linear EIS. Then a ``weakly" nonlinear
response is induced, with analytical predictions from the model
possible via a perturbation expansion. We will refer to this technique
as nonlinear EIS (NLEIS) \cite{McDonald2012}. It was first widely applied in the context of solid-oxide fuel cells \cite{Wilson2006,Wilson2007,Kadyk2009,Kadyk2011,Xu2011,Xu2013}, proving useful for model selection, i.e.\ identifying inconsistencies between proposed models and experiment. It has since been considered for many other electrochemical systems, as discussed in the review by Fasmin \& Srinivasan \cite{Fasmin2017}. %

There has been limited application of NLEIS to lithium-ion batteries. Murbach \& Schwartz \cite{Murbach2017}, using a DFN-type physical model, calculated the second and third harmonics by an expansion in current amplitude, but this required numerical solution of the spatial system of ODEs due to the model complexity. They showed that the higher harmonics are sensitive to the symmetry of the charge transfer reaction and to several parameters of interest, e.g.\ solid-state and electrolyte diffusivities. The authors also showed qualitative similarity between measured and predicted nonlinear impedances and that the second harmonic may be sensitive to cell ageing \cite{Murbach2018}. %

There have been several other studies where moderate or large amplitude sinusoidal currents have been applied to lithium-ion batteries \cite{Harting2017,Harting2018,Wolff2018,Wolff2019}. In particular, nonlinear frequency response analysis (NFRA) is a term used to refer to a particular type of NLEIS where there is no restriction on the input amplitude, and only the modulus (magnitude) of the harmonics is retained. No restriction on input amplitude means a stronger nonlinear response, however, the model analysis is purely numerical, with direct simulation in the time domain necessary before transformation to the frequency domain, which is computationally expensive. The aforementioned ``weakly nonlinear" analysis, on the other hand, can be done directly in the frequency domain. In addition, the practice in NFRA of discarding the harmonics' phase and hence potentially half of the nonlinear information does not seem justified. NFRA on lithium-ion batteries was explored experimentally by Harting et al.\ \cite{Harting2017,Harting2018}, and signatures of degradation processes (e.g.\ lithium plating) were able to be identified. Wolff et al.\ \cite{Wolff2018} considered NFRA applied to a DFN-type model and performed a parameter sensitivity analysis, and they \cite{Wolff2019} investigated the signatures of several models of SEI. Large amplitude multisine signals have recently been considered by Fan et al.\ \cite{Fan2021a,Fan2021b} who also performed a sensitivity analysis on a DFN model, giving support to the notion of asymmetry of charge transfer kinetics reported by others \cite{Murbach2018}.



Previously, NLEIS analysis of battery models has been applied only to relatively complex physical models with many parameters, such as the DFN model, making the evaluation and interpretation of the model impedances difficult. In this paper, we consider instead the NLEIS response of a  reduced-order physical model, the single-particle model (SPM), which is the simplest physical model that considers each electrode separately and captures the key battery dynamics and overpotentials, e.g.\ Butler\textendash Volmer reaction kinetics, solid-state diffusion of lithium, and double-layer capacitance. This allows us to derive analytical formulae for the impedances up to the second harmonic, and interpret each contribution in terms of physical processes, extending the established language of EIS Nyquist plots to include nonlinear harmonics. The  nonlinearities are identified explicitly, originating from: (i) asymmetric Butler\textendash Volmer kinetics; (ii) concentration dependence of the exchange current density; (iii) open circuit potential functions; (iv) solid-state diffusion. We then consider the implications of this new nonlinear information for parameter estimation, extending the structural identifiability analysis of a single-particle model by Bizeray et al.\ \cite{Bizeray2019}, which was limited to linear impedances. This is complemented with parameter estimates from synthetic and from experimentally measured NLEIS data using a commercial pouch cell. Our results show that the model fits the impedance data accurately, capturing many of the key features of each harmonic and determining all model parameters excluding open circuit potentials. Lastly, the model is independently validated in the time domain, showing improvements in accuracy of voltage predictions over conventional methods.

\section{Model formulation}

\subsection{The Single-Particle Model (SPM)}

\begin{figure}
    \centering
    \includegraphics[width=0.8\textwidth]{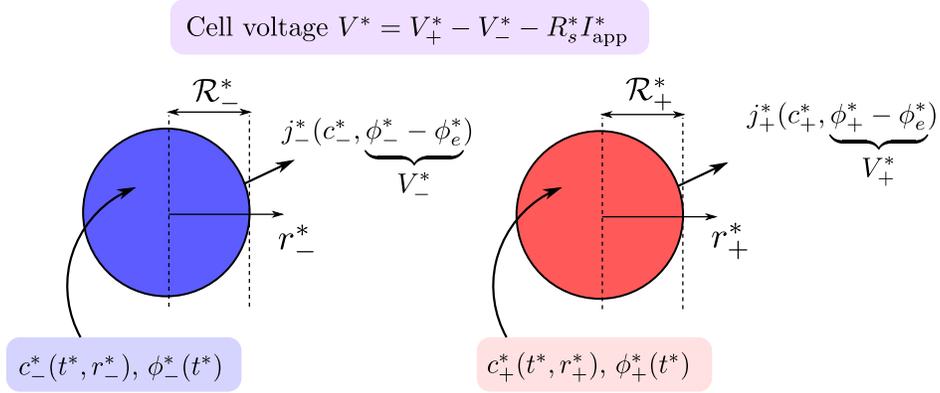}
    \caption{Single particle model schematic with a representative particle for each electrode. Not depicted is a resistor in series with the two electrodes.}
    \label{fig:SPM_schematic}
\end{figure}

The model of a lithium-ion cell we consider is the single-particle
model (SPM) \cite{Atlung1979,Ning2004,Bizeray2019}. Here we summarise the model assumptions and governing equations. The transport
of lithium ions in the electrolyte is assumed to be fast,
so that they remain at a constant uniform concentration $c_{e}^{*}$.
Throughout, asterisks denote dimensional quantities. In each
electrode, the active particles that make up
the porous electrode medium are all assumed identical in size and
shape. Thus they behave identically, and only a model of a single
representative particle is necessary. A schematic of the model geometry is shown in Fig.\ \ref{fig:SPM_schematic}. 
Let the subscript $\pm$ denote
that a quantity applies to, or is defined in, the positive and negative
electrode, respectively. Mass transport of lithium within spherical
particles of radius $\mathcal{R}_{\pm}^{*}$ is modelled here by
spherically-symmetric nonlinear diffusion, 
\begin{align}
\frac{\partial c_{\pm}^{*}}{\partial t^{*}} & = -\frac{1}{r_{\pm}^{*2}}\frac{\partial}{\partial r_{\pm}^{*}}\left(r_{\pm}^{*2}N_r^*\right),\qquad\text{for }0<r_{\pm}^{*}<\mathcal{R}_{\pm}^{*},\label{eq:dimensional_diffusion_eq} \\
N_r^* &= -D_{\pm}^{*}(c_{\pm}^{*})\frac{\partial c_{\pm}^{*}}{\partial r_{\pm}^{*}},
\end{align}
where $c_{\pm}^{*}(r_{\pm}^{*},t^{*})$ is the concentration of lithium
in the electrode material, and $D_{\pm}^{*}(c_{\pm}^{*})$ is the
nonlinear solid-state diffusion coefficient that is a function of concentration. 
The boundary conditions are regularity at the particle
centres,
\begin{equation}
\frac{\partial c_{\pm}^{*}}{\partial r_{\pm}^{*}}=0,\qquad\text{at }r_{\pm}^{*}=0,\label{eq:dimensional_centre_BC}
\end{equation}
and a surface flux, i.e., lithium (de)intercalation into (out of)
the particles,
\begin{equation}
-D_{\pm}^{*}(c_{\pm}^{*})\frac{\partial c_{\pm}^{*}}{\partial r_{\mathrm{\pm}}^{*}}=j_{\pm}^{*},\qquad\text{at }r_{\mathrm{\pm}}^{*}=\mathcal{R}_{\mathrm{\pm}}^{*},\label{eq:dimensional_surface_BC}
\end{equation}
which is modelled by nonlinear Butler\textendash Volmer
kinetics,
\begin{align}
&j_{\pm}^{*}  =\frac{i_{0,\pm}^{*}(c_{\mathrm{\pm}}^{*})}{2F^{*}}\left[\exp\left(\frac{F^{*}(1-\beta_{\pm})}{R_{g}^{*}T^{*}}\eta_{\mathrm{\pm}}^{*}\right)-\exp\left(-\frac{F^{*}\beta_{\pm}}{R_{g}^{*}T^{*}}\eta_{\mathrm{\pm}}^{*}\right)\right]\qquad\text{at }r_{\mathrm{\pm}}^{*}=\mathcal{R}_{\mathrm{\pm}}^{*},\label{eq:dimensional_Butler_Volmer}\\
&i_{0,\pm}^{*}(c_{\mathrm{\pm}}^{*})  =m_{\mathrm{\pm}}^{*}(c_{\mathrm{\pm}}^{*})^{\beta_{\pm}}(c_{e}^{*})^{1-\beta_{\pm}}(c_{\pm,\text{max}}^{*}-c_{\pm}^{*})^{1-\beta_{\pm}}. \label{eq:dimensional_exchange_current_density}
\end{align}
Here $F^{*}$ is Faraday's constant, $R_{g}^{*}$ is the universal
gas constant, $T^{*}$ is the temperature (assumed constant), $i_{0,\pm}^{*}$
is the exchange current density (with a dependence on the surface value of $c_{\mathrm{\pm}}^{*}$, with evaluation on $r_\pm^*=\mathcal{R}_{\pm}^{*}$ assumed),
$m_{\pm}^{*}$ is a reaction rate coefficient, $c_{\pm,\text{max}}^{*}$
is the maximum lithium concentration in the positive/negative electrode
material. The cathodic transfer coefficient $\beta_{\pm}$
(and corresponding anodic transfer coefficient $1-\beta_{\pm}$) is not assumed to be the typical value of
$1/2$, but may take any value between 0 and 1, i.e., $\beta_{\pm}\in(0,1)$. Electrical
conductivity in the electrode materials is assumed large, meaning
the electric potential in each is uniform, only a function of time
$t^{*}$, and expressed relative to the potential in the electrolyte,
$V_{\pm}^{*}(t^{*})=\phi_{\pm}^{*}(t^{*})-\phi_{e}^{*}$. Then $\eta_{\pm}^{*}(t^{*})=V_{\pm}^{*}(t^{*})-U_{\pm}^{*}(c_{\pm}^{*}|_{r_{\pm}^{*}=R_{\pm}^{*}})$
is the reaction overpotential, where the open circuit potential (OCP)
is given by $U_{\pm}^{*}$, a function of the surface concentration.
The evaluation of $c_{\pm}^{*}$ at $r_{\pm}^{*}=\mathcal{R}_{\pm}^{*}$ in the current density and overpotential expressions will, for brevity,
be understood from here on and not stated explicitly. 

The nonconstant diffusion coefficient $D_\pm^*(c_\pm^*)$ will be left as a general function of $c_\pm^*$ in our analysis, but for our results we will interpret it as originating from a diffusive flux driven by chemical potential gradients, such as in concentrated solution theory. Therefore we take the radial lithium flux to be \cite{Horner2021} 
\begin{align}
    N_r^* = \frac{{D}_{\mathrm{Li},\pm}^*F^*c_\pm^*}{R_g^*T^*}\frac{\partial U_\pm^*}{\partial r_\pm^*}, \label{eq:radial-flux} 
\end{align}
which has been shown to give more accurate behaviour than a Fickian diffusion model when parameterized using galvanostatic intermittent titration technique (GITT). Here ${D}_{\mathrm{Li},\pm}^*$ is the constant diffusivity of the Li species in each electrode material, and the flux (\ref{eq:radial-flux}) gives rise to  (\ref{eq:dimensional_diffusion_eq}) with the concentration-dependent diffusivity
\begin{align}
    D_\pm^*(c_\pm^*) &= -\frac{{D}_{\mathrm{Li},\pm}^*F^*}{R_g^*T^*}\frac{\mathrm{d}U_\pm^*}{\mathrm{d}c_\pm^*} c_\pm^*.
\end{align}
Note that, as per this equation, the dependence on $c_\pm^*$ is completely determined, and depends on the gradient of $U_\pm^*$.

Finally, conservation of charge relates both of the surface lithium
flux densities to the macroscopic applied circuit current $I_{\text{app}}^{*}$.
Let $a_{\pm}^{*}$ be the surface area per unit volume, related to
the electrode volume fraction $\epsilon_{\pm}$ for spherical particles
by $a_{\pm}^{*}=3\epsilon_{\pm}/\mathcal{R}_{\pm}^{*}$, then the interfacial current density is related to the applied current via the differential equation\begin{equation}
C_{\mathrm{dl},\pm}^{*}\frac{\mathrm{d}V_{\pm}^{*}}{\mathrm{d}t^{*}}=\pm\frac{I_{\text{app}}^{*}}{\mathcal{A}^{*}L_{\pm}^{*}a_{\pm}^{*}}-F^{*}j_{\pm}^{*},\label{eq:dimensional_capacitance}
\end{equation}
where $I_{\text{app}}^{*}>0$ corresponds to a charging current, $C_{\mathrm{dl},\pm}^{*}$ is the double-layer capacitance at the surface of each electrode,
$L_{\pm}^{*}$ is the through-cell thickness of each electrode,
and $\mathcal{A}^{*}$ is the area of the electrodes in the transverse
directions. 

We include double-layer capacitance here due to its relevance in
modelling the frequency domain behaviour, in particular at high
frequencies. However, a model neglecting double-layer capacitance has
computational advantages and is particularly useful in
time-domain simulations of scenarios where short timescales are less
important, e.g., charging or discharging. We may neglect capacitance by
setting $C_{\mathrm{dl},\pm}^{*}=0$ in
(\ref{eq:dimensional_capacitance}), reducing it to the algebraic
equation 
\begin{equation}
j_{\pm }^{*} = \pm \frac{I_{\text{app}}^{*}}{F^{*}\mathcal{A}^{*}L_{\pm}^{*} a_{\pm }^{*}}.
\label{eq:dimensional_I_app}
\end{equation}
If the applied current is prescribed, then substituting
(\ref{eq:dimensional_I_app}) into the boundary condition
(\ref{eq:dimensional_surface_BC}) allows
(\ref{eq:dimensional_diffusion_eq})-(\ref{eq:dimensional_surface_BC}) 
to be solved for $c_{\mathrm{\pm}}^{*}$, and (\ref{eq:dimensional_Butler_Volmer})
subsequently inverted to give the potentials in each electrode (relative to
the electrolyte potential) 
\begin{equation}
V_{\pm}^{*}=U_{\pm}^{*}(c_{\pm}^{*})+\eta_{\pm}^{*}, \label{eq:dimensional_Vs}
\end{equation}
With the potentials $V_{\pm}^{*}$ given either by the solution of (\ref{eq:dimensional_capacitance}) (when capacitance is included) or the explicit expression (\ref{eq:dimensional_Vs}) (when capacitance is neglected), the terminal voltage for the cell is then
\begin{align}
V^{*}(t^{*}) & =\phi_{\mathrm{+}}^{*}-\phi_{-}^{*}+R_{s}^{*}I_{\text{app}}^{*} = V_{+}^{*}-V_{-}^{*}+R_{s}^{*}I_{\text{app}}^{*}.\label{eq:dimensional_V}
\end{align}
A series resistance, $R_{s}^{*}$, is  included to account for
linear resistances arising from, e.g., current collector contacts,
and ionic and electronic conduction in the electrodes and electrolyte.

\subsection{Scaling and nondimensional SPM}
\label{sec:scaling}

To reduce the number of parameters required and establish the minimal set of
physical parameter groups necessary to use the model, we proceed to
nondimensionalise it. The internal variables $c_{\pm}^{*}$
and $r_{\pm}^{*}$ are scaled by the maximum lithium concentrations
$c_{\pm,\mathrm{max}}^{*}$ and particle radii $\mathcal{R}_{\pm}^{*}$, respectively. The
external variables, i.e., those directly measurable via the cell terminals
such as the current $I_{\mathrm{app}}^{*}$, voltage $V^{*}$, and
time $t^{*}$, are scaled with typical values $J^{*}$, $\Phi^{*}$,
$\tau^{*}$. It is convenient for these values to be known \emph{a
priori}, that is, not chosen from some internal property of the cell
itself. We choose the potential scale to be the thermal voltage, $\Phi^{*}=F^{*}/(R_{g}^{*}T^{*})$,
and $J^{*} = 1$ A, $\tau^{*} = 1$ s.

The nondimensional variables are defined as
\begin{align}
c_{\pm} & =c_{\pm}^{*}/c_{\pm,\mathrm{max}}^{*}, & r & =r_{\pm}^{*}/\mathcal{R}_{\pm}^{*}, & t &=t^{*}/\tau^{*}, & D_{\pm}(c_{\pm}) & = \frac{\tau^*D_{\pm}^*(c_{\pm}^*)}{\mathcal{R}_\pm^{*2}}\label{eq:Scaling_1}
\end{align}
\begin{align}
I & =I_{\mathrm{app}}^{*}/J^{*}, & (V,V_{\pm},U_{\pm},\eta_{\pm}) & =(V^{*},V_{\pm}^{*},U_{\pm}^{*},\eta_{\pm}^{*})/\Phi^{*}, & j_{\pm} &= \frac{\tau^{*}j_{\pm}^{*}}{c_{\pm,\mathrm{max}}^{*}\mathcal{R}_{\pm}^{*}}, & &\label{eq:Scaling_2}
\end{align}
resulting in the nondimensional model equations
\begin{align}
\frac{\partial c_{\pm}}{\partial t} & =\frac{1}{r^{2}}\frac{\partial}{\partial r}\left(D_{\pm}(c_{\pm})r^{2}\frac{\partial c_{\pm}}{\partial r}\right), & \text{for }0<r<1, \label{eq:c_equation}\\
\frac{\partial c_{\pm}}{\partial r} & =0, & \text{at }r=0,\\
-D_{\pm}(c_{\pm})\frac{\partial c_{\pm}}{\partial r_{\mathrm{\pm}}} & =j_{\pm}, & \text{at }r=1,\\
c_{\pm} & =c_{\pm,\mathrm{init}} & \text{at }t=0, \label{eq:init}
\end{align}and
\begin{equation}
C_{\pm}\frac{\mathrm{d}V_{\pm}}{\mathrm{d}t}=\pm I(t)-\frac{j_{\pm}}{\xi_{\pm}}, \label{eq:capacitance_ODE_nondim}
\end{equation}
with surface fluxes, overpotentials, and terminal voltage given by
\begin{align}
	j_{\pm} &= \frac{2\xi_{\pm}}{R_{\pm}(c_{\pm})}E_{\pm}(\eta_{\pm}), \label{eq:j}\\
    \eta_{\pm} &= V_{\pm} - U_{\pm}(c_{\pm}), \label{eq:eta}\\ 
    V & =V_{+}-V_{-}+R_{s}I(t), \label{eq:V}
\end{align}
and $C_{\pm}$ defined later in (\ref{eq:nondim_params}).
For convenience we have introduced the notation
\begin{align}
E_{\pm}(\eta) & =\frac{1}{2}\left[\exp\left((1-\beta_{\pm})\eta\right)-\exp\left(-\beta_{\pm}\eta\right)\right], \label{eq:E}\\
R_{\pm}(c) & =\frac{2\chi_{\pm}}{c^{\beta_{\pm}}(1-c)^{1-\beta_{\pm}}},\label{eq:R}
\end{align}
where $\chi_{\pm}$ is defined below in (\ref{eq:nondim_params}). If the reactions are symmetric, i.e.\ $\beta_{\pm}=1/2$, then $E_{\pm}(\eta)$  reduces to $\sinh (\eta/2)$.
Here $R_{\pm}$ is the reciprocal of the (nondimensional)
exchange current density, and can be interpreted as a nonlinear reaction resistance, with a  dependence on the surface concentration $c_{\pm}$ (evaluation at $r_{\pm}=\mathcal{R}_{\pm}$ understood). This dependence has the nonlinear, but assumed known, functional form in (\ref{eq:R}) with parameters $\beta_{\pm}$ and $\chi_{\pm}$. The factor of 1/2 in (\ref{eq:j}) is necessary for $R_{\pm}$ to correspond to the well-known charge transfer resistance if the model is linearized\textemdash see the Nonlinear Impedance section. 

If we neglect double-layer capacitance, then we may set $C_{\pm}=0$ in (\ref{eq:capacitance_ODE_nondim}), which reduces to 
\begin{equation}
j_{\pm}=\pm\xi_{\pm}I(t),
\end{equation}
allowing $j_{\pm}$ to be eliminated from the model. In this case, substituting for $j_{\pm}$ in (\ref{eq:j}) and rearranging for the overpotential gives 
\begin{align}
\eta_{\pm} & =E_{\pm}^{-1}\left[\pm\frac{1}{2}R_{\pm}(c_{\pm})I(t)\right], \label{eq:eta_no_C}
\end{align}
where $E_{\pm}^{-1}$  denotes the inverse of $E_{\pm}$, which has no convenient closed form representation in general (note that if $\beta_{\pm}=1/2$ then $E_{\pm}^{-1}(x)=2\sinh^{-1} x$). The electrode potentials $V_{\pm}$ and terminal voltage $V$ are then calculated from (\ref{eq:eta})-(\ref{eq:V}).

The remaining nondimensional parameter groups are
\begin{align}
D_{\pm}(c_{\pm}) & =  -\frac{\tau^{*}}{\tau_{\mathrm{d},\pm}^{*}}\frac{\mathrm{d}U_\pm}{\mathrm{d}c_\pm} c_\pm, & \xi_{\pm} & =\frac{J^{*}\tau^{*}}{3Q_{\mathrm{th},\pm}^{*}}, & \chi_{\pm} & =\frac{R_{\mathrm{ct,typ},\pm}^{*}J^{*}}{\Phi^{*}}, \\
C_{\pm} &= \frac{ C_{\mathrm{dl},\pm}^{*}\Phi^{*}\mathcal{A}^{*}L_{\pm}^{*}a_{\pm}^{*}}{J^{*}\tau^{*}}, & 
R_{s} &= \frac{R_{s}^{*}J^{*}}{\Phi^{*}},\label{eq:nondim_params}
\end{align}
which are defined in terms of the following physically meaningful
quantities of each electrode
\begin{align}
\tau_{\mathrm{d},\pm}^{*} & =\frac{\mathcal{R}_{\pm}^{*2}}{{D}_{\mathrm{Li},\pm}^*}= \text{solid-state diffusion timescales}\\
R_{\mathrm{ct,typ,\pm}}^{*} & =  \frac{\Phi^{*}}{\mathcal{A}^{*}L_{\pm}^{*} a_{\pm }^{*}m_{\pm}^{*}(c_{e}^{*})^{1-\beta_{\pm}}c_{\pm,\mathrm{max}}^*}=\text{typical charge transfer resistances} \label{eq:typical_CT_resistance}\\
Q_{\mathrm{th},\pm}^{*} & = F^{*}\epsilon_{\pm}c_{\pm,\mathrm{max}}^{*}L_{\pm}^{*}\mathcal{A}^{*}=\text{theoretical electrode capacities}
\end{align}
To recognise that $R_{\mathrm{ct,typ,\pm}}^{*}$ is an order-of-magnitude estimate of charge transfer resistance, notice that $\mathcal{A}^{*}L_{\pm}^{*} a_{\pm }^{*}$ is the total active surface area in each electrode, and $m_{\pm}^{*}(c_{e}^{*})^{1-\beta_{\pm}}c_{\pm,\mathrm{max}}^*$ is the typical exchange current density, i.e.\ magnitude of $i_{0,\pm}^*$ from  \ref{eq:dimensional_exchange_current_density}. Thus, the denominator in (\ref{eq:typical_CT_resistance}) is the typical reaction current.

\subsection{Model parameters}

Here we summarize the model parameters necessary to use the nondimensional
SPM (\ref{eq:c_equation})-(\ref{eq:R}).

\subsubsection{OCPs and electrode balancing}

These are quantities that relate to the rest-state of the cell, and
should be determined by investigations at equilibrium. They
are: 
\begin{itemize}
\item $U_{\pm}(c_{\pm})$, the electrode OCPs (relative to a Li/Li$^{+}$ reference)
as a function of stoichiometry $c_{\pm}\in[0,1]$;
\item $\xi_{\pm}$, the ratio of the theoretical electrode capacities $Q_{\mathrm{th},\pm}^{*}$ to the reference scale $J^{*}\tau^{*}$; 
\item $c_{\pm,\mathrm{init}}$, the initial electrode stoichiometries. These
can be expressed in terms of the depth-of-discharge (DoD) of the cell, or $Q=Q^{*}/Q_{\mathrm{cap}}^{*}\in[0,1]$, where $Q^*$ is the discharge capacity and $Q_{\mathrm{cap}}^{*}$ is the rated cell capacity. Then 
$c_{\pm,\mathrm{init}}$ can be calculated using the linear relations 
\begin{align}
c_{\pm} & =c_{\pm}^{0\%}+(c_{\pm}^{100\%}-c_{\pm}^{0\%})Q, \label{eq:c-Q_relation}
\end{align}
where $c_{\pm}^{0\%}$ and $c_{\pm}^{100\%}$ are the stoichiometries at 0\% and 100\% DoD (i.e.\ $Q=0$ and $Q=1$). Note that these can be related to the theoretical electrode capacities, and hence $\xi_\pm$, via
\begin{align}
  c_{\pm}^{100\%}-c_{\pm}^{0\%} & = \pm\frac{Q_{\mathrm{cap}}^{*}}{Q_{\mathrm{th},\pm}^{*}} = \pm\frac{3Q_{\mathrm{cap}}^{*}}{J^*\tau^*}\xi_\pm, \label{eq:ksi_stoichiometry_relation} 
\end{align}
therefore only one of the stoichiometries in each electrode, say $c_{\pm}^{0\%}$, must be provided independently.
\end{itemize}
In general, this amounts to requiring the knowledge of 2 functions ($U_{\pm}(c_{\pm})$) and 4 constants ($\xi_{\pm}$, $c_{\pm}^{0\%}$) related to the balancing of the electrodes. These
are typically found invasively, by disassembling a similar cell and
reassembling into two half cells and then measuring the OCVs\textemdash, or
by the use of reference electrodes. 
For the experimental data used in this study, the $U_{\pm}$ are taken to be known as a function
of DoD (or $Q$), measured using a minimally invasive reference electrode
\cite{Bizeray2019,McTurk2015}\textemdash see the Methods section. 
We note, however, that the use of a reference electrode does not allow the electrode OCPs to be determined outside the normal operation range of the full cell. To do so, half-cell measurements would be required.

\subsubsection{Dynamical parameters}

With the electrode balancing parameters determined, the remaining
9 parameter groups necessary for the dynamic modelling of
the cell (nonzero currents) consist of 4 per electrode and 1 series resistance, as follows:
\begin{itemize}
\item $\tau_{\mathrm{d},\pm} = \tau_{\mathrm{d},\pm}^* / \tau^*$, typical diffusion timescale (scaled by the reference timescale $\tau^* = 1$ s);
\item $\chi_{\pm}$, typical nondimensional charge transfer resistances;
\item $\beta_{\pm}$, cathodic charge transfer coefficients;
\item $C_{\pm}$, nondimensional double-layer capacitances;
\item $R_{s}$, nondimensional series resistance.
\end{itemize}
This list of parameters can be written as a vector,
\begin{align}
\bm{\theta} & =(\tau_{\mathrm{d},+}\,\,\,\chi_+\,\,\,\beta_+\,\,\,C_+\,\,\,\tau_{\mathrm{d},-}\,\,\,\chi_-\,\,\,\beta_-\,\,\,C_-\,\,\,R_{s})^{T}. \label{eq:dynamical-params}
\end{align}
Structural identifiability and estimation of these parameters from
NLEIS data will  be explored as part of the results of this paper, in the Results section. 

\subsection{Model nonlinearities}
\label{sec:Model-nonlinearities}

Since NLEIS is a nonlinear extension of EIS, it is useful to list the various sources of nonlinearity present in the model, namely
\begin{itemize}
    \item the OCPs, i.e.\ the nonlinear functions $U_{\pm}(c_{\pm})$ which encode much of the thermodynamic behaviour of the electrodes, and
    \item the Butler\textendash Volmer reaction kinetics. These contribute nonlinearities in two different ways: (i) the exponential dependence of the reaction current on overpotential, given by $E_\pm(\eta_\pm)$ in (\ref{eq:E}); (ii) the dependence of the exchange current density, or equivalently the charge transfer resistance $R_\pm(c_\pm)$ in (\ref{eq:R}), on lithium concentration. Finally,
    \item nonlinear diffusion in the particles, i.e.\ the dependence of the diffusion coefficient on lithium concentration.
\end{itemize}
Each of these nonlinearities contributes to the measured nonlinear impedance, and in the following section we will derive explicit formulae for the impedance, making each distinct contribution apparent.

\section{Nonlinear impedance}
\label{sec:Nonlinear-impedances}

\subsection{General description of NLEIS}

The technique of (galvanostatic) NLEIS involves the application
of a sinusoidal current, $I(t;\omega,\hat{I})=\hat{I}\mathrm{e}^{\mathrm{i}\omega t}+\hat{I}\mathrm{e}^{-\mathrm{i}\omega t}=\text{Re}\{2\hat{I}\mathrm{e}^{\mathrm{i}\omega t}\}$ to a device under test, where, by choice of phase, $\hat{I}$ is real and positive, giving the real representation $I=2\hat{I}\cos \omega t$ (it is more convenient to work with $\hat{I}$, rather than the real amplitude $2\hat{I}$). Then, the nonlinear voltage response of the system, $V(t;\omega,\hat{I})$, is measured. After the decay of any initial transients, $V$ will
in general be periodic in $t$ with period $P=2\pi/\omega$ matching
the input current, and hence have the Fourier representation
\begin{align}
V(t;\omega,\hat{I}) &= \sum_{n=-\infty}^{\infty}\widehat{V}_{n}(\omega,\hat{I})\mathrm{e}^{\mathrm{i}n\omega t},\qquad\text{where }\widehat{V}_{-n}=\overline{\widehat{V}_{n}},
\label{eq:Fourier-series-V}
\end{align}
and the bar denotes complex conjugation. The component at frequency $\omega$, matching that of the input current, is referred to as the fundamental. However, due to nonlinearities in the system, all higher harmonics, i.e.\ at frequencies $2\omega,3\omega,\ldots$,
are in general also present. The Fourier coefficients
$\widehat{V}_{n}(\omega,\hat{I})$ depend on the excitation amplitude
$\hat{I}$. Thus, we express each Fourier coefficient as a Taylor series expansion for small current amplitude
$\hat{I}\ll1$, and it can be shown \cite{Murbach2018,Fasmin2017} that in general,
\begin{equation}
\widehat{V}_{n}(\omega,\hat{I})=\sum_{r=0}^{\infty}\hat{I}^{n+2r}Z_{n}^{(n+2r)}(\omega)=\hat{I}^{n}Z_{n}^{(n)}+\hat{I}^{n+2}Z_{n}^{(n+2)}+O(\hat{I}^{n+4}),\label{eq:V-expansion-1}
\end{equation}
where $Z_{n}^{(n+2r)}(\omega)$, $n,r=0,1,2,\ldots$ are independent
of $\hat{I}$. The subscript refers to the frequency mode or harmonic in the Fourier series expansion, and the superscript is the index in the Taylor series expansion in powers of $\hat{I}$. That is, a superscript $(m)$ denotes a coefficient of $\hat{I}^m$. Then $Z_{0}^{(0)}$ is simply the rest (open circuit) voltage, $Z_{1}^{(1)}$
is the usual linear impedance, and all other $Z_{n}^{(n+2r)}$ with
$n\geq 2$ or $r\geq 1$ are 
\emph{nonlinear} impedances.

For a given harmonic $n$ in the voltage, $\hat{I}^{n}Z_{n}^{(n)}$ is
the leading-order response, and so $Z_{n}^{(n)}$ is the leading-order
impedance for that harmonic and can be isolated from
$\widehat{V}_{n}$ via 
\begin{equation}
    Z_{n}^{(n)}(\omega) = \lim_{\hat{I}\to 0} \frac{\widehat{V}_{n}(\omega,\hat{I})}{\hat{I}^{n}}.
\end{equation}
The higher order impedances $Z_{n}^{(n+2r)}$, $r\geq 1$, can be
isolated sequentially in a similar way, provided all lower-order terms
for that harmonic are first subtracted from $\widehat{V}_{n}$. We will
focus on the leading-order impedances in this paper,
e.g.\ $Z_{1}^{(1)}$ and $Z_{2}^{(2)}$, however $Z_{0}^{(2)}$ is also easily derived
 in this analysis. 

The expansion in current amplitude $\hat{I}\ll1$ can be performed
analytically on the SPM, solving for each $Z_{n}^{(n+2r)}$, order-by-order, in a systematic fashion. Formally, all variables $X\in\{c_{\pm},\eta_{\pm},j_{\pm},V_{\pm}\}$ have a series expansion identical to (\ref{eq:Fourier-series-V})-(\ref{eq:V-expansion-1}). Up to $O(\hat{I}^{2})$, which is the highest order we will consider here, these take the form
\begin{equation}
X(t;\omega,\hat{I})=\sum_{n=-\infty}^{\infty}\widehat{X}_{n}(\omega,\hat{I})\mathrm{e}^{\mathrm{i}n\omega t}\qquad\text{where }\widehat{X}_{-n}=\overline{\widehat{X}_{n}},
\end{equation}
with Fourier coefficients $\widehat{X}_{n}$, as follows:
\begin{align}
\text{Zero frequency }(0): &  & \widehat{X}_{0} & =\widehat{X}_{0}^{(0)}+\hat{I}^{2}\widehat{X}_{0}^{(2)}+O(\hat{I}^{4}),\\
\text{Fundamental/1st harmonic }(\omega): &  & \widehat{X}_{1} & =\hat{I}\widehat{X}_{1}^{(1)}+O(\hat{I}^{3}),\\
\text{2nd harmonic }(2\omega): &  & \widehat{X}_{2} & =\hat{I}^{2}\widehat{X}_{2}^{(2)}+O(\hat{I}^{4}),\\
\text{Higher harmonics }(n\omega,n\geq3): &  &  & \vdots
\end{align}

\subsection{SPM neglecting double-layer capacitance}
\label{sec:Nonlinear-impedances-no-C}

First, we consider the SPM with double-layer capacitance neglected, i.e., equations (\ref{eq:c_equation})-(\ref{eq:init}), (\ref{eq:E})-(\ref{eq:eta_no_C}). In this case there is only a single path that the current can take through the cell, and the impedance formulae are useful in expressing the full formulae with capacitance included. We will denote the impedances without capacitance by a lowercase $z$ to distinguish them from the full impedances with capacitance included, denoted by $Z$, which will be stated afterwards.

The details of the expansion are given in section  S1 of the supplementary material, 
and the resulting impedances for the full cell model can be written in terms of half-cell impedances $z_{n,\pm}^{(m)}=\widehat{V}_{n,\pm}^{(m)}$, found from the expansion of the half-cell voltages $V_\pm$.

At leading order in the current amplitude, there is only a zero frequency response, $z_{0,\pm}^{(0)}=U_{\pm}^{(0)}=U_\pm(c_{\pm}^{(0)})$, the OCP at the stoichiometry $c_{\pm}^{(0)}$ for that state of charge.

At first order, $O(\hat{I})$, there is a response at fundamental frequency given by\footnote{The $\pm$ prefactor is due to the fact that a positive (charging) cell current corresponding to a charging current in the positive electrode but \emph{discharging} current in the negative one.}
\begin{align}
    z_{1,\pm}^{(1)} &= \pm\left[ R_{\pm}^{(0)} + Z^{\mathrm{W}}_{1,\pm}(\omega)\right],\label{eq:z_11_pm}
\end{align}
with two contributions: charge transfer resistance $R_{\pm}^{(0)}$ and so-called finite Warburg impedance due to the diffusion of lithium within a spherical particle,
\begin{align}
    Z^{\mathrm{W}}_{1,\pm}(\omega) &= U_{\pm}^{(0)\prime}\frac{\xi_{\pm}}{D_{\pm}^{(0)}}H_{1}\left(\frac{\omega}{D_{\pm}^{(0)}}\right), \label{eq:Z_W_1}
\end{align}
where a prime denotes a derivative with respect to $c_{\pm}$,
\begin{equation}
U_{\pm}^{(0)\prime}=\left.\frac{\mathrm{d}U_{\pm}}{\mathrm{d}c_{\pm}}\right|_{c_{\pm}^{(0)}},
\end{equation}
and a
superscript (0) denotes the leading order (zero current) quantity
at $c_{\pm}^{(0)}$. Note that $D_\pm^{(0)} = - U_{\pm}^{(0)\prime}c_{\pm}^{(0)} / \tau_{\mathrm{d},\pm}$, but for more generality and brevity we will leave formulae in terms of $D_\pm^{(0)}$. The function $H_{1}(\omega)$ encodes the frequency dependence of the transport process at this order, and is given by
\begin{align}
H_{1}(\omega)&=\frac{\tanh\sqrt{\mathrm{i}\omega}}{\tanh\sqrt{\mathrm{i}\omega}-\sqrt{\mathrm{i}\omega}},\label{eq:H_1}
\end{align}
with low and high $\omega$ asymptotic behaviours (differential capacitance and diffusion in a half-space, respectively)
\begin{align}
    H_1 (\omega) &\sim -\frac{3}{\mathrm{i}\omega}-\frac{1}{5}, & \text{as }\omega &\to 0,\\
    H_1 (\omega) &\sim \frac{1}{1-\sqrt{\mathrm{i}\omega}} \sim -\frac{1}{\sqrt{\mathrm{i}\omega}}, & \text{as }\omega &\to \infty. \label{eq:H_1_high_omega}
\end{align}
Equation (\ref{eq:z_11_pm}) for the fundamental is a commonly used impedance for batteries (and many other electrochemical systems) that appears in linear EIS studies, e.g.\ \cite{Bizeray2019}.

At second order, $O(\hat{I}^2)$, there is a response at twice the excitation frequency, $2\omega$, found to be
\begin{align}
    z_{2,\pm}^{(2)}&= Z^{\mathrm{kin}}_{\mathrm{asym},\pm} + Z^{\mathrm{kin}}_{\mathrm{CD},\pm} + Z^{\mathrm{OCP}}_{\pm} + Z^{\mathrm{W}}_{2,\pm}, \label{eq:z22_pm}
\end{align}
with four contributions,
\begin{align}
    Z^{\mathrm{kin}}_{\mathrm{asym},\pm} &= (\beta_{\pm}-1/2)\left(R_{\pm}^{(0)}\right)^{2}, \label{eq:Z_kin_asym}\\
    Z^{\mathrm{kin}}_{\mathrm{CD},\pm}(\omega) &= R_{\pm}^{(0)\prime}\left(\frac{\xi_{\pm}}{D_{\pm}^{(0)}}\right)H_{1}\left(\frac{\omega}{D_{\pm}^{(0)}}\right), \label{eq:Z_kin_CD}\\
    Z^{\mathrm{OCP}}_{\pm}(\omega) &= \frac{1}{2}U_{\pm}^{(0)\prime\prime}\left(\frac{\xi_{\pm}}{D_{\pm}^{(0)}}\right)^{2}\left[H_{1}\left(\frac{\omega}{D_{\pm}^{(0)}}\right)\right]^{2}, \label{eq:Z_OCP}\\    
    Z^{\mathrm{W}}_{2,\pm}(\omega) &= U_{\pm}^{(0)\prime}\left(\frac{-D_{\pm}^{(0)\prime}}{D_{\pm}^{(0)}}\right)\left(\frac{\xi_{\pm}}{D_{\pm}^{(0)}}\right)^{2}H_{2}\left(\frac{\omega}{D_{\pm}^{(0)}}\right),  \label{eq:Z_W_2}
\end{align}
These impedances, (\ref{eq:Z_kin_asym})-(\ref{eq:Z_W_2}), each correspond to the different nonlinearities in the model, listed in the Model Formulation section. 
The first and second terms, (\ref{eq:Z_kin_asym})-(\ref{eq:Z_kin_CD}), are due to nonlinearities in the kinetics. The kinetic term $Z^{\mathrm{kin}}_{\mathrm{asym},\pm}$ represents the charge transfer asymmetry, i.e.\ $\beta_{\pm}\neq 1/2$, and is purely real and independent of frequency. The kinetic term $Z^{\mathrm{kin}}_{\mathrm{CD},\pm}$ is due to the concentration dependence of the exchange current density (equivalently, the charge transfer resistance $R_{\pm}(c_\pm)$, see (\ref{eq:R})), which is clear from the presence of the derivative $R_{\pm}^{(0)\prime}$.  
The third term, $Z^{\mathrm{OCP}}_{\pm}$, represents the nonlinearity of the OCP with respect to stoichiometry, indicated by the second derivative appearing, $U_{\pm}^{(0)\prime\prime}$. Finally, the term $Z^{\mathrm{W}}_{2,\pm}$ is due to nonlinear diffusion of lithium within the particles, i.e.\ that the diffusivity depends on concentration, indicated by $D_{\pm}^{(0)\prime}$. Therefore, it could be interpreted as a second-order finite Warburg impedance, cf.\ (\ref{eq:Z_W_1}). Given our assumption on the functional dependence of $D_\pm^{(0)}$ (see (\ref{eq:nondim_params})), we may substitute
\begin{align}
    D_{\pm}^{(0)\prime} & =-\frac{1}{\tau_{\mathrm{d},\pm}} \left(U_{\pm}^{(0)\prime\prime}c_\pm^{(0)} + U_{\pm}^{(0)\prime}\right),
\end{align}
into (\ref{eq:Z_W_2}). At this order, the frequency dependence of the diffusion process is encoded in $H_2(\omega)$, which is the solution of a linear ODE (see section S2 of the supplementary material),
found numerically. However, the high frequency behaviour of $H_2$ may be extracted and is found to be
\begin{align}
    H_2 (\omega) & \sim \frac{1}{\mathrm{i}\omega}\left(1-\frac{1}{\sqrt{2}}\right), & \text{as }\omega &\to \infty,    \label{eq:H_2_high_omega}
\end{align}
which is just a real multiple of $(H_1)^2$ as $\omega \to \infty$ to leading order.
If the diffusion is linear, a common simplification, then $D_{\pm}^{(0)\prime}=0$ and the term $Z^{\mathrm{W}}_{2,\pm}(\omega)$  vanishes.

In addition to the response at $2\omega$, there is a nonlinear response at zero frequency, where many of the same terms arise:
\begin{align}
    z_{0,\pm}^{(2)} &= 2\mathrm{Re}\left[
    Z^{\mathrm{kin}}_{\mathrm{asym},\pm} + Z^{\mathrm{kin}}_{\mathrm{CD},\pm} +
    Z^{\mathrm{OCP}}_{0,\pm} + 
    Z^{\mathrm{W}}_{0,\pm}
    \right],\label{eq:z02_pm}
\end{align}
The terms are the same as in $z_{2,\pm}^{(2)}$, (\ref{eq:z22_pm}), except for the OCP and Warburg terms, which are now
\begin{align}
    Z^{\mathrm{OCP}}_{0,\pm}(\omega) &= \frac{1}{2}U_{\pm}^{(0)\prime\prime}\left(\frac{\xi_{\pm}}{D_{\pm}^{(0)}}\right)^{2}\left|H_{1}\left(\frac{\omega}{D_{\pm}^{(0)}}\right)\right|^{2}, \\
    Z^{\mathrm{W}}_{0,\pm}(\omega) &= U_{\pm}^{(0)\prime}\left(\frac{-D_{\pm}^{(0)\prime}}{D_{\pm}^{(0)}}\right)\left(\frac{\xi_{\pm}}{D_{\pm}^{(0)}}\right)^{2}H_{0}\left(\frac{\omega}{D_{\pm}^{(0)}}\right),
\end{align}
where a different function, $H_0(\omega)$ appears in the place of $H_2(\omega)$\textemdash see section S2 of the supplementary material. 
Notice that $z_{0,\pm}^{(2)}$ is real; this is because it represents a correction to the time average of the response, and hence is not oscillatory.

The nonlinear impedances at $O(\hat{I}^{3})$ and higher can be readily calculated, but with diminishing returns on their usability. Their amplitude is inherently smaller, implying lower signal-to-noise ratios, with significant increase in analytical complexity, so we do not consider them here.


The impedances for the full cell, including both electrodes and a
series resistance, are straightforwardly given in terms of the half-cell impedances (\ref{eq:z_11_pm}), (\ref{eq:z22_pm}), (\ref{eq:z02_pm}), using (\ref{eq:V}):
\begin{align}
z_{0}^{(0)} & =z_{0,+}^{(0)}-z_{0,-}^{(0)}, & z_{1}^{(1)} & =z_{1,+}^{(1)}-z_{1,-}^{(1)}+R_{s},\label{eq:z00}\\
z_{2}^{(2)} & =z_{2,+}^{(2)}-z_{2,-}^{(2)}, & z_{0}^{(2)} & =z_{0,+}^{(2)}-z_{0,-}^{(2)},\label{eq:z02}
\end{align}
Note that the series resistance $R_{s}$ only enters the
$z_{1}^{(1)}$ expression as it is assumed to be a linear resistor.

The fundamental impedance $z_{1}^{(1)}$
is the usual linear one, measured through conventional EIS, with dependence only on the series resistance $R_{s}$
and electrode quantities $U_{+}^{(0)\prime}$, $U_{-}^{(0)\prime}$, $D_{+}^{(0)}$, $D_{-}^{(0)}$, $\chi_{+}$, $\chi_{-}$, $\beta_{+}$, $\beta_{-}$. The second order impedances, $z_{2}^{(2)}$ and $z_{0}^{(2)}$, additionally depend on the higher derivatives $U_{+}^{(0)\prime\prime}$, $U_{-}^{(0)\prime\prime}$, $D_{+}^{(0)\prime}$, $D_{-}^{(0)\prime}$. They also have explicit dependence on $\beta_{+}$ and $\beta_{-}$ via the kinetic asymmetry terms $Z^{\mathrm{kin}}_{\mathrm{asym},+}$ and $Z^{\mathrm{kin}}_{\mathrm{asym},-}$.

\subsection{SPM including double-layer capacitance}

We now extend the formulae of the previous section to the more general case of the SPM that includes double-layer capacitance, (\ref{eq:c_equation})-(\ref{eq:R}). One may repeat the calculations of the previous section, starting from the full equations, but the resulting impedance formulae may be expressed concisely in terms of those already derived. The calculation of the half-cell impedances, now employing
an uppercase $Z$ to signify the inclusion of capacitance, is given in section S1.2 of the supplementary material, 
and yields
\begin{align}
Z_{1,\pm}^{(1)}=\,\, & \frac{z_{1,\pm}^{(1)}(\omega)}{1\pm C_{\pm}\mathrm{i}\omega z_{1,\pm}^{(1)}(\omega)}=\pm\left(\frac{1}{\left(C_{\pm}\mathrm{i}\omega\right)^{-1}}+\frac{1}{R_{\pm}^{(0)} + Z^{\mathrm{W}}_{1,\pm}(\omega)}\right)^{-1}, \label{eq:Z11_pm}\\
Z_{2,\pm}^{(2)}=\,\, & \frac{z_{2,\pm}^{(2)}(\omega)}{\left[1\pm C_{\pm}\mathrm{i}2\omega z_{1,\pm}^{(1)}(2\omega)\right]\left[1\pm C_{\pm}\mathrm{i}\omega z_{1,\pm}^{(1)}(\omega)\right]^{2}}, \label{eq:Z22_pm}\\
Z_{0,\pm}^{(2)}=\,\, & \frac{z_{0,\pm}^{(2)}(\omega)}{\left|1\pm C_{\pm}\mathrm{i}\omega z_{1,\pm}^{(1)}(\omega)\right|^{2}},\label{eq:Z02_pm}
\end{align}
where $z_{n,\pm}^{(m)}$ are the corresponding half-cell impedance formulae assuming no capacitance
($C_{\pm}=0$), given by (\ref{eq:z_11_pm}), (\ref{eq:z22_pm}) and (\ref{eq:z02_pm}).
As there is now more than one path the current can take through each electrode, these formulae are no longer just a sum of individual impedances as in (\ref{eq:z_11_pm}), (\ref{eq:z22_pm}) and (\ref{eq:z02_pm}), but they are nonetheless still explicit. 

The linear impedances $Z_{1,\pm}^{(1)}$
have the clear interpretation as being the harmonic mean of $(C_{\pm}\mathrm{i}\omega)^{-1}$ and $z_{1,\pm}^{(1)}(\omega)$. This corresponds to a capacitor in parallel with a charge transfer resistance that is in series with a (finite) Warburg element\textemdash this is exactly the Randles circuit. The nonlinear impedances do
not have such a clear analogy with traditional circuit elements. The
expressions (\ref{eq:Z11_pm})-(\ref{eq:Z02_pm}) arise from
nonlinearities in the charge transfer reaction and diffusion process
when they are placed in parallel with a capacitance, and therefore had
to be derived mathematically rather than from physical intuition. 

\begin{figure}
    \centering
    \includegraphics[width=0.45\textwidth]{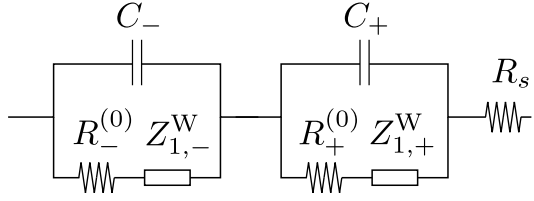}
    \caption{Equivalent circuit diagram representing the linear response of the full cell model, consisting of a Randles circuit for each electrode along with a series resistor $R_s$.} 
    \label{fig:circuit_diagram}
\end{figure}

The impedances for a full-cell are  given, using (\ref{eq:V}), by
\begin{align}
Z_{0}^{(0)} & =Z_{0,+}^{(0)}-Z_{0,-}^{(0)}, & Z_{1}^{(1)} & =Z_{1,+}^{(1)}-Z_{1,-}^{(1)}+R_{s},\label{eq:Z00}\\
Z_{2}^{(2)} & =Z_{2,+}^{(2)}-Z_{2,-}^{(2)}, & Z_{0}^{(2)} & =Z_{0,+}^{(2)}-Z_{0,-}^{(2)}.\label{eq:Z02}
\end{align}
The fundamental full-cell impedance then has the familiar equivalent circuit analogy shown in Fig.\ \ref{fig:circuit_diagram}, consisting of two Randles circuits (one for each electrode) placed in series.

\subsection{Simplifications: Separation of capacitance and diffusion timescales}
Although the impedance formulae (\ref{eq:Z11_pm})-(\ref{eq:Z02_pm}) are already explicit, some further simplification is possible using common assumptions about the physical timescales in the system.
There are two prominent timescales in (\ref{eq:Z11_pm})-(\ref{eq:Z02_pm}), corresponding to capacitance and diffusion effects. These are given by $R_{\pm}^{(0)}C_{\pm}$ and $1/D_{\pm}^{(0)}$, respectively. Note that these timescales have been scaled by a known reference timescale $\tau^{*}$. Here $1/D_{\pm}^{(0)} = \tau_{\mathrm{d},\pm}/(-U_\pm^{(0)\prime}c_\pm^{(0)})$ can be thought of as the \emph{local} diffusion timescale, at this DoD, which is more convenient to use here instead of $\tau_{\mathrm{d},\pm}$ directly. Typically the capacitance timescale ($O(10^{-2})$ s) is much shorter than the diffusive one ($O(10^{4})$ s), hence we can consider the well-separated limit of $R_{\pm}^{(0)}C_{\pm}\ll 1/D_{\pm}^{(0)}$. 

Taking $R_{\pm}^{(0)}C_{\pm}\ll 1/D_{\pm}^{(0)}$ in (\ref{eq:Z11_pm})-(\ref{eq:Z02_pm}) gives different results depending on the frequency $\omega$. If the frequency is sufficiently low, $\omega =O(D_{\pm}^{(0)})$ (region (I)), then diffusion will dominate and the effects of capacitance will be negligible, i.e.\ $C_\pm$ may be set to zero. If $\omega =O(D_{\pm}^{(0)}/(R_{\pm}^{(0)}C_{\pm}))$ (region (II)), the opposite is true and we may take the high $\omega$ limits of $z_{n,\pm}^{(m)}(\omega)$, which are real and given by
\begin{align}
    z_{1,\pm}^{(1)}(\omega) &\to  \pm R_{\pm}^{(0)},\qquad \omega \to \infty, \label{eq:z11_pm_high_omega}\\
    z_{2,\pm}^{(2)}(\omega) &\to  (\beta_{\pm}-1/2)(R_{\pm}^{(0)})^2,\qquad \omega \to \infty, \label{eq:z22_pm_high_omega}\\
    z_{0,\pm}^{(2)}(\omega) &\to  2(\beta_{\pm}-1/2)(R_{\pm}^{(0)})^2,\qquad \omega \to \infty.\label{eq:z02_pm_high_omega}
\end{align}
Then approximations of (\ref{eq:Z11_pm})-(\ref{eq:Z02_pm}) in both regions take the form

\begin{align}
Z_{1,\pm}^{(1)}\sim\,\, & 
    \begin{dcases}
        z_{1,\pm}^{(1)}(\omega) & \text{(I), }\omega=O(D_{\pm}^{(0)}), \\
        \frac{\pm R_{\pm}^{(0)}}{1+R_{\pm}^{(0)}C_{\pm}\mathrm{i}\omega} &  \text{(II), }\omega =O\left(\frac{D_{\pm}^{(0)}}{R_{\pm}^{(0)}C_{\pm}}\right),\\
    \end{dcases}
    \label{eq:Z11_pm_regions}
\end{align}
\begin{align}
Z_{2,\pm}^{(2)}\sim\,\, & 
    \begin{dcases}
        z_{2,\pm}^{(2)}(\omega) & \text{(I), }\omega=O(D_{\pm}^{(0)}), \\
        \frac{(\beta_{\pm}-1/2)(R_{\pm}^{(0)})^2}{\left[1+2R_{\pm}^{(0)}C_{\pm}\mathrm{i}\omega \right]\left[1+R_{\pm}^{(0)}C_{\pm}\mathrm{i}\omega \right]^{2}}, &  \text{(II), }\omega =O\left(\frac{D_{\pm}^{(0)}}{R_{\pm}^{(0)}C_{\pm}}\right),\\
    \end{dcases}
    \label{eq:Z22_pm_regions}
\end{align}
\begin{align}
Z_{0,\pm}^{(2)}\sim\,\, & 
\begin{dcases}
        z_{0,\pm}^{(2)}(\omega) & \text{(I), }\omega=O(D_{\pm}^{(0)}), \\
        \frac{2(\beta_{\pm}-1/2)(R_{\pm}^{(0)})^2}{\left|1+R_{\pm}^{(0)}C_{\pm}\mathrm{i}\omega \right|^{2}}, &  \text{(II), }\omega =O\left(\frac{D_{\pm}^{(0)}}{R_{\pm}^{(0)}C_{\pm}}\right),\\
    \end{dcases}
    \label{eq:Z02_pm_regions}
\end{align}

\begin{figure}
    \centering
    \includegraphics[width=0.5\textwidth]{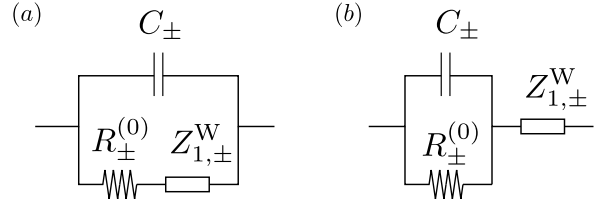}
    \caption{Equivalent circuit diagrams representing the linear response of each electrode, but corresponding to different impedance formulae. (a) Formula (\ref{eq:Z11_pm}), with no assumption on the diffusion and capacitance timescales. (b) Composite (approximate) formula (\ref{eq:Z11_pm_comp}), where the timescales are assumed well-separated and the Warburg impedance acts in series with the kinetics.} 
    \label{fig:circuit_diagram_full_vs_composite}
\end{figure}

There is a narrow region between Region I and region II where both capacitance and diffusive effects act with similar small strength. This narrow region occurs for frequencies 
\begin{align}
\omega &\approx \left(\frac{-{U_{\pm}^{(0)}}' \xi_{\pm}\sqrt{D_{\pm}^{(0)}}}{\left(R_{\pm}^{(0)}\right)^2 C_{\pm}}\right)^{2/3}.
\label{eq:omega_overlap}
\end{align}
We can now create an approximation valid for all the frequencies by
forming a composite approximation made by summing the region I and II
approximations and subtracting the leading order solution in the
overlap region. This overlap solution is found by taking the $\omega
\to \infty$ limit from region I (see
(\ref{eq:z11_pm_high_omega})-(\ref{eq:z02_pm_high_omega})) or the
$\omega \to 0$ limit from region II (at the fundamental frequency,
this is simply the charge transfer resistance $R_{\pm}^{(0)}$). The
resulting composite solutions are 
\begin{align}
Z_{1,\pm}^{(1),\mathrm{comp}} &\sim \pm\left[ \frac{R_{\pm}^{(0)}}{1+R_{\pm}^{(0)}C_{\pm}\mathrm{i}\omega} + Z^{\mathrm{W}}_{1,\pm}(\omega)\right], \label{eq:Z11_pm_comp}\\
Z_{2,\pm}^{(2),\mathrm{comp}} &\sim \frac{ Z^{\mathrm{kin}}_{\mathrm{asym},\pm}}{\left[1+2R_{\pm}^{(0)}C_{\pm}\mathrm{i}\omega \right]\left[1+R_{\pm}^{(0)}C_{\pm}\mathrm{i}\omega \right]^{2}} + Z^{\mathrm{kin}}_{\mathrm{CD},\pm} + Z^{\mathrm{OCP}}_{\pm} + Z^{\mathrm{W}}_{2,\pm}, \label{eq:Z22_pm_comp}\\
Z_{0,\pm}^{(2),\mathrm{comp}} &\sim  
\frac{2Z^{\mathrm{kin}}_{\mathrm{asym},\pm}}{\left|1+R_{\pm}^{(0)}C_{\pm}\mathrm{i}\omega \right|^{2}} + 2\mathrm{Re}\left[
     Z^{\mathrm{kin}}_{\mathrm{CD},\pm} +
    Z^{\mathrm{OCP}}_{0,\pm} + 
    Z^{\mathrm{W}}_{0,\pm}
    \right].\label{eq:Z02_pm_comp}
\end{align}
Notice that these are close in form to the expressions (\ref{eq:z_11_pm}), (\ref{eq:z22_pm}) and (\ref{eq:z02_pm}), but with capacitance modifying only the first term. Looking at the linear impedance (\ref{eq:Z11_pm_comp}), it now appears that the Warburg term is no longer in \emph{parallel} with the capacitance (as in Fig.\ \ref{fig:circuit_diagram_full_vs_composite}$(a)$) but in \emph{series} (as in Fig.\ \ref{fig:circuit_diagram_full_vs_composite}$(b)$). Since they each dominate at different frequency scales, the diffusion process interacts minimally with the kinetics. The same is true for the nonlinear impedances (\ref{eq:Z22_pm_comp})-(\ref{eq:Z02_pm_comp}), where the terms that depend on the diffusion process separate out, leaving only the constant kinetic term $Z^{\mathrm{kin}}_{\mathrm{asym},\pm}$ (due to reaction asymmetry) in ``parallel" with the capacitance. This simplifies the structure and interpretation of impedances, since the contribution of each nonlinearity can be more easily assessed.

\section{Methods}
\label{sec:Methods}

Having introduced the model and analysed it in the frequency domain,  we now detail the numerical, experimental and parameterization methods employed in the remainder of the paper to produce the results given in the Results section. 

\subsection{Numerical solution in the time domain for generation of synthetic data}
\label{sec:Numerical-methods}

Here we describe the numerical methods for solving the SPM
in the time domain and the procedure for generating synthetic
impedance data from a known set of parameter values---this is useful for parameter
identifiability and error sensitivity analysis (the Results section) because `ground truth' parameters are known exactly. Synthetic data was generated by solving
the full nonlinear model equations (\ref{eq:c_equation})-(\ref{eq:R})
in the time domain numerically given a sinusoidal input current of dimensional
frequency $\omega^*$ and amplitude $2\hat{I}^*$ (or peak-to-peak
amplitude $4\hat{I}^*$). This was then converted to the frequency
domain with a Fourier transform. The radial
dimension within the electrode particles was discretized using a
finite volume scheme, and the time integration was performed using the
adaptive explicit stiff ODE solver \texttt{ode15s} in MATLAB. To ensure sufficient relaxation of
initial transients and convergence of the output voltage to a
periodic profile, 20 periods were simulated and only the final 2 were
selected to be fast Fourier transformed (FFT) with 20 equispaced
samples per period. This procedure was performed for 30 frequencies
logarithmically spaced from $\omega^*=$ \SI{e-4}{Hz} to \SI{e2}{Hz}, i.e.\ 5 frequencies per decade. The parameter set used corresponds to a LCO/LiC$_6$ cell, with values modified from
\cite{Marquis2019,Murbach2018}, summarized in Tables
S1, S2 of the supplementary material. 
Simulations were performed at three DoDs
(30\%, 50\% and 70\%) with a fixed current amplitude of \SI{50}{mA}. 

Zero-mean Gaussian noise was added to the simulated voltages in the frequency domain ($n=1,2$)
\begin{align}
    \widehat{V}_{n}^{*,\mathrm{data}} &= \widehat{V}_{n}^{*} + \varepsilon_n^*, & 
    \varepsilon_n^* &= \varepsilon_{n,r}^* + \mathrm{i}\varepsilon_{n,i}^*,\quad \mbox{where } \varepsilon_{n,r}^*,\varepsilon_{n,i}^* \sim N(0,\sigma_n^{*2}), \label{eq:synthetic-data-noise}
\end{align}
and the standard deviations were chosen to give similar signal-to-noise ratios to those seen in our experiments, i.e.\ $\sigma_1^*=\sigma_2^*=10^{-7}$ V. The synthetic impedances were calculated from
\begin{align}
    Z_{n}^{*,\mathrm{data}} &= \frac{\widehat{V}_{n}^{*,\mathrm{data}}}{(\hat{I}^*)^{n}}, \qquad n=1,2.
\end{align}


\subsection{Experimental measurements}
\label{sec:Experimental-methods}

Experiments were undertaken using a bespoke test rig to create an NLEIS dataset to parameterize the model. All measurements were collected in the time domain, then post-processed in MATLAB. A bipolar power source/sink (Kikusui PBZ60-6.7) was connected in series with an NMC/graphite pouch cell (Kokam SLPB533459H4, \SI{740}{mAh} \cite{KokamDataSheet}) and a \SI{0.1}{\ohm} precision shunt resistor as shown in Fig.\ S1 in the supplementary material. 
Voltage measurements were made across the test battery and the shunt respectively using a 24-bit analog-to-digital converter (National Instruments 9239). Two cells were tested.

Prior to NLEIS experiments, the discharge capacity of each cell was measured at \SI{19}{\celsius} and found to be \SI{720}{mAh} and \SI{729}{mAh} respectively. For capacity measurements, first a constant-current charge of \SI{500}{mA} to \SI{4.2}{V} was applied followed by a constant-voltage charge until the current dropped below \SI{50}{mA}. Then, after a \SI{30}{\minute} rest period, a constant-current discharge at \SI{500}{mA} to a voltage of \SI{2.7}{V} was conducted followed by a constant-voltage discharge to a current below \SI{50}{mA}.  

Nonlinear EIS measurements were collected at 10\% DoD to 90\% DoD (equivalently 90\% SOC to 10\% SOC), in steps of 10\%. At each DoD, zero-mean sinusoidal currents of \SI{1}{A} peak-to-peak were applied and voltage measurements taken at frequencies from \SI{0.1}{Hz} to \SI{10}{kHz}, with 10 measurements per decade. The length of the measurement was taken as the larger of \SI{2}{s} or 20 cycles at each frequency. For more details on the procedures, including thermal limits, data acquisition and processing, see section S4 of the supplementary material. 

The measured voltage and current signals were each multiplied by a Hann window function, and then fast-Fourier transformed using MATLAB's built-in FFT function, giving the components 
$\widehat{V}_{n}(\omega,\hat{I})$ of the voltage as 
per (\ref{eq:Fourier-series-V}).

\subsection{OCP functions and electrode balancing}

\begin{figure}
    \centering
    \includegraphics[width=1.\textwidth]{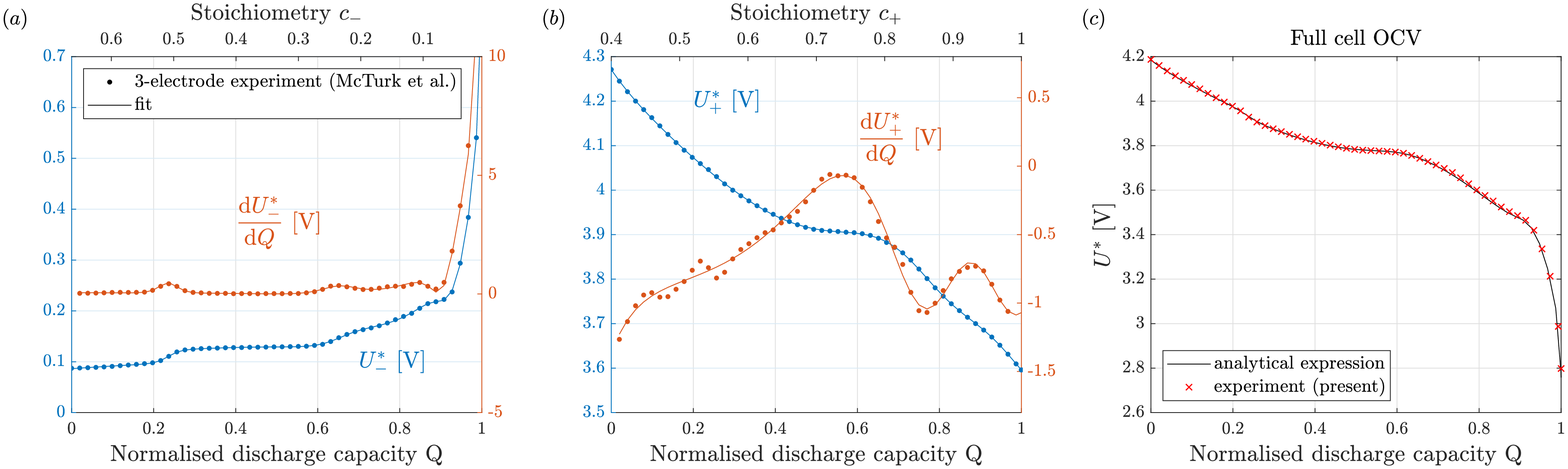}\\

    \caption{$(a)$ Negative electrode OCP, $(b)$ positive electrode OCP, and $(c)$ full-cell OCV for a Kokam SLPB533459H4 \SI{740}{mAh} pouch cell. Dot markers are 3-electrode data from \cite{McTurk2015}, solid lines are  analytical expressions (Eqs.\ (S4.1), (S4.2) in supplementary material) 
    that were fitted to individual electrode OCP data. Panel $(c)$ compares these fits to the measured full-cell OCV for the present cells from which NLEIS data was collected.}
    \label{fig:kokam_OCPs}
\end{figure}

Individual electrode OCPs $U_\pm^*$ for the Kokam SLPB533459H4 cells were determined from 3-electrode measurements by McTurk et al.\ \cite{McTurk2015}, who inserted a minimally invasive lithium reference electrode into cells of the same model. This data for $U_\pm^*$ is given in terms of normalised discharge capacity $Q=\text{DoD}/100$ with $Q \in [0, 1]$. For use with our model, analytical expressions (S4.5 of supplementary material) 
were fitted to the data, shown in Figs.\ \ref{fig:kokam_OCPs}($a$)-($b$). Fig.\ \ref{fig:kokam_OCPs}($c$) demonstrates that the resulting full-cell OCV agrees with the measured OCV of the cells used in this paper. %

To balance the electrodes, the stoichiometry limits $c_\pm^{0\%}$ and $c_\pm^{100\%}$ corresponding to 0\% and 100\% DoD ($Q=0$ and 1) were chosen using knowledge of the electrode chemistries. The resulting stoichiometry limits (and values of $\xi_\pm$ that follow from (\ref{eq:ksi_stoichiometry_relation})) are given later in Table \ref{tab:Param-groups-validation}. Then $U_\pm^*$ is expressed in terms of $c_\pm$ using (\ref{eq:c-Q_relation}). When it is required, the first derivative of $U_\pm^*$ is calculated from the analytical fits, but the second derivative is calculated numerically from the OCP data. The OCP expressions and further details of the fitting procedures and electrode balancing are given in section S4.5 of the supplementary material. 

\subsection{Parameter estimation}
\label{sec:Parameter-estimation-methods}
To identify parameters and analyse their uniqueness, we consider fitting the (dimensional) model impedances $Z_n^{*,(n)}=\Phi^* Z_n^{(n)} / (J^*)^n$ to measured or simulated data $Z_n^{*,\mathrm{data}}$ across a range of frequencies at a given DoD. The parameter estimation problem takes the form
\begin{align}
    \hat{\bm{\theta}} &=\underset{\bm{\theta}}{\mathrm{argmin}}\,l(\bm{\theta}), \label{eq:theta_min_problem}
\end{align}
where $\bm{\theta}$ is the vector of parameters to fit, and $l$ is a loss function describing the goodness of fit. It is not obvious what loss function to use for NLEIS data since more than one impedance (i.e.\ $Z_1^{*}$ and $Z_2^{*}$) is fitted simultaneously, and higher harmonics are typically smaller in magnitude but also more sensitive to some parameters. A straightforward choice of the total squared error across both harmonics would not account for this difference in scale or magnitude. Therefore, the relative weighting of the error across each harmonic merits consideration. 

We take the approach of maximum likelihood estimation (MLE), assuming zero-mean Gaussian measurement noise added to the voltage harmonics, as in (\ref{eq:synthetic-data-noise}). Notably, we assume in general that the noise variances of the first and second harmonics are not the same. The details can be found in section S5 of the supplementary material, 
and the resulting MLE for $\bm{\theta}$ given NLEIS data $Z_{1,j}^{*,\mathrm{data}}$, $Z_{2,j}^{*,\mathrm{data}}$ at frequencies $\omega_j, j=1,2,\ldots,N_\omega$, is reduced to
\begin{align}
    \hat{\bm{\theta}}_{\mathrm{NLEIS}} &= \underset{\bm{\theta}}{\mathrm{argmin}}\,(l_1(\bm{\theta}) + l_2(\bm{\theta})),
\end{align}
where
\begin{align}
    l_{1}(\bm{\theta}) &= 
    \log\left(\sum_{j=1}^{N_\omega}\left|Z_1^{*(1)}(\omega_j;\bm{\theta})-Z_{1,j}^{*,\mathrm{data}} \right|^2\right),\\
    l_{2}(\bm{\theta}) &= 
    \log\left(\sum_{j=1}^{N_\omega}\left|Z_2^{*(2)}(\omega_j;\bm{\theta})-Z_{2,j}^{*,\mathrm{data}} \right|^2\right),
\end{align}
are the (negative) log-likelihoods, up to a scaling factor, from each harmonic. We will refer to their sum as the total loglikelihood $l_{12}(\bm{\theta}) = l_1 + l_2$. The form of $l_{12}$, which is the sum of the logarithm of the squared error of each harmonic, originates from the MLE process and has the advantage of being independent of the scale of each harmonic, as desired. That is, rescaling $(Z_{n}^{*n},Z_{n}^{*,\mathrm{data}}) \mapsto (aZ_{n}^{*n},aZ_{n}^{*,\mathrm{data}})$ for any $a$ only changes $l_{12}$ by an additive constant, leaving $\mathrm{argmin}_{\bm{\theta}}\:l_{12}$ unchanged. %
Given only linear EIS data, consisting of $Z_{1,j}^{*,\mathrm{data}}$, the corresponding MLE is
\begin{align}
    \hat{\bm{\theta}}_{\mathrm{EIS}} &= \underset{\bm{\theta}}{\mathrm{argmin}}\,l_1(\bm{\theta}),
\end{align}
where the loss function is simply $l_1$.

The minimization problem (\ref{eq:theta_min_problem}) was solved using nonlinear optimization routines (local and global) in MATLAB employing bound constraints as described in section S5 of the supplementary material. 


\section{Results}
\label{sec:Results}

\subsection{Analysis of model impedances}
\label{sec:Results-model-impedances}
We begin with an illustration of the form of the exact model impedances, up to second order, given by the formulae (\ref{eq:Z11_pm}), (\ref{eq:Z22_pm}) and (\ref{eq:Z02_pm}). The dependence of $Z_{2,\pm}^{(2)}$ in particular on the various model parameters will be demonstrated. Then, the simplified (composite) formulae (\ref{eq:Z11_pm_comp}), (\ref{eq:Z22_pm_comp}) and (\ref{eq:Z02_pm_comp}) will be compared to the aforementioned exact results.

Typical Nyquist plots of the impedances $Z_{1,+}^{(1)}$, $Z_{2,+}^{(2)}$ and $Z_{0,+}^{(2)}$ are shown in Fig.\ \ref{fig:results_Z11_Z22_Z02}, using the parameter values in Tables S1, S2 in the supplementary material. 
(Only the positive electrode impedances are shown---negative electrode ones have a similar structure.) In Fig.\ \ref{fig:results_Z11_Z22_Z02}$(a)$, $Z_{1,+}^{(1)}$ consists of a high frequency semi-circular kinetic arc, and a low frequency diffusive tail with a capacitive effect ($Z_{1,+}^{(1)}=O(1/\omega)$) as $\omega \to 0$. That is, the same as the well-understood Randles circuit with finite-space diffusion. 
In Fig.\ \ref{fig:results_Z11_Z22_Z02}$(b)$, we see the second harmonic
impedance $Z_{2,+}^{(2)}$ also consists of a high frequency kinetic
arc and diffusive tail, but this arc appears ``spiral'' in
nature rather than circular. Equation (\ref{eq:Z22_pm_regions})
(region II) describes this and, excluding the factor of
$(1+2R_\pm^{(0)}C_\pm \mathrm{i} \omega)$ (which affects mainly the
shape close to the origin), shows that $Z_{2,+}^{(2)}$ 
is proportional to the square of the semi-circular kinetics in $Z_{1,\pm}^{(1)}$, resulting in a \emph{cardioid}-like shape. Lastly, the second-order correction at the zeroth harmonic, $Z_{0,+}^{(2)}$, contains similar model information as $Z_{2,+}^{(2)}$ but with no imaginary component. In Fig.\ \ref{fig:results_Z11_Z22_Z02}$(c)$, we see $Z_{0,+}^{(2)}=O(\omega^{-2})$ at high frequencies (square of double-layer capacitive effects) but also begin to see this behaviour at low frequencies (square of electrode ``differential capacitive" effects). However, in intermediate regions $Z_{0,+}^{(2)}$ appears relatively flat and the various effects are difficult to delineate.

\begin{figure}
    \centering
    \includegraphics[width=1.0\textwidth]{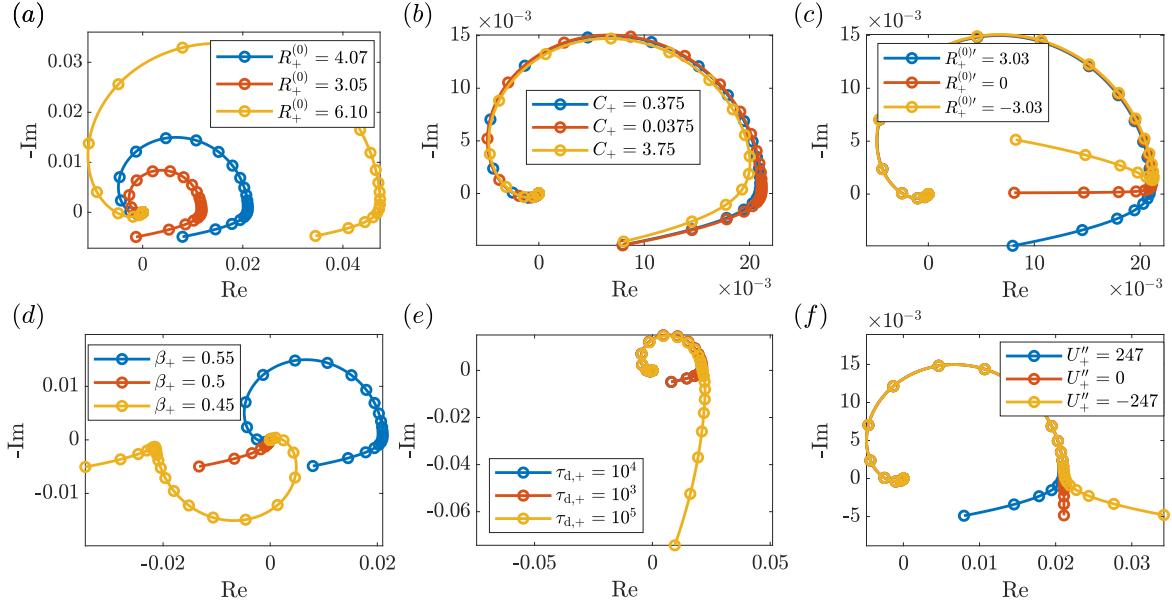}\\
    \caption{Dependence of the second harmonic impedance $Z_{2,+}^{(2)}$ on quantities $(a)$ $R_+^{(0)}$, $(b)$ $C_+$, $(c)$ $R_+^{(0)\prime}$, $(d)$ $\beta_+$, $(e)$ $\tau_{\mathrm{d},+}$, $(f)$ $U_+^{(0)\prime\prime}$. For each, we used the value in Table S2 
    (at $\mathrm{DoD}=30\%$), shown in blue, and 2 other representative values.}
    \label{fig:results_Z22_param_dependence}
\end{figure}

Next, we demonstrate in Fig.\ \ref{fig:results_Z22_param_dependence} how the various features of $Z_{2,+}^{(2)}$ (and therefore $Z_{2,-}^{(2)}$) depend on quantities in the model. Starting from the values in Tables S1, S2 
and at 30\% DoD, we varied $R_+^{(0)}$, $C_+$, $R_+^{(0)\prime}$, $\beta_+$, $\tau_{\mathrm{d},+}$, and $U_+^{(0)\prime\prime}$ across a representative range (considering both positive and negative values for the derivatives $R_+^{(0)\prime}$, $U_+^{(0)\prime\prime}$). From Fig.\ \ref{fig:results_Z22_param_dependence} we observe that:
\begin{enumerate}
    \item[$(a)$] $R_+^{(0)}$ affects the size of the kinetic arc (its ``width" on the real axis is $Z^{\mathrm{kin}}_{\mathrm{asym},+} = (\beta_{+}-1/2)(R_{+}^{(0)})^2$);
    \item[$(b)$] $C_+$ affects the frequency dependence of the kinetic arc, but not its shape (just as for $Z_{1,+}^{(1)}$);
    \item[$(c)$] $R_+^{(0)\prime}$ affects the initial direction and magnitude of the diffusion tail (due to $Z^{\mathrm{kin}}_{\mathrm{CD},+}$) emanating from the kinetic arc;
    \item[$(d)$] $\beta_+$ affects the size (via the magnitude $|\beta_+-1/2|$) and orientation (via $\mathrm{sgn}(\beta_+-1/2$)) of the kinetic arc (see point $(a)$ above). Important to note that as $\beta_+\to 1/2$ the totality of the arc shrinks to the origin. Thus $\beta_+\neq 1/2$ is \emph{necessary} to observe an arc at all (this was first pointed out by Murbach et al.\ \cite{Murbach2017});
    \item[$(e)$] $\tau_{\mathrm{d},+}$ affects the length and shape of the diffusion tail (via the terms $Z^{\mathrm{kin}}_{\mathrm{CD},+}$ and $Z^{\mathrm{OCP}}_{+}$);
    \item[$(f)$] $U_+^{(0)\prime\prime}$ affects whether the diffusion tail ultimately diverges to $\mathrm{Re}=-\infty$ (when $U_+^{(0)\prime\prime}>0$) or $\mathrm{Re}=+\infty$ (when $U_+^{(0)\prime\prime}<0$) as $\omega\to 0$.
\end{enumerate}
It is clear that the possible behaviour of $Z_{2,+}^{(2)}$ is much more varied than $Z_{1,+}^{(1)}$, with visible signatures of many quantities in the model that are not visible with EIS, e.g.\ $\beta_+$, $R_+^{(0)\prime}$, and $U_+^{(0)\prime\prime}$. In addition, dependence of $Z_{2,+}^{(2)}$ on $R_+^{(0)}$, $C_+$ and $\tau_{\mathrm{d},+}$ gives further information that may be used to improve their identifiability over traditional linear EIS.

\begin{figure}
    \centering
    \includegraphics[width=0.9\textwidth]{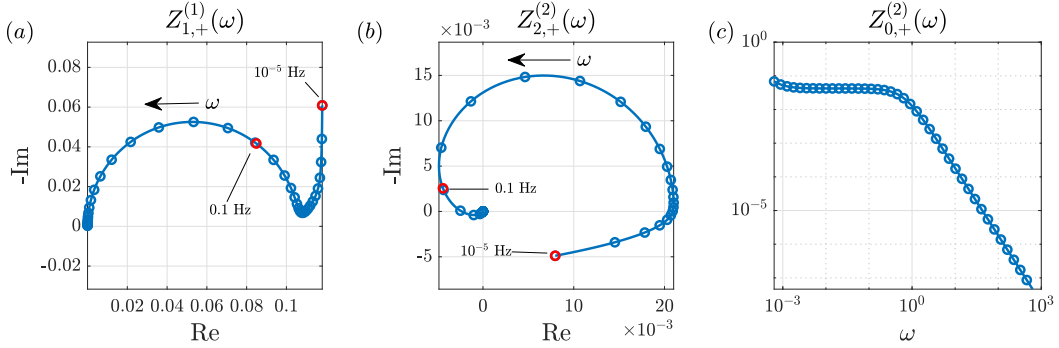}\\
    \caption{Nyquist plot of the fundamental (linear) impedance $Z_{1,+}^{(1)}$, and the second-order nonlinear impedances $Z_{2,+}^{(2)}$ (second harmonic) and $Z_{0,+}^{(2)}$ (zero frequency mode), as calculated from formulae (\ref{eq:Z11_pm}), (\ref{eq:Z22_pm}) and (\ref{eq:Z02_pm}). Parameters in Table S2 
    at $\mathrm{DoD}=30\%$ (i.e.\ $c_+=0.80$). Frequency range is $10^{-4}\leq \omega/2\pi \leq 10^{2}$ [Hz], with 5 values (marked) per decade. Only the impedances for positive electrode shown.}
    \label{fig:results_Z11_Z22_Z02}
\end{figure}

Finally, we remark on the accuracy of the simplified (composite) expressions (\ref{eq:Z11_pm_comp}), (\ref{eq:Z22_pm_comp}) and (\ref{eq:Z02_pm_comp}), which assume that the capacitive timescale $R_\pm^{(0)}C_\pm$ is much shorter than the diffusive one, $\tau_{\mathrm{d},\pm}=1/D_\pm^{(0)}$. Comparing these to the exact impedances for a full-cell, we find them practically indistinguishable\textemdash across all frequencies and each harmonic\textemdash with mean relative errors less than 0.03\% for the fundamental and 0.64\% for the second harmonic. Details are given in section S6 of the supplementary material. 
Consequently, these composite expressions are a useful and simple substitute for the full expressions in practice.

\subsection{Structural parameter identifiability}
\label{sec:Results-structural-ident}

Nine dynamical parameter groups (\ref{eq:dynamical-params}) are sufficient
to fully parameterize the nonlinear SPM
(\ref{eq:c_equation})-(\ref{eq:R}). However, these parameters are not
necessarily able to be determined uniquely (or at all) from the model
structure. This problem is referred to as structural parameter
identifiability \cite{Bellman1970}, and we will consider the
improvements that NLEIS provides relative to linear EIS
\cite{Bizeray2019}. 

Structural identifiability is typically defined for linear systems, but here we will use a nonlinear extension. Consider a set of nonlinear impedances $\{Z_1^{(1)}(\omega,\bm{\theta}),Z_2^{(2)}(\omega,\bm{\theta}),\ldots\}$, characterising a nonlinear model in the frequency domain, for parameters $\bm{\theta}\in \Theta$. If the system of equations with $\bm{\theta}, \tilde{\bm{\theta}}\in \Theta$
\begin{align}
    Z_1^{(1)}(\omega,\bm{\theta}) & = Z_1^{(1)}(\omega,\tilde{\bm{\theta}}), \label{eq:struct-ident-1}\\
    Z_2^{(2)}(\omega,\bm{\theta}) & = Z_2^{(2)}(\omega,\tilde{\bm{\theta}}), \label{eq:struct-ident-2}\\
    & \vdots \nonumber
\end{align}
for all $\omega \in \mathbb{R}$, has: (i) a unique solution for $\bm{\theta}$, then the model is globally identifiable; (ii) a finite number of solutions for $\bm{\theta}$, it is locally identifiable; (iii) an infinite number of solutions for $\bm{\theta}$, it is unidentifiable.

\subsubsection{Single electrode}

First, we will consider the parameter identifiability of a single electrode (positive or negative) and 
restrict ourselves to the accurate simplified formulae (\ref{eq:Z11_pm_comp})-(\ref{eq:Z22_pm_comp}). For a single ($\pm$) electrode, the relevant unknown parameters are $\chi_\pm, C_\pm, \beta_\pm, \tau_{\mathrm{d},\pm}$, but for convenience we will consider \begin{align}
    R_\pm^{(0)} &= \frac{2\chi_{\pm}}{(c_\pm^{(0)})^{\beta_{\pm}}(1-c_\pm^{(0)})^{1-\beta_{\pm}}}, \label{eq:R_pm_0}
\end{align}
in place of $\chi_\pm$, and discuss identifiability of $\chi_\pm$ subsequently.

For a single electrode, if only $Z_{1,\pm}^{(1)}$ is given, (\ref{eq:Z11_pm_comp}) shows that $R_\pm^{(0)}, C_\pm$ have unique solutions (globally identifiable), and so does $\tau_{\mathrm{d},\pm}$ provided $U_\pm^{(0)\prime}\neq 0$. If $U_\pm^{(0)\prime}=0$, i.e.\ if the OCP slope is exactly zero, the diffusive contribution to the impedance vanishes and $\tau_{\mathrm{d},\pm}$ is unidentifiable\textemdash as per Bizeray et al.\ \cite{Bizeray2019} for Fickian diffusion. However, the charge transfer coefficient $\beta_\pm$ does not appear in $Z_{1,\pm}^{(1)}$ and is therefore always unidentifiable. In this case $\chi_\pm$ cannot be identified from (\ref{eq:R_pm_0}), leaving only $R_\pm^{(0)}$ identified. 

However, if the second harmonic is also considered, i.e.,
$\{Z_{1,\pm}^{(1)},Z_{2,\pm}^{(2)}\}$ is given, then more information
on each parameter is provided. In particular, $\beta_\pm$ (and hence
$\chi_\pm$) becomes globally identifiable since it appears explicitly
in $Z_{2,\pm}^{(2)}$. Regarding the diffusion timescale
$\tau_{\mathrm{d},\pm}$: if $U_\pm^{(0)\prime}=0$ (hence
$\tau_{\mathrm{d},\pm}$ unidentifiable from $Z_{1,\pm}^{(1)}$ alone)
but $U_\pm^{(0)\prime\prime}\neq 0$, then the terms
$Z_{\pm}^{\mathrm{OCP}}$ and $Z_{\pm,2}^{\mathrm{W}}$ in
$Z_{2,\pm}^{(2)}$ do not vanish and in fact make
$\tau_{\mathrm{d},\pm}$ identifiable. However, if
$U_\pm^{(0)\prime\prime}= 0$ also, then $\tau_{\mathrm{d},\pm}$
remains unidentifiable. 

\subsubsection{Full two-electrode cell}

Given the impedances $\{Z_{1,\pm}^{(1)},Z_{2,\pm}^{(2)}\}$ for a
single electrode, the full cell impedances are given by
(\ref{eq:Z00})-(\ref{eq:Z02}) and they inherit each electrode's
structural identifiability since no parameters appear in impedances
from \emph{both} electrodes. However, they may have additional
identifiability issues arising from cell symmetries, i.e., ``electrode
swapping'', or inability to decide whether a parameter refers to the
positive or negative electrode. 

If the full-cell linear impedance $Z_{1}^{(1)} = Z_{1,+}^{(1)} -
Z_{1,-}^{(1)} + R_s$ is given, the series resistance $R_s$ is unique
(globally identifiable) and can be extracted from the high frequency
limit $\omega \to \infty$, where $Z_{1}^{(1)} \to R_s$. However, the
identifiable kinetic parameters now have two solutions: if 
$(R_-^{(0)}, C_-, R_+^{(0)}, C_+)$ is a solution then so is
$(R_+^{(0)}, C_+, R_-^{(0)}, C_-)$. Hence they are only
\emph{locally} identifiable. A similar symmetry exists for the
diffusion timescales $\tau_{\mathrm{d},\pm}$ but only for special
values of the OCP slopes. Substituting (\ref{eq:Z11_pm_comp}) into
condition (\ref{eq:struct-ident-1}), and keeping only the Warburg terms,
gives (recall $D_{\pm}^{(0)} = -U_{\pm}^{(0)\prime}c_{\pm}^{(0)} /
\tau_{\mathrm{d,\pm}}$): 
\begin{align}
    \frac{\xi_{+}U_{+}^{(0)\prime}}{D_{+}^{(0)}}H_{1}\left(\frac{\omega}{D_{+}^{(0)}}\right)
    + \frac{\xi_{-}U_{-}^{(0)\prime}}{D_{-}^{(0)}}H_{1}\left(\frac{\omega}{D_{-}^{(0)}}\right) 
    & = 
    \frac{\xi_{+}U_{+}^{(0)\prime}}{\tilde{D}_{+}^{(0)}}H_{1}\left(\frac{\omega}{\tilde{D}_{+}^{(0)}}\right)
    + \frac{\xi_{-}U_{-}^{(0)\prime}}{\tilde{D}_{-}^{(0)}}H_{1}\left(\frac{\omega}{\tilde{D}_{-}^{(0)}}\right).  \label{eq:tau_d_ident_1}   
\end{align}
If the OCP slopes are nonzero and satisfy $\xi_{+}U_{+}^{(0)\prime} = \xi_{-}U_{-}^{(0)\prime}$, this reduces to
\begin{align}
    \frac{1}{D_{+}^{(0)}}H_{1}\left(\frac{\omega}{D_{+}^{(0)}}\right)
    + \frac{1}{D_{-}^{(0)}}H_{1}\left(\frac{\omega}{D_{-}^{(0)}}\right) 
    & = 
    \frac{1}{\tilde{D}_{+}^{(0)}}H_{1}\left(\frac{\omega}{\tilde{D}_{+}^{(0)}}\right)
    + \frac{1}{\tilde{D}_{-}^{(0)}}H_{1}\left(\frac{\omega}{\tilde{D}_{-}^{(0)}}\right),  \label{eq:tau_d_ident_2}
\end{align}
The left hand side has the symmetry $D_-^{(0)} \mapsto D_+^{(0)}$, $D_+^{(0)} \mapsto D_-^{(0)}$, meaning there are always two solutions for $(D_+^{(0)}, D_-^{(0)})$ and hence $(\tau_{\mathrm{d,+}}, \tau_{\mathrm{d,-}})$. This symmetry in terms of $\tau_{\mathrm{d,\pm}}$ is
\begin{align}
    \tau_{\mathrm{d,+}} & \mapsto \frac{U_{+}^{(0)\prime}c_{+}^{(0)}}{U_{-}^{(0)\prime}c_{-}^{(0)}} \tau_{\mathrm{d,-}}, &
    \tau_{\mathrm{d,-}} & \mapsto \frac{U_{-}^{(0)\prime}c_{-}^{(0)}}{U_{+}^{(0)\prime}c_{+}^{(0)}} \tau_{\mathrm{d,+}}.
\end{align}
This electrode symmetry of diffusion timescales in $Z_{1}^{(1)}$ was pointed out by Bizeray \emph{et al.} \cite{Bizeray2019}, which we extend here to the case of nonlinear diffusion.

If the full-cell second harmonic is also considered, i.e.,
$\{Z_{1,}^{(1)},Z_{2}^{(2)}\}$ is given, the above symmetries are removed in
most cases. Substituting (\ref{eq:Z22_pm_comp}) into condition
(\ref{eq:struct-ident-2}), as $H_1$, $H_1^2$ and $H_2$ are distinct
functions of $\omega$, the diffusion symmetry $D_-^{(0)} \mapsto
D_+^{(0)}$, $D_+^{(0)} \mapsto D_-^{(0)}$ only holds (using similar
arguments to (\ref{eq:tau_d_ident_1})-(\ref{eq:tau_d_ident_2})) if
$\xi_{+}^2U_{+}^{(0)\prime\prime} = -\xi_{-}^2
U_{-}^{(0)\prime\prime}$ and $\xi_{+}^2U_{+}^{(0)\prime}/c_+^{(0)} =
-\xi_{-}^2 U_{-}^{(0)\prime}/c_-^{(0)}$. But we already require
$\xi_{+}U_{+}^{(0)\prime} = \xi_{-}U_{-}^{(0)\prime}$ from
$Z_{1}^{(1)}$ which, if $U_{\pm}^{(0)\prime} \not = 0$, reduces the latter to $\xi_{+}/c_+^{(0)} =
-\xi_{-}/c_-^{(0)}$, and this is not possible since $\xi_+$ and $\xi_-$
are both positive. If actually $U_{\pm}^{(0)\prime} = 0$, then the symmetry can persist so long as $\xi_{+}^2U_{+}^{(0)\prime\prime} = -\xi_{-}^2
U_{-}^{(0)\prime\prime}$.

If the diffusion symmetry is removed, then the terms
$Z_{\pm,\mathrm{CD}}^{\mathrm{kin}}$ (see (\ref{eq:Z_kin_CD})) prevent
the symmetry in $R_\pm^{(0)}, C_\pm$ too as they depend on both
kinetic and diffusive parameters. However, in the special case of $R_-^{(0)\prime} = R_+^{(0)\prime} = 0$, i.e.\ both exchange current densities are independent of concentration (at this DoD), then $Z_{\pm,\mathrm{CD}}^{\mathrm{kin}}$ vanishes and a symmetry in the kinetic parameters does persist, given by $R_\pm^{(0)}  \mapsto R_\mp^{(0)}$,  $C_\pm  \mapsto C_\mp, \beta_\pm  \mapsto 1-\beta_\mp$.

In summary, including the second harmonic $Z_{2}^{(2)}$, makes
$\beta_\pm$ (and hence $\chi_\pm$) globally identifiable, and improves
the identifiability of the diffusion timescales
$\tau_{\mathrm{d},\pm}$ when an OCP is flat  ($U_{\pm}^{(0)\prime}=
0$) but has nonzero curvature ($U_{\pm}^{(0)\prime\prime}\neq 0$). In
addition, it removes the electrode swapping symmetries in general,
except when both OCVs are flat ($U_+^{(0)\prime}=U_-^{(0)\prime}=0$) or
 both exchange current densities are independent of
concentration ($R_-^{(0)\prime} = R_+^{(0)\prime} = 0$).


\subsection{Parameter estimation from synthetic data}
\label{sec:Results-estimation-synthetic-data}

With the structural identifiability from the first two harmonics analysed, we move to analyse the practical identifiability, i.e.\ parameter estimation from data. In this section, we consider noisy synthetic data from a known parameter set\textemdash see the Methods section 
for the generation method and parameter values. The parameter estimation algorithm from EIS data ($Z_1^{*,\mathrm{data}}$ only) and NLEIS data ($Z_1^{*,\mathrm{data}}$ and $Z_2^{*,\mathrm{data}}$) was given in the Methods section 
and results in estimators $\hat{\bm{\theta}}_{\mathrm{EIS}}$, $\hat{\bm{\theta}}_{\mathrm{NLEIS}}$ for the parameter groups.

\begin{figure}
    \centering
    \includegraphics[width=1.\textwidth]{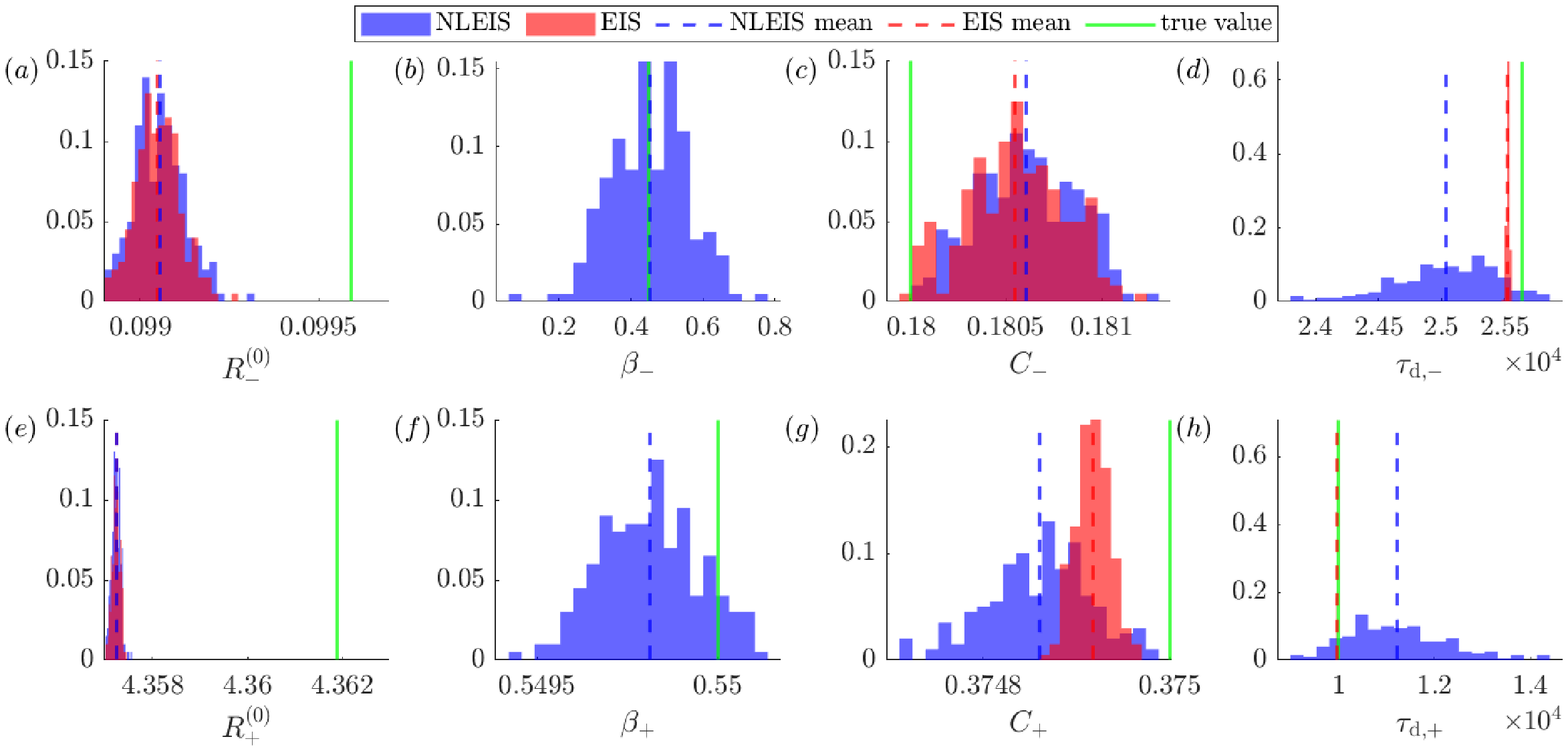}\\
    \includegraphics[width=0.82\textwidth]{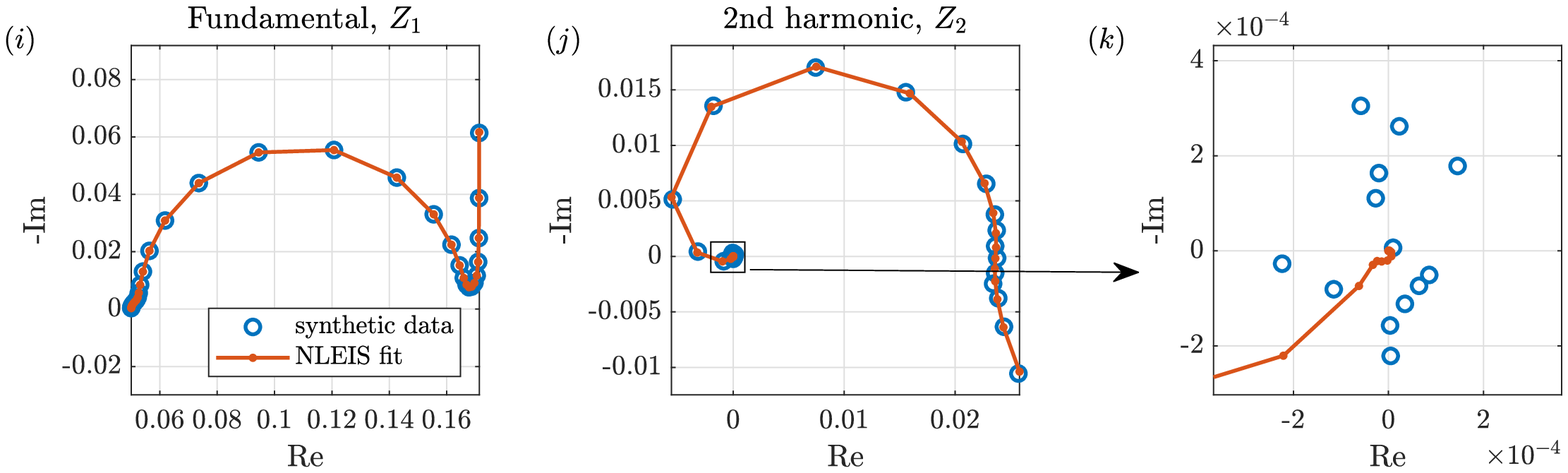}\\
    \caption{(50\% DoD) Parameter distributions (histograms) for $\hat{\bm{\theta}}_{\mathrm{NLEIS}}$ from repeated fitting to synthetic NLEIS data, showing sensitivity to error $(a)-(h)$. Distributions for fitting $\hat{\bm{\theta}}_{\mathrm{EIS}}$ to just EIS data (fundamental only) also shown. Dashed lines are mean values, with true ones shown in green. Estimates for $\beta_{\pm}$ not shown for EIS as they are not identifiable. Example impedance fit shown in $(i)$-$(k)$, including blown-up plot $(k)$ of $Z_2$ near the origin.}
    \label{fig:results_error_sensitivity_DoD50}
\end{figure}

\begin{figure}
    \centering
    \includegraphics[width=1.\textwidth]{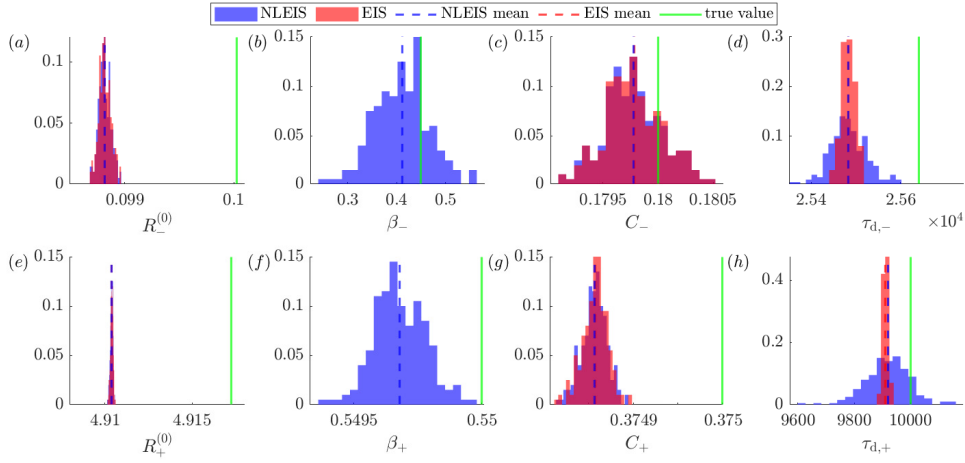}\\
    \caption{(70\% DoD) Parameter distributions (histograms) for $\hat{\bm{\theta}}_{\mathrm{NLEIS}}$ from repeated fitting to synthetic NLEIS data, showing sensitivity to error. See caption for Fig.\ \ref{fig:results_error_sensitivity_DoD50}.}
    \label{fig:results_error_sensitivity_DoD70}
\end{figure}

To test the consistency and noise sensitivity of our approach we repeatedly applied the estimation algorithm to synthetic data but each time with separate samples drawn from the noise distribution (given in
(\ref{eq:synthetic-data-noise})). This resampling, known as
``bootstrapping" \cite{BootstrappingTextbook93}, allows the construction of the distribution of
the estimators $\hat{\bm{\theta}}_{\mathrm{EIS}}$,
$\hat{\bm{\theta}}_{\mathrm{NLEIS}}$. The distributions (histograms)
and their means for 200 resamples are shown and compared in
Figs. \ref{fig:results_error_sensitivity_DoD50} (with data at 50\% DoD) and
\ref{fig:results_error_sensitivity_DoD70} (at 70\% DoD). Note that the transfer coefficient $\beta_\pm$ is unidentifiable from EIS data and thus only estimates
from NLEIS data are shown. Also, although NLEIS data can be used to estimate the typical non-dimensional charge transfer resistance $\chi_\pm$,
EIS data cannot, so to compare methods we plot the charge transfer
resistances $R_\pm^{(0)}$ instead. 

 Figs. \ref{fig:results_error_sensitivity_DoD50} and
 \ref{fig:results_error_sensitivity_DoD70} show that the mean and
 variance of the kinetic parameter estimators $R_\pm^{(0)},C_\pm$ are similar in all cases. The NLEIS-based estimator of $\beta_+$ has low variance,
 but that for $\beta_-$ has high variance. This is due to the resistance of the
 negative electrode ($R_-^{(0)}$) being much lower than that of the
 positive, hence its kinetic cardioid spiral in $Z_2^*$ is much
 smaller in magnitude, and affected more by noise\textemdash see
 Fig.\ \ref{fig:results_error_sensitivity_DoD50}$(j)$-$(k)$. The
 discrepancy between the means and the true values (i.e.\ estimator
 bias) is due to the small-current-amplitude approximation we used to
 derive the model impedances. We determined this by fitting to synthetic data generated using smaller current amplitudes and found that the bias reduced. This bias thus represents model error rather
 than noise, as the data generation used a finite amplitude. 

The diffusion timescales $\tau_{\mathrm{d},\pm}$ are more influenced
by noise, with the NLEIS estimates showing larger variance than the EIS
ones. However, this variance is comparable to the model
error---the EIS estimates of $\tau_{\mathrm{d},\pm}$ can be overly
confident (low variance), despite having a similar model error to
NLEIS (Fig.\ \ref{fig:results_error_sensitivity_DoD70} for 70\% DoD). 


\subsubsection*{Log-likelihood analysis of diffusion timescales}

\begin{figure}
    \centering
    \includegraphics[width=1.1\textwidth]{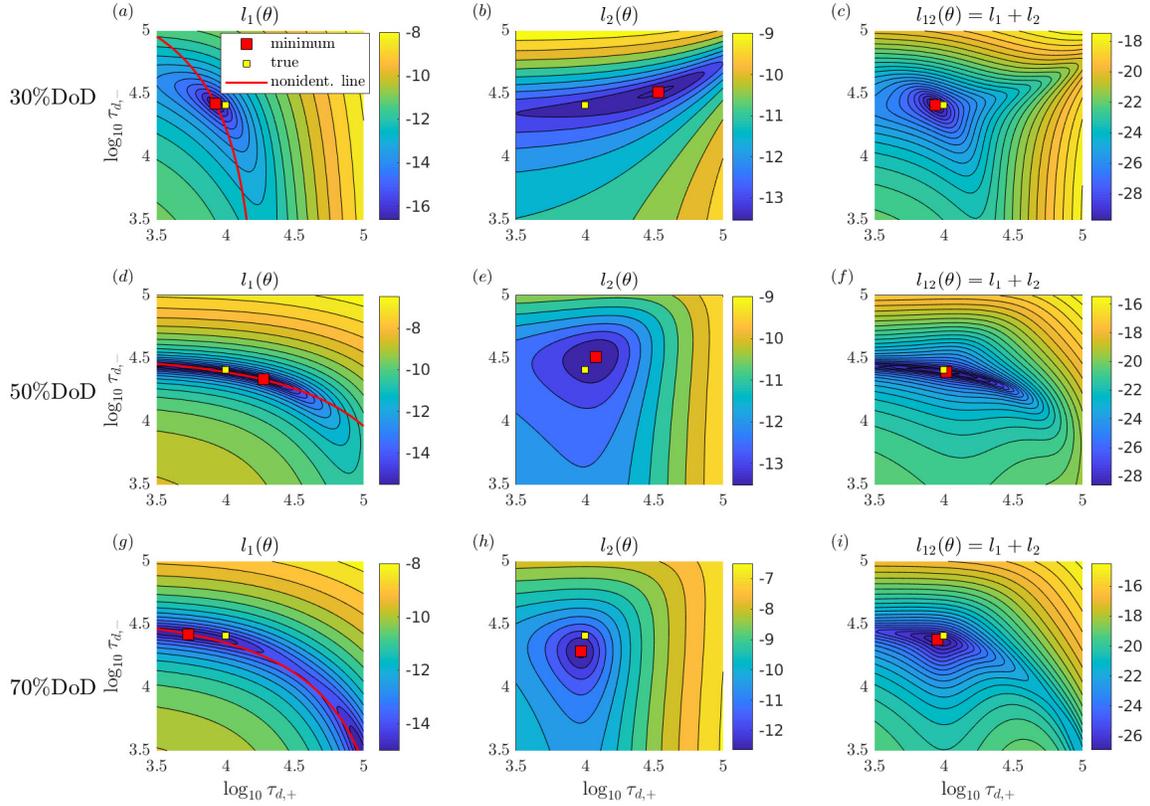}
       
    \caption{Synthetic data: Log-likelihood plotted in the $\tau_{d,+}$ and $\tau_{d,-}$ plane (other parameters fixed at true values). DoD (indicated) increases from top row to bottom row. Left column is log-likelihood $l_1$ for $Z_{1}^{(1)}$ only; middle column is log-likelihood $l_2$ for $Z_{2}^{(2)}$ only; right column is the total log-likelihood $l_{12}=l_1 + l_2$. For 30\%, 50\%, 70\% DoD we have $-U_-^{(0)\prime}\approx 0.4,17.9,1.6$, and $-U_+^{(0)\prime}\approx 18.0,6.5,2.2$. Red line is (\ref{eq:tau_d_unident}). }
    \label{fig:Synthetic-tau-tau-known-ddU}
\end{figure}

A clearer understanding of the practical identifiability of diffusion time $\tau_{\mathrm{d},\pm}$ can be reached by looking at the log-likelihoods (i.e.\ the cost functions associated with the parameter estimation problem) for each harmonic directly, $l_1(\bm{\theta})$, $l_2(\bm{\theta})$ in the $(\tau_{\mathrm{d},+}, \tau_{\mathrm{d},-})$ plane. Fig.\ \ref{fig:Synthetic-tau-tau-known-ddU} shows these (and the full log-likelihood $l_{12} = l_1 + l_2$) at 3 DoDs, with kinetic parameters fixed at their true values.

The shape of the $l_1$ landscape (Fig.\ \ref{fig:Synthetic-tau-tau-known-ddU}$(a,d,g)$) depends on OCP slopes $U_\pm^{(0)\prime}$, as discussed in detail by Bizeray et al.\  \cite{Bizeray2019} (for Fickian diffusion, but the structure is qualitatively the same). When one of the OCP slopes $U_\pm^{(0)\prime}$ is small, a valley in $l_1$ appears through the minimum, showing poor identifiability of the corresponding $\tau_{\mathrm{d},\pm}$.  Of the results shown, the negative electrode OCP slope $|U_-^{(0)\prime}|$ is largest ($\approx 17.9$) at 50\% DoD, where $\tau_{\mathrm{d},-}$ identifiability is best, and smallest ($\approx 0.36$) at 30\% DoD, where it is worst. But the positive electrode OCP slope $|U_+^{(0)\prime}|$ is largest ($\approx 18.0$) at 30\% DoD and decreases at 50\% DoD ($\approx 6.5$) and 70\% DoD ($\approx 2.2$), decreasing $\tau_{\mathrm{d},+}$ identifiability. At 70\% DoD,  $|U_-^{(0)\prime}|\approx |U_+^{(0)\prime}|$, giving electrode symmetry (see the Structural Parameter Identifiability section), 
and 2 local minima in $l_1$. The red line on the $l_1$ plots is the relationship between $\tau_{\mathrm{d},-}$ and $\tau_{\mathrm{d},+}$ if both slopes $U_-^{(0)\prime}, U_+^{(0)\prime}\to 0$, in which case (\ref{eq:tau_d_ident_1}) reduces to, at leading order,
\begin{align}
    A_+ \sqrt{\tau_{\mathrm{d},+}} + 
    A_- \sqrt{\tau_{\mathrm{d},-}} &= \mbox{const.} & \mbox{where }A_\pm &= \xi_\pm \sqrt{\frac{|U_\pm^{(0)\prime}|}{c_\pm^{(0)}}}. \label{eq:tau_d_unident}
\end{align}
This gives a visual guide to the predominant structure of the $l_1$ landscape.

For the log-likelihood $l_2(\bm{\theta})$ of the second harmonic
(Fig.\ \ref{fig:Synthetic-tau-tau-known-ddU}$(b,e,h)$), its landscape
depends on $U_\pm^{(0)\prime}, U_\pm^{(0)\prime\prime}$ and 
$R_\pm^{(0)\prime}$ and so is more difficult to predict from the
formulae, but crucially it is distinct from the landscape of
$l_1$. Thus, when it is combined with $l_1$ to produce the total log-likehood
$l_{12}$, the identifiability of the pair $(\tau_{\mathrm{d},+},
\tau_{\mathrm{d},-})$ is improved. For example, at 50\% and 70\% DoD,
identifiability of $\tau_{\mathrm{d},+}$ is improved, and the second
minimum (at 70\%) eliminated. In each case, including $l_2$ moves the
global minimum closer to the true value. We make an additional remark that the minima of $l_1$ in Fig.\ \ref{fig:Synthetic-tau-tau-known-ddU} are further from the true value than the estimates in Fig.\ \ref{fig:results_error_sensitivity_DoD50}. This is because, in the latter, the kinetic parameters were fitted  simultaneously which allowed a better estimate of the diffusion times (i.e.\ reduced the bias due to the small current amplitude expansion). As kinetic parameters were fixed at their true values in Fig.\ \ref{fig:Synthetic-tau-tau-known-ddU}, this bias is increased for the diffusion times. Using fitted kinetic parameters in Fig.\ \ref{fig:Synthetic-tau-tau-known-ddU} instead, the minima move closer to the true values. However, this highlights the high sensitivity of linear EIS estimates to errors in the kinetic parameters.

\begin{figure}
    \centering
    \includegraphics[width=1.1\textwidth]{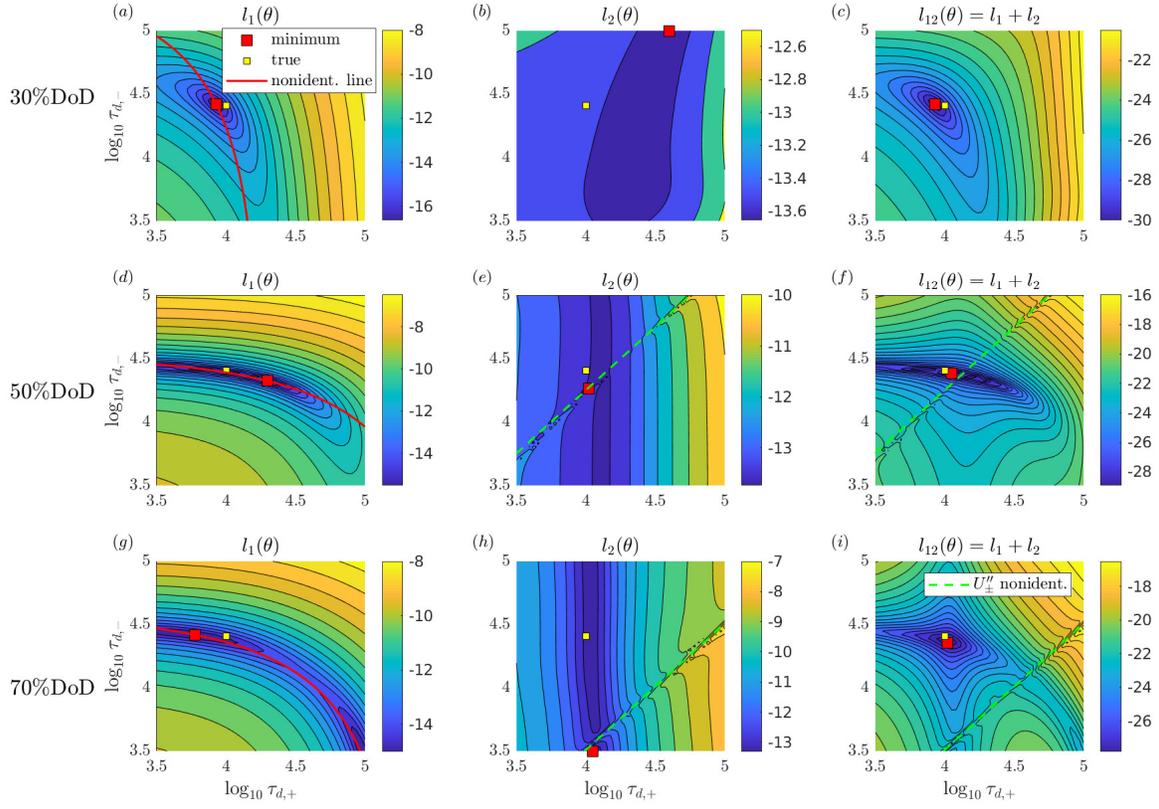}
       
    \caption{Synthetic data: Log-likelihood plotted in the $\tau_{d,+}$ and $\tau_{d,-}$ plane. Same as Fig.\ \ref{fig:Synthetic-tau-tau-known-ddU}, but at each ($\tau_{d,+},\tau_{d,-}$) pair, we minimize $l_2$ over $U_+^{(0)\prime\prime}$ and $U_-^{(0)\prime\prime}$. Dashed line is $\tau_{\mathrm{d},-} = (c_+^{(0)}U_+^{(0)\prime})\tau_{\mathrm{d},+}/(c_-^{(0)}U_-^{(0)\prime})$ (or $D_- = D_+$) where $U_\pm^{(0)\prime\prime}$ are unidentifiable.}
    \label{fig:Synthetic-tau-tau-fit-ddU}
\end{figure}

Lastly, relevant for fitting to experimental data in the next section,
we consider the case where OCP curvature $U_\pm^{(0)\prime\prime}$ is not known,
due to measurement error and difficulties associated with
approximating second derivatives numerically from noisy
data. In this case, Fig.\ \ref{fig:Synthetic-tau-tau-fit-ddU} shows $l_1, l_2,
l_{12}$ where, for each pair of values $(\tau_{\mathrm{d},+},
\tau_{\mathrm{d},-})$, we have minimised over 
$U_-^{(0)\prime\prime}$ and $U_+^{(0)\prime\prime}$---this corresponds to
constructing ``profile likelihoods'' where $U_-^{(0)\prime\prime},
U_+^{(0)\prime\prime}$ are nuisance parameters that are not of
interest \cite{CoxLikelihoodBook, Simpson2020}. Note that  $l_1$ is unchanged from Fig.\ \ref{fig:Synthetic-tau-tau-known-ddU} as it does not depend on $U_\pm^{(0)\prime\prime}$. Even here, with no knowledge of $U_\pm^{(0)\prime\prime}$, the second harmonic provides enough information to improve identifiability over the fundamental alone.

In Fig.\ \ref{fig:Synthetic-tau-tau-fit-ddU}, the dashed overlaid line
corresponds to $\tau_{\mathrm{d},-} =
(c_+^{(0)}U_+^{(0)\prime})\tau_{\mathrm{d},+}/(c_-^{(0)}U_-^{(0)\prime})$
(or $D_-^{(0)} = D_+^{(0)}$) along which the pair $U_-^{(0)\prime\prime},
U_+^{(0)\prime\prime}$ are unidentifiable. This was verified by
investigating the $(U_-^{(0)\prime\prime}, U_+^{(0)\prime\prime})$
plane when  $(\tau_{\mathrm{d},+}, \tau_{\mathrm{d},-})$ are on this
dashed line. However, this only introduces small numerical
irregularities which are irrelevant if far from the global minimum. 


\subsection{Parameter estimation from experimental data}

We have explored the parameter estimation approach and advantages of NLEIS using synthetic data, and now turn to experimental data collected from an NMC cathode/graphite anode cell\textemdash see the Experimental Methods section 
for details. We consider data collected at 10\%, 20\%, ..., 90\% DoD, at each of which we consider EIS ($Z_1^*$ only) and NLEIS ($Z_1^*$ and $Z_2^*$) model fits.

\begin{figure}
    \centering
    \includegraphics[width=0.95\textwidth]{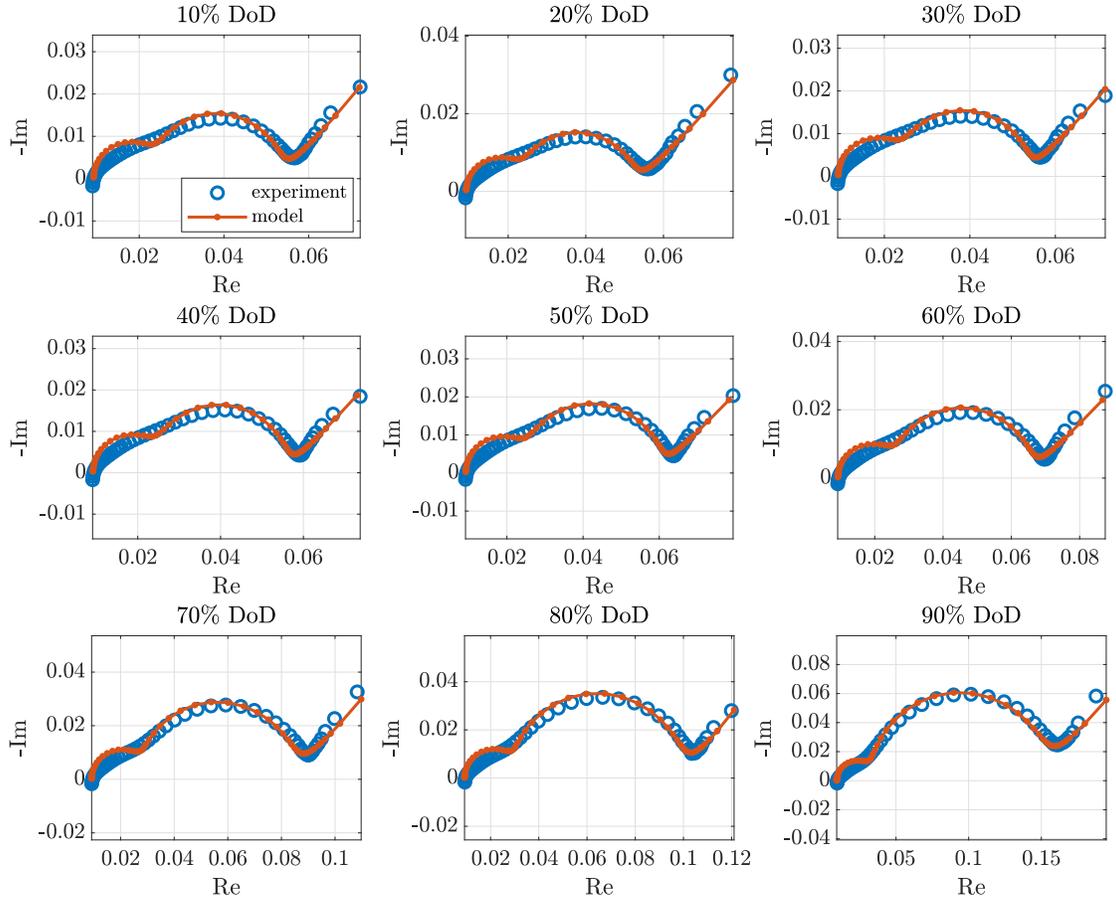}
    \caption{Fits of the model fundamental impedance $Z_{1}^{(1)*}$ to corresponding experimental fundamental impedance data. Units are in Ohms (VA$^{-1}$). Fits performed at each DoD in the range 10\%, 20\%,... 90\% are shown, for the frequency range $f^*=0.01-500$ Hz. Parameter estimates are given in Fig. \ref{fig:results_EIS_NLEIS_param_fits}.}
    \label{fig:results_Z11_fits_to_experiment}
\end{figure}

\begin{figure}
    \centering
    \includegraphics[width=0.95\textwidth]{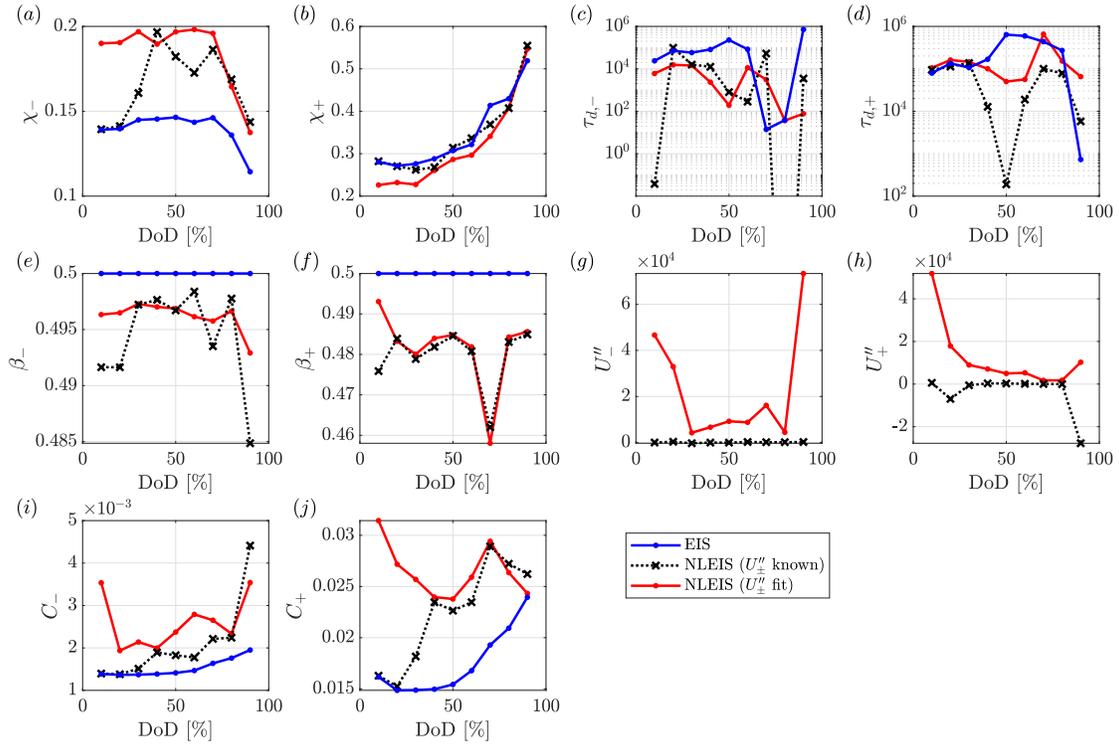}
    \caption{Parameter values from fitting to experimental EIS data ($Z_{1}^{*}$ only) or NLEIS data ($Z_{1}^{*}$ and $Z_{2}^{*}$) at different DoDs, from 10\% to 90\%. Kinetic parameters are in the left two columns ($a,b,e,f,i,j$), with diffusion timescales and OCP parameters in the remaining two; $\beta_+$ and $\beta_-$ were taken as 0.5 in the EIS fits.}
    \label{fig:results_EIS_NLEIS_param_fits}
\end{figure}

The fits to EIS data are shown first, in
Fig.\ \ref{fig:results_Z11_fits_to_experiment}, with the associated parameter
estimates in Fig.\ \ref{fig:results_EIS_NLEIS_param_fits}. The model is
able to fit the semi-circular reaction kinetics well, although we do
not expect to capture the very high frequency behaviour as our model
does not include phenomena such as transport through SEI
layers, relevant for $\omega^* \gtrsim 1$ kHz. We can extract
the series resistance as the high frequency intercept with the real
axis, giving $R_s^* =$ \SI{9.2}{\milli\ohm} (and dimensionless $R_s =
0.358$). As discussed in the Structural Parameter Identifiability section, estimating kinetics from
EIS data suffers from an inability to determine which electrode the
parameters correspond to. Hence, we made the assumption that the
larger charge transfer resistance (radius of semi-circle) was due to
the positive electrode, but this was arbitrary and did not affect the goodness of fit. Note also that $\beta_-$ and $\beta_+$ cannot be
determined at all, so we set them equal to 1/2, as is typically done
implicitly in other studies. 

The diffusion tails are fitted reasonably well. However the corresponding
timescales show extreme variation with DoD, up to 4 orders of magnitude\textemdash as is usually seen when estimating diffusion parameters from EIS or
GITT data \cite{Ecker2015, Schmalstieg2018, Chen2020}. As we will see,
NLEIS can be a more sensitive tool for probing the validity of the underlying diffusion model.

\begin{figure}
    \centering
    \includegraphics[width=0.95\textwidth]{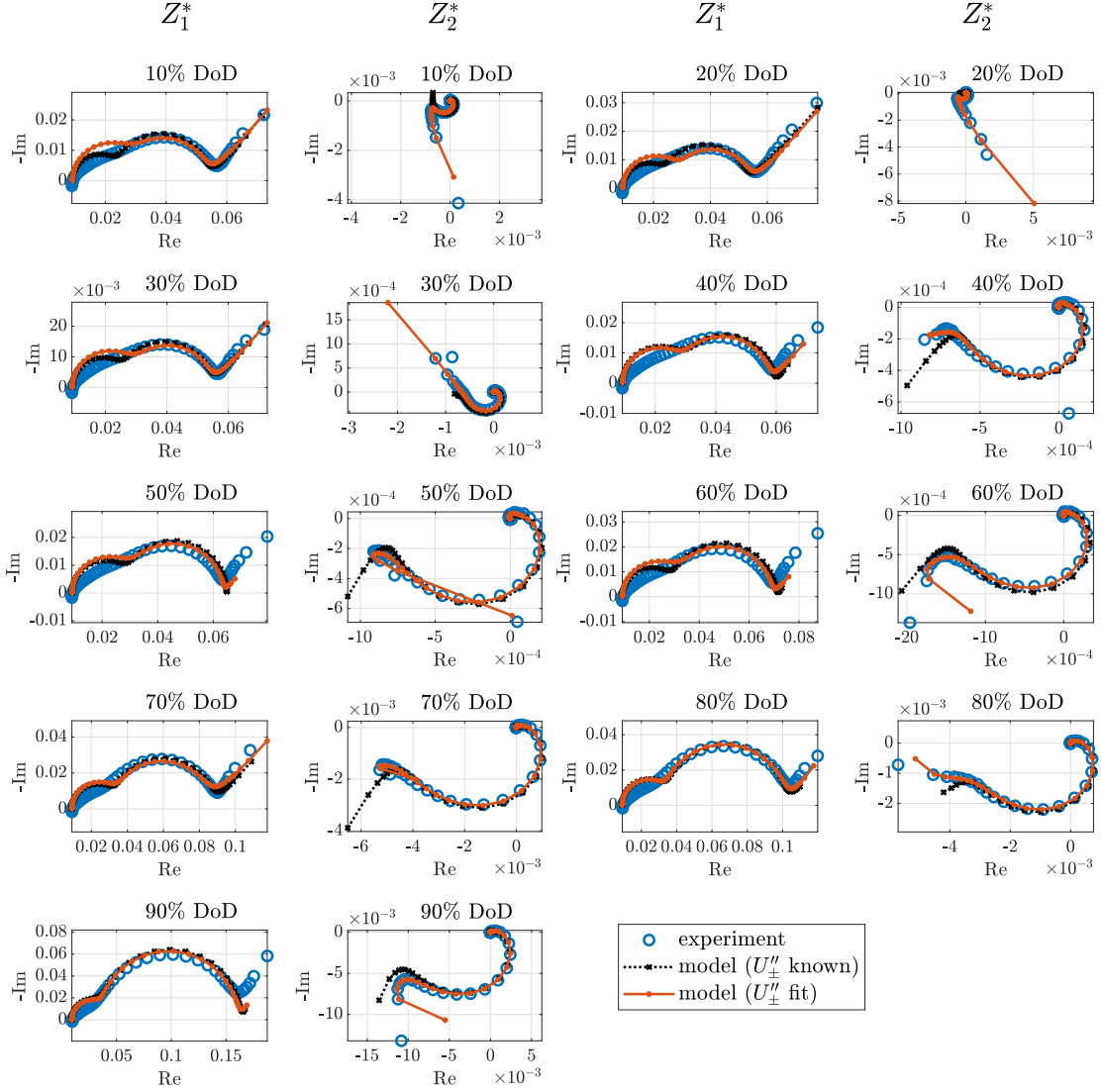}
    \caption{Simultaneous fits of the model impedances $Z_{1}^{(1)*}$ (fundamental) and $Z_{2}^{(2)*}$ (second harmonic) to corresponding experimental data. $Z_{1}^{*},Z_{2}^{*}$ units are in VA$^{-1}$ (Ohms) and VA$^{-2}$. Two cases, where $U_{\pm}''$ is either fixed (assumed known) or chosen as a fitting parameter are shown. Fits performed at each DoD in the range 10\%, 20\%,... 90\% separately, increasing from top left to the bottom right panel. Parameter estimates are given in Fig. \ref{fig:results_EIS_NLEIS_param_fits}.}
    \label{fig:results_Z11_Z22_fits_to_experiment}
\end{figure}

\begin{figure}
    \centering
    \includegraphics[width=0.85\textwidth]{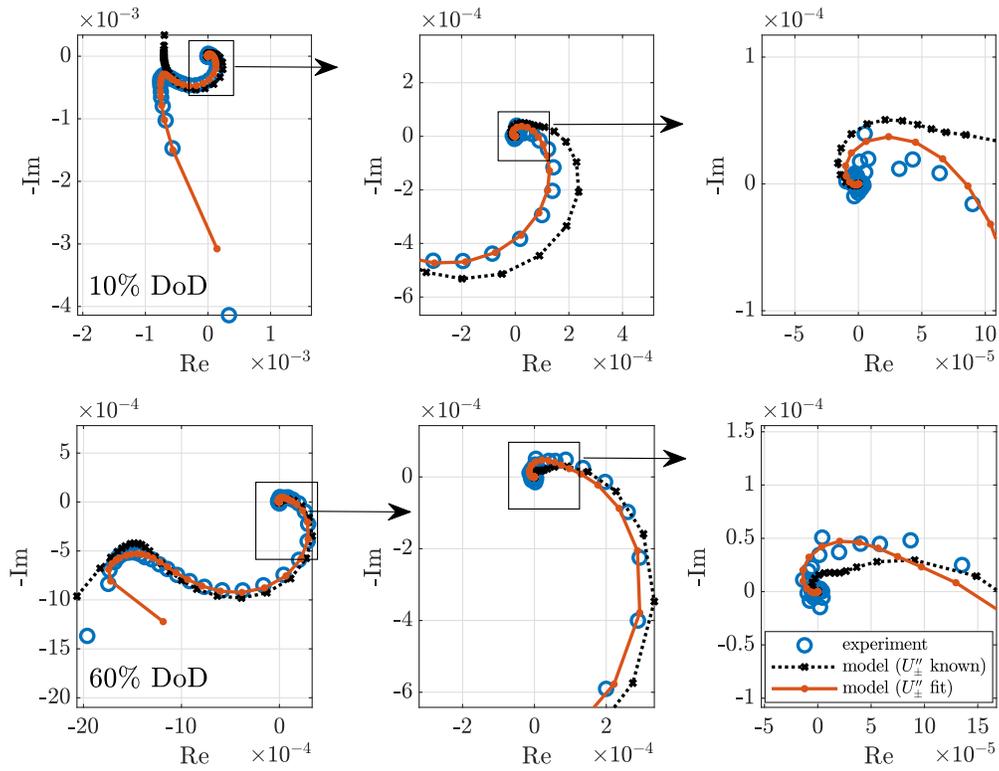}
    \caption{Successive zoom-ins on selected $Z_{2}^{*}$ impedances in Fig. \ref{fig:results_Z11_Z22_fits_to_experiment}. The top row is at 10\% DoD, and the bottom row is at 60\% DoD.}
    \label{fig:results_Z11_Z22_fits_zoom}
\end{figure}

Next, fits to NLEIS data ($Z_1^*$ and $Z_2^*$ simultaneously) are shown in Fig.\ \ref{fig:results_Z11_Z22_fits_to_experiment} (with parameters in Fig.\ \ref{fig:results_EIS_NLEIS_param_fits}). We consider two cases: (i) using values of $U_\pm^{(0)\prime\prime}$ calculated from Fig.\ \ref{fig:kokam_OCPs}, and; (ii) fitting  $U_\pm^{(0)\prime\prime}$, assuming they are not known \emph{a priori}, or are inaccurate. In both cases, the reaction kinetics in each harmonic are captured very well for each DoD, as shown by the semi-circles in $Z_1^*$ and cardioid spirals in $Z_2^*$. The resulting values for $\beta_\pm, C_\pm, \chi_\pm$ also do not vary appreciably across DoD except $\chi_+$, suggesting the functional form of the exchange current density (\ref{eq:dimensional_exchange_current_density}) is less appropriate for that electrode. The charge transfer coefficients $\beta_\pm$ both only slightly deviate from the symmetric value of 1/2 for these cells. Here $\beta_+$ deviates most from 1/2 and is responsible for the predominant cardioid in $Z_2^*$. The cardioid from the negative electrode is small and only visible by zooming in closer to the origin\textemdash see Fig.\ \ref{fig:results_Z11_Z22_fits_zoom}.

Regarding the low frequency diffusion tails, the model fits are less
convincing. When $U_\pm^{(0)\prime\prime}$ is assumed known, the tail
in $Z_2^*$ may point in the opposite direction (10\%, 20\%, 50\% DoD)
or have the wrong length (30\%, 40\%). Also, the tail in $Z_1^*$ is
now too short in some cases (40\%, 50\%, 60\%). However, if OCP curvatures
$U_\pm^{(0)\prime\prime}$ are allowed to be fitted instead, the fit to
the tail in $Z_2^*$ is significantly improved. Also, the tail in $Z_1^*$ and the fit to the
kinetic portions are improved; see the zoom-ins of
Fig.\ \ref{fig:results_Z11_Z22_fits_zoom}. This suggests, even if only
the \emph{kinetic} parameters are of interest, fitting
$U_\pm^{(0)\prime\prime}$ may be better than prescribing inaccurate
values. The resulting diffusion timescales
(Fig.\ \ref{fig:results_EIS_NLEIS_param_fits}$(c,d)$ also show the
least variation with DoD for NLEIS data fits when $U_\pm^{(0)\prime\prime}$ is fitted
(red), an improvement over the EIS data fits (blue), and hence closer to the model assumption that it is does not vary with DoD.



\subsubsection*{Log-likelihood analysis of diffusion timescales}
\begin{figure}
    \centering
    \includegraphics[width=1.\textwidth]{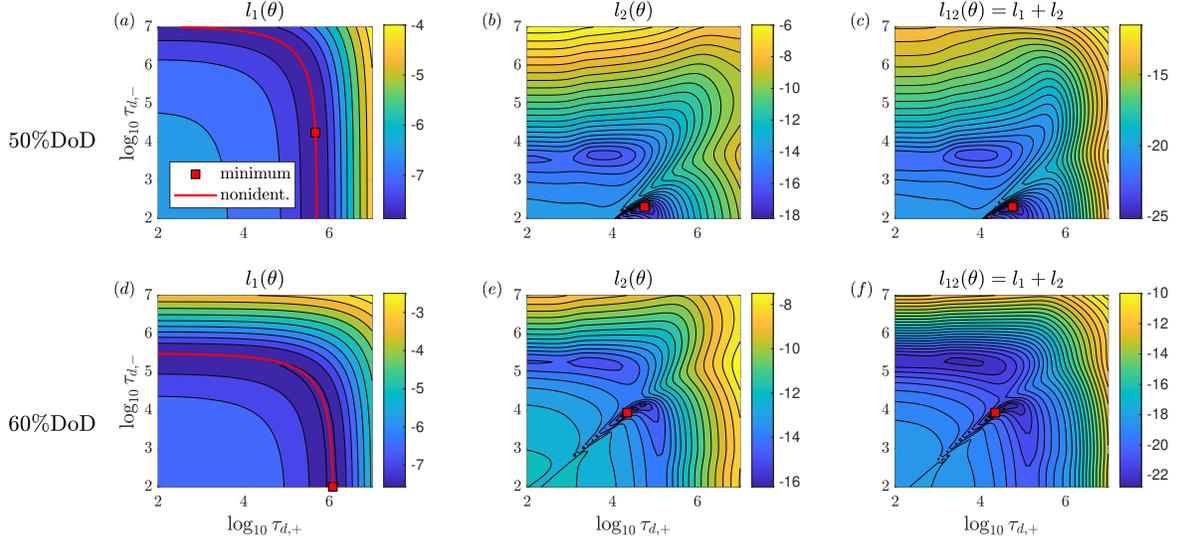}
    \caption{Log-likelihoods in the diffusion timescale plane $\tau_{\mathrm{d},+}$-$\tau_{\mathrm{d},-}$, using experimental data. Same as Fig.\ \ref{fig:Synthetic-tau-tau-known-ddU} but now at each point we have minimised over $U_-^{\prime\prime}$ and $U_+^{\prime\prime}$.}
    \label{fig:results_tau_tau_fit_ddU}
\end{figure}

The fits of diffusion times to experimental data can be further probed by examining the underlying likelihood landscape in the $(\tau_{\mathrm{d},+}, \tau_{\mathrm{d},-})$ plane. As we did for synthetic data, we plot here
the log-likelihoods in Fig.\ \ref{fig:results_tau_tau_fit_ddU}, minimising over $U_\pm^{(0)\prime\prime}$. 
The DoDs chosen (50\% and 60\%) are ones at which the timescale estimates from EIS vs.\ NLEIS differ significantly. The rest of the parameters are fixed at their estimated values from Fig.\ \ref{fig:results_EIS_NLEIS_param_fits}.

The $l_1$ landscape ($Z_1^*$) clearly shows unidentifiability given by
the red line, i.e.\ (\ref{eq:tau_d_unident}). Here this is likely due to
insufficient data at lower frequencies,  as (\ref{eq:tau_d_unident})
also holds if only the high frequency approximation
(\ref{eq:H_1_high_omega}) is taken in the Warburg impedance. In the
experiments, we were limited to a minimum allowable frequency of \SI{0.01}{Hz}
 by the Kikusui PBZ60-6.7 power supply. This was not an issue when
estimating from synthetic data
(Fig.\ \ref{fig:Synthetic-tau-tau-fit-ddU}) as we were able to generate
data at arbitrarily low frequencies\textemdash observe that the diffusion
tail in Fig.\ \ref{fig:results_error_sensitivity_DoD50}($i$) becomes
close to vertical at the lowest frequencies. This unidentifiability
valley was therefore not present, with one (or 2) local minima present
instead. 

The $l_2$ landscape ($Z_2^*$) in
Fig.\ \ref{fig:results_tau_tau_fit_ddU} has its minima considerably far
from where $l_1$ predicts they should be. Further, the depth of $l_2$
minima overwhelms those in $l_1$ when combined in $l_{12}=l_1 +
l_2$. This differs significantly from the behaviour for synthetic data
where the model exactly matches that used to generate the data,
i.e.\ there is no model error, only approximation error due to the
expansion in $\hat{I}$, and noise. In that case, the results
showed that the fundamental is more dominant, with deeper minima in
$l_1$ than $l_2$.  

This inconsistency between the minima of $l_1$ and $l_2$ suggests that there may be deficiencies in the diffusion model even though the model is able to to fit the data very well. Here we considered a nonlinear diffusion model, but to alleviate these issues one may need to consider other approaches, e.g.\ those incorporating phase-field dynamics (such as Cahn\textendash Hilliard) which have shown to better represent lithiation in some electrode materials, e.g.\ graphite \cite{Ferguson2014}. On the other hand, it may be that a single representative particle for each electrode is inadequate, or that large-scale geometric effects are important, or the inclusion of transport in the electrolyte is necessary. These results suggest NLEIS could be  a promising tool for model selection. Other dynamics that may also relate to this inconsistency are those in the electrolyte, e.g.\ diffusion and migration of lithium, which have been neglected here. 


\subsection{Model validation in the time domain}
\label{sec:Results-validation}
We have parameterized the nonlinear SPM using experimental NLEIS
(frequency-domain) data and now proceed to validate the model performance and
parameters by numerical solution in the time domain. We considered a
current demand based on an electric vehicle drive cycle (see
\cite{Zhao2017}) with measurements taken on the same model of Kokam
cell. Current was specified at 1 second intervals, with peaks reaching
\SI{4.4}{A} (or C-rates of 6), and initiated at 0\% DoD. The model
(\ref{eq:c_equation})-(\ref{eq:R}) was solved using this profile as an input, which
was interpolated using piecewise-cubic Hermite interpolating
polynomials (PCHIP) to ensure no overshooting of the data. Further
details of numerical methods are given in the Methods. 

We compared the model performance using two parameter sets: estimates from EIS data, and estimates from NLEIS data (see Fig.\ \ref{fig:results_EIS_NLEIS_param_fits}). We used NLEIS parameters for the case where $U_\pm^{\prime\prime}$ were considered unknown. For $\beta_\pm,C_\pm,\tau_{\mathrm{d},\pm}$, we chose single representative values by averaging over the full range of DoDs (with the average of $\tau_{\mathrm{d},\pm}$ taken on $\log\tau_{\mathrm{d},\pm}$). For $\chi_\pm$, we converted Fig.\ \ref{fig:results_EIS_NLEIS_param_fits} data from DoD to stoichiometry $c_\pm$ and interpolated (PCHIP) to give the functions $\chi_+(c_+),\chi_-(c_-)$.\footnote{$\chi_\pm$ were originally assumed constant in the model, but making them depend on $c_\pm$ accounts for errors in the functional form of $R_\pm(c_\pm)$, (\ref{eq:R}).} The parameter values used are summarized in Table \ref{tab:Param-groups-validation}.

\begin{figure}
    \centering
    \includegraphics[width=1\textwidth]{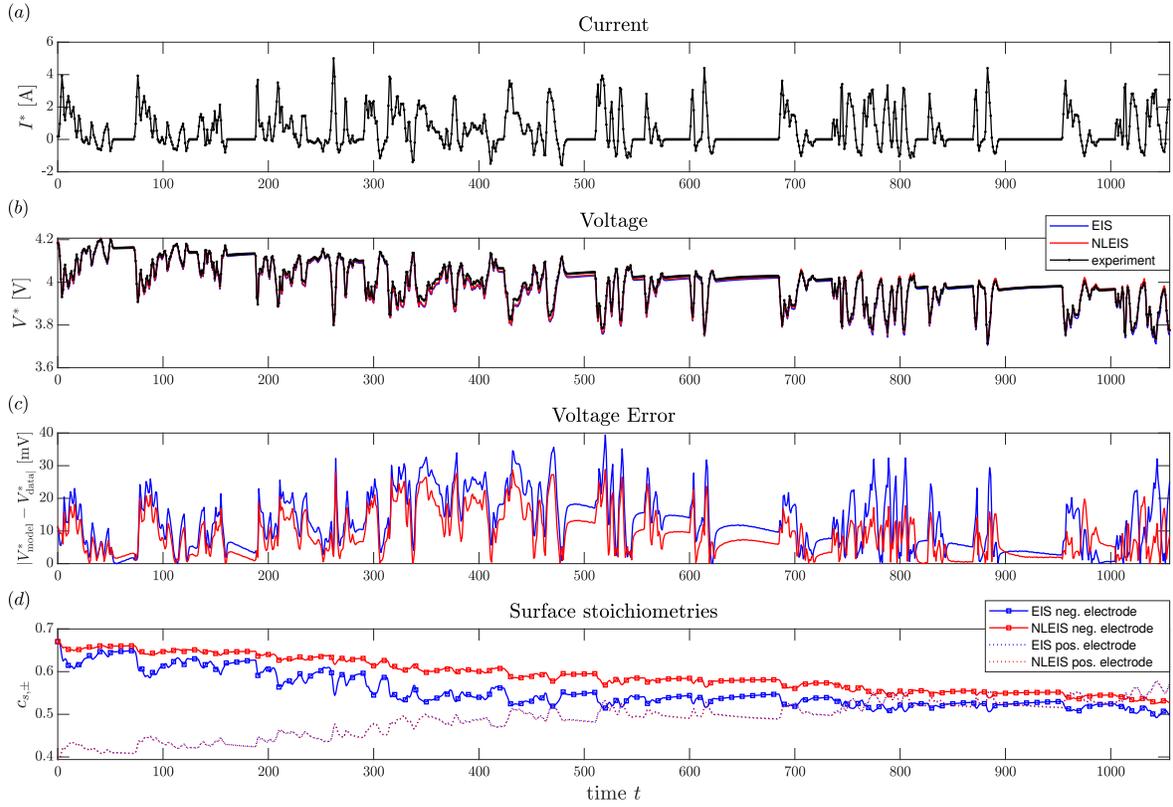}
    \caption{Model validation in the time domain, corresponding to a drive cycle of prescribed current $(a)$. The terminal voltage from experiment \cite{Zhao2017} and two model simulations, using parameter values (Table \ref{tab:Param-groups-validation}) from fitting to EIS or NLEIS data, are in $(b)$. Also shown are $(c)$ the absolute voltage error, and $(d)$ surface stoichiometries $c_{s,\pm}$ in each electrode for each simulation.}
    \label{fig:results_time_domain_validation_1}
\end{figure}

Comparison between the model and experiment is shown in
Fig.\ \ref{fig:results_time_domain_validation_1}. In general, the model results
with either parameter set (EIS or NLEIS) agree excellently with the measured 
data over the full duration ($\approx 20$ min), with no parameter
tuning performed.\footnote{The series resistance $R_s^*$ was changed
  before simulation to 20 m$\Omega$ (taken from Fig.\ 4 of
  \cite{Bizeray2019})) since the drive cycle measurements were
  performed using a different test rig, but no subsequent changes to
  $R_s^*$ were made.} Even the EIS predictions show an RMSE of only
\SI{15.2}{mV} and no overall drift in error as time increases. This is a
marked improvement over the linear SPM of Bizeray \emph{et al.}
\cite{Bizeray2019} where, for the same cells and validation data, they
required the linearized OCPs to be adjusted to correct for
drift in the DoD . Our model exhibited no discernable drift and we considered a
longer drive cycle simulation  than
Bizeray \emph{et al.} \cite{Bizeray2019} ($\approx 17$ mins versus $9$ mins). 

The predictions from NLEIS, however, agree even better with the measured voltage data, with a lower RMSE of \SI{11.1}{mV}. We conclude that NLEIS can provide better estimates of the reaction kinetic parameters ($\chi_\pm$, $\beta_\pm$, and $C_\pm$). Looking at the stoichiometries on the surface of electrode particles (Fig.\ \ref{fig:results_time_domain_validation_1}$(d)$) we see appreciable differences only in the negative electrode, but this has minimal effect on the voltage since $U_-(c_-)$ is mostly flat over the range of $c_-$ operated in. Hence, diffusion timescale estimates from either method are sufficient for good voltage accuracy in the time domain, but predictions of internal states may differ greatly, which will have consequences for degradation modelling. 

\begin{table}[]
\centering
\footnotesize
\begin{tabular}{@{}lllll@{}}
\toprule
\multirow{2}{*}{Parameter groups} & \multicolumn{2}{c}{Negative (-)}                 & \multicolumn{2}{c}{Positive (+)}                 \\ 
                                  &  EIS estimate  & NLEIS estimate                  & EIS estimate           & NLEIS estimate          \\ \midrule 
$\chi_\pm$                        & Fig.\ \ref{fig:results_EIS_NLEIS_param_fits}$(a)$ & Fig.\ \ref{fig:results_EIS_NLEIS_param_fits}$(a)$ & Fig.\ \ref{fig:results_EIS_NLEIS_param_fits}$(b)$  & Fig.\ \ref{fig:results_EIS_NLEIS_param_fits}$(b)$ \\
$\beta_\pm$                       & 0.5                     & 0.496                  & 0.5                    & 0.482                   \\
$C_\pm$                           & $1.528\times 10^{-3}$  & $2.589\times 10^{-3}$   & $1.747\times 10^{-2}$  & $2.645\times 10^{-2}$ \\ 
$\tau_{\mathrm{d},\pm}$           & $1.571\times 10^{4}$  & $1.570\times 10^{3}$     & $1.227\times 10^{5}$    & $1.185\times 10^{5}$   \\ \midrule
$\xi_\pm$                         & \multicolumn{2}{c}{$8.126\times 10^{-5}$}        & \multicolumn{2}{c}{$7.414\times 10^{-5}$}       \\
$c_\pm^{0\%}$                     & \multicolumn{2}{c}{0.664}                        & \multicolumn{2}{c}{0.40}                         \\
\bottomrule
\end{tabular}
\caption{Parameter group values estimated from EIS and NLEIS data, which are used for model validation in Fig.\ \ref{fig:results_time_domain_validation_1}. The $\chi_\pm$ parameters are interpolated on the values in Fig.\ \ref{fig:results_EIS_NLEIS_param_fits}$(a,b)$ for different DoDs (NLEIS estimates used are the red markers). From our impedance measurements the series resistance is $R_s^*=9.2$ mV ($R_s=0.358$), but for the model validation (Fig.\ \ref{fig:results_time_domain_validation_1}) we use $R_s^*=20$ mV and $R_s=0.779$.}
\label{tab:Param-groups-validation}
\end{table}

\section{Conclusions}
In this paper, we presented a comprehensive study on the application of NLEIS to a standard physical model of a lithium-ion battery: the single particle model. We derived a suite of analytical formulae for the fundamental harmonic and second harmonic impedances resulting from this model, and then interpreted each contribution to the formulae physically. Signatures of various processes and model nonlinearities were identified in the higher harmonics, in particular: (i) asymmetric charge transfer in Butler\textendash Volmer kinetics; (ii) concentration dependence of the exchange current density; (iii) nonlinearity of the electrode OCPs; and (iv) nonlinearity of solid-state diffusion (i.e.\ concentration dependence of the diffusivity). As a result, we demonstrated how the Nyquist plot of the second harmonic impedance may be interpreted in a similar manner to that of the fundamental, but built from different features and therefore containing additional information about the system and its parameters. This information was then exploited for parameter estimation from experimental data that was collected from commercial cells.

From our nonlinear impedance formulae, we explored how the second harmonic impedance in combination with the usual fundamental impedance  can be used to improve parameter identifiability, both structurally and practically, over the use of the fundamental alone (EIS). Structurally, NLEIS can identify new parameters not possible from EIS, e.g.\ charge transfer coefficients (usually assumed to be $1/2$ for Li-ion batteries), and improve the identifiability of the remaining parameters. In particular, positive-negative electrode symmetries are eliminated, allowing all parameters to be associated with the correct electrode, and the curvature (if nonzero) of the OCP can help identify a diffusion timescale which would otherwise be unidentifiable from EIS (e.g., due to a ``flat'' OCP change with respect to stoichiometry).

To investigate the impact of NLEIS on practical identifiability, we implemented maximum likelihood parameter estimation. First, we demonstrated the procedure using synthetic data and bootstrap resampling, showing that one can robustly determine all the model parameter groups (kinetic and diffusive) from NLEIS data. The second harmonic has a lower signal-to-noise ratio resulting in larger variances for some parameters, but these uncertainties are comparable in magnitude to model error, e.g.\ due to the small excitation amplitude approximation.

Finally, our parameter estimation procedure was applied to experimental NLEIS data that was collected from a commercial cell. We could fit the impedance data very well, capturing many of the key features of both harmonics. All the kinetic parameters were determined, and the charge transfer coefficients were found to deviate slightly from $1/2$, albeit remaining within 10\%. However, the diffusion timescale estimates from EIS and NLEIS data differed greatly and the latter, via analysis of the likelihood landscape, suggested a discrepancy in the diffusion model\textemdash a discrepancy otherwise not revealed from EIS data alone. Nonetheless, validation of the model and its parameterization from either EIS or NLEIS data was conducted in the time domain with a drive cycle experiment, where an improvement in accuracy was observed for the NLEIS parameterization, showing that the technique is beneficial when the model is used in practice.

Areas of further work could consider the impacts of several physical mechanisms that were neglected here, such as transport through SEI layers and in the electrolyte. To keep a similar level of model complexity, one could consider the latter by extending the current analysis to the single particle model with electrolyte (SPMe) \cite{Marquis2019}. The electrolyte introduces additional diffusive effects which may address the difficulties observed here with fitting the diffusion tails in a consistent manner. Other solid-state transport models, such as phase-field models \cite{Ferguson2014}, could also be considered, and NLEIS used as a model selection tool. In addition, extending the minimum frequency measured experimentally could improve diffusion time estimates. Finally, metrics to quantify model parsimony could be considered, i.e.\ to analyse whether there are too many (or too few) parameters, and to compare different model structures.

\section{Acknowledgements}
This research was funded in whole, or in part, by the EPSRC Faraday Institution
Multiscale Modelling project (EP/S003053/1, grant number FIRG025). The authors also wish to thank Robert Timms and Steven Psaltis for helpful discussions. For the purpose of Open Access, the authors have applied a CC BY public copyright licence to any Author Accepted Manuscript (AAM) version arising from this submission. 

\bibliographystyle{elsarticle-num} 
\bibliography{NLEIS-refs-no-URLs}

\end{document}


\begin{frontmatter}



\title{Supplementary material:
Nonlinear electrochemical impedance spectroscopy for lithium-ion battery model parameterisation}



\author[inst1,inst2]{Toby L. Kirk}

\affiliation[inst1]{organization={Mathematical Institute, University of Oxford},
            addressline={Andrew Wiles Building, Woodstock Road}, 
            city={Oxford},
            postcode={OX2 6GG},
            country={UK}}

\author[inst2,inst3]{Adam Lewis-Douglas}
\author[inst2,inst3]{David Howey}
\author[inst1,inst2]{Colin P. Please}
\author[inst1,inst2]{S. Jon Chapman}

\affiliation[inst2]{organization={The Faraday Institution},
            addressline={Quad One, Becquerel Avenue, Harwell Campus}, 
            city={Didcot},
            postcode={OX11 0RA}, 
            country={UK}}
\affiliation[inst3]{organization={Battery Intelligence Lab, Department of Engineering Science, University of Oxford},
            city={Oxford},
            postcode={OX1 3PJ}, 
            country={UK}}


\end{frontmatter}


\vspace{-2cm}

\section{Derivation of nonlinear impedance formulae}
\subsection{Neglecting double-layer capacitance}
\label{sec:Derivation-no-C}

To derive the nonlinear impedance formulae when double-layer capacitance is neglected, we consider the SPM given by (\ref{eq:c_equation})-(\ref{eq:init}), (\ref{eq:E})-(\ref{eq:eta_no_C}). Given a sinusoidal current of the form $I(t) = \hat{I}\mathrm{e}^{\mathrm{i}\omega t} + \hat{I}\mathrm{e}^{-\mathrm{i}\omega t}$, first consider an expansion for small current amplitude. Note, by choice of phase, $\hat{I}$ is real and positive so this limit corresponds to $\hat{I}\ll 1$. The analysis is identical for both electrodes, so we drop the $\pm$ subscripts for brevity.

\subsubsection*{First-order solution}
Substituting the expansion $c_\pm = c = c^{(0)} + c^{(1)}\hat{I} + c^{(2)}\hat{I}^2 + O(\hat{I}^3)$ into (\ref{eq:c_equation})-(\ref{eq:init}), we find at leading order that $c^{(0)}=c_{\mathrm{init}}$, and at first order:
\begin{align}
    \frac{\partial c^{(1)}}{\partial t} & =D^{(0)}\nabla^2 c^{(1)}, & \text{for }0<r<1, \label{eq:c_1_equation}\\
    \frac{\partial c^{(1)}}{\partial r} & =0, & \text{at }r=0,\\
    -D^{(0)}\frac{\partial c^{(1)}}{\partial r} & =\pm\xi (\mathrm{e}^{\mathrm{i}\omega t} + \mathrm{e}^{-\mathrm{i}\omega t}), & \text{at }r=1,
\end{align}
where for brevity we write $\nabla^2 = \frac{1}{r^{2}}\frac{\partial}{\partial r}r^{2}\frac{\partial}{\partial r}$ and it is understood that the gradient operator throughout is purely radial. The problem is linear and forced by $\mathrm{e}^{\mathrm{i}\omega t}$ and its complex conjugate, and hence the solution is also of this frequency, $c^{(1)} = \hat{c}_1^{(1)} \mathrm{e}^{\mathrm{i}\omega t} + \mbox{c.c.}$, where c.c. denotes complex conjugate of the preceding term, and $\hat{c}_1^{(1)}$ satisfies
\begin{align}
    \mathrm{i} \omega \hat{c}_1^{(1)} & =D^{(0)}\nabla^2 \hat{c}_1^{(1)}, & \text{for }0<r<1, \\
    \frac{\mathrm{d} \hat{c}_1^{(1)}}{\mathrm{d} r} & =0, & \text{at }r=0,\\
    -D^{(0)}\frac{\mathrm{d} \hat{c}_1^{(1)}}{\mathrm{d} r} & =\pm\xi, & \text{at }r=1.
\end{align}
This is solved using standard techniques to give
\begin{align}
\hat{c}_{1}^{(1)}(r,\omega) &= \pm\frac{\xi}{D^{(0)}}\frac{\sinh(\sqrt{\mathrm{i}\omega/D^{(0)}} r)}{r(\sinh(\sqrt{\mathrm{i}\omega/D^{(0)}})-\sqrt{\mathrm{i}\omega/D^{(0)}}\cosh(\sqrt{\mathrm{i}\omega/D^{(0)}}))},\\
\hat{c}_{1}^{(1)}(1,\omega) & = \pm\frac{\xi}{D^{(0)}} H_1(\omega/D^{(0)}) = \pm\frac{\xi}{D^{(0)}}\frac{\tanh(\sqrt{\mathrm{i}\omega/D^{(0)}} )}{\tanh(\sqrt{\mathrm{i}\omega/D^{(0)}})-\sqrt{\mathrm{i}\omega/D^{(0)}}},
\end{align}
where $H_1(\omega)$ is the first-order surface concentration transfer function (\ref{eq:H_1}).

With $c$ found to first order, substituting $c = c^{(0)} + \hat{c}_1^{(1)} \hat{I} \mathrm{e}^{\mathrm{i}\omega t} + \mbox{c.c.}$ into the electrode voltage $V = U(c) + \eta$ where $\eta$ is given by (\ref{eq:eta_no_C}), and expanding in $\hat{I}$ up to first order,
\begin{align}
    V =& U^{(0)} + U^{(0)\prime} (\hat{c}_1^{(1)} \hat{I} \mathrm{e}^{\mathrm{i}\omega t} + \mbox{c.c.}) \pm R^{(0)} (\hat{I} \mathrm{e}^{\mathrm{i}\omega t} + \mbox{c.c.}) + O(\hat{I}^2),
\end{align}
where we have used that $E^{-1}(x) = 2x + O(x^2)$ for $|x|\ll1$. From the Fourier series (\ref{eq:Fourier-series-V}) of $V$, we can read off the coefficients $V_0 = U^{(0)} + O(\hat{I}^2)$ and $V_1 = \overline{V_{-1}} = U^{(0)\prime} \hat{c}_1^{(1)} \hat{I} \pm R^{(0)} \hat{I} + O(\hat{I}^2)$, and from (\ref{eq:V-expansion-1}), the impedances (using lowercase $z$ to denote that capacitance is neglected):
\begin{align}
    z_0^{(0)} &= U^{(0)}, & z_1^{(1)} &= U^{(0)\prime} \hat{c}_1^{(1)}  \pm R^{(0)}. \label{eq:z00_z11_derivation}
\end{align}
giving (\ref{eq:z_11_pm}) in the paper.

\subsubsection*{Second-order solution}
Substituting $c_\pm = c = c^{(0)} + c^{(1)}\hat{I} + c^{(2)}\hat{I}^2 + O(\hat{I}^3)$ into (\ref{eq:c_equation})-(\ref{eq:init}), we find at $O(\hat{I}^2)$:
\begin{align}
    \frac{\partial c^{(2)}}{\partial t} & =D^{(0)}\nabla^2 c^{(2)} + D^{(0)\prime}\nabla \cdot (c^{(1)}\nabla c^{(1)}), & \text{for }0<r<1, \label{eq:c_2_equation}\\
    \frac{\partial c^{(2)}}{\partial r} & =0, & \text{at }r=0,\\
    -D^{(0)}\frac{\partial c^{(2)}}{\partial r} & = D^{(0)\prime}c^{(1)}\frac{\partial c^{(1)}}{\partial r}, & \text{at }r=1, \label{eq:c_2_BC}
\end{align}
with forcing terms from the previous order appearing in the equation (\ref{eq:c_2_equation}) and on the surface (\ref{eq:c_2_BC}), due to nonlinear diffusion. Using (\ref{eq:c_1_equation}), then substituting the solution for $c_1$ into the forcing term in (\ref{eq:c_2_equation}) gives
\begin{align}
   D^{(0)\prime}\nabla \cdot (c^{(1)}\nabla c^{(1)}) &= D^{(0)\prime}\left[ (\nabla c^{(1)})^2 + \frac{c^{(1)}}{D^{(0)}} \frac{\partial c^{(1)}}{\partial t} \right], \\
   &=  D^{(0)\prime}\left[
    2 |\nabla \hat{c}_1^{(1)}|^2 + \left\{ (\nabla \hat{c}_1^{(1)})^2 + \frac{\mathrm{i}\omega}{D^{(0)}} (\hat{c}_1^{(1)})^2 \right\}\mathrm{e}^{\mathrm{i}2\omega t} + \mbox{c.c.}
    \right].
\end{align}
Similarly, the forcing in (\ref{eq:c_2_BC}) is
\begin{align}
   D^{(0)\prime}c^{(1)}\frac{\partial c^{(1)}}{\partial r} &= D^{(0)\prime}\left[
    2 \mathrm{Re}\left\{ \overline{\hat{c}_1^{(1)}} \frac{\mathrm{d}\hat{c}_1^{(1)}}{\mathrm{d} r}  \right\}
    + \hat{c}_1^{(1)} \frac{\mathrm{d}\hat{c}_1^{(1)}}{\mathrm{d} r}\mathrm{e}^{\mathrm{i}2\omega t} + \mbox{c.c.}
    \right]
\end{align}
Both of these forcings, as they are products of terms of frequency $\omega$, contain all possible binary products of the exponentials $\mathrm{e}^{\mathrm{i}\omega t}, \mathrm{e}^{-\mathrm{i}\omega t}$, which produces terms of frequency $2\omega$ (from $\mathrm{e}^{\mathrm{i}\omega t}\times \mathrm{e}^{\mathrm{i}\omega t}$), but also terms of zero frequency ($\mathrm{e}^{\mathrm{i}\omega t}\times \mathrm{e}^{-\mathrm{i}\omega t}$), i.e.\ real and constant. Therefore, the solution for $c^{(2)}$ takes the form
\begin{align}
    c^{(2)} &= \hat{c}_0^{(2)} + \hat{c}_2^{(2)} \mathrm{e}^{\mathrm{i}2\omega t} + \mbox{c.c.}
\end{align}
where 
\begin{align}
    \hat{c}_0^{(2)}(r,\omega) &= -\frac{2D^{(0)\prime}}{D^{(0)}} \left(\frac{\xi}{D^{(0)}}\right)^2 M_0\left(r, \frac{\omega}{D^{(0)}}\right), &  \hat{c}_0^{(2)}(1,\omega) & = -\frac{2D^{(0)\prime}}{D^{(0)}} \left(\frac{\xi}{D^{(0)}}\right)^2 H_0\left(\frac{\omega}{D^{(0)}}\right), \\
    \hat{c}_2^{(2)}(r,\omega) &= -\frac{D^{(0)\prime}}{D^{(0)}} \left(\frac{\xi}{D^{(0)}}\right)^2 M_2\left(r, \frac{\omega}{D^{(0)}}\right), &  \hat{c}_2^{(2)}(1,\omega) & = -\frac{D^{(0)\prime}}{D^{(0)}} \left(\frac{\xi}{D^{(0)}}\right)^2 H_2\left(\frac{\omega}{D^{(0)}}\right),     
\end{align}
and $M_0(r,\omega)$, $M_2(r,\omega)$ are the solutions of the linear ODEs 
in \ref{sec:H2_H0} for $n=0,2$, with $H_0(\omega)$, $H_2(\omega)$ being their values on the boundary $r=1$.

Given $c$ to second order, substituting $c = c^{(0)} + (\hat{c}_1^{(1)} \hat{I}\mathrm{e}^{\mathrm{i}\omega t} + \mbox{c.c.})  + (\hat{c}_0^{(2)} +  \hat{c}_2^{(2)}\mathrm{e}^{\mathrm{i}2\omega t}+ \mbox{c.c.})\hat{I}^2  + O(\hat{I}^3)$ into $V = U(c) + \eta$ where $\eta$ is given by (\ref{eq:eta_no_C}), and expanding in $\hat{I}$ up to $O(\hat{I}^2)$,
\begin{align}
    V =& U^{(0)} + U^{(0)\prime} (\hat{c}_1^{(1)} \hat{I} \mathrm{e}^{\mathrm{i}\omega t} + \mbox{c.c.}) 
     \pm R^{(0)} (\hat{I} \mathrm{e}^{\mathrm{i}\omega t} + \mbox{c.c.})\\
    &+ \frac{1}{2} U^{(0)\prime\prime} (2\hat{I}^2 |\hat{c}_1^{(1)}|^2  + \hat{I}^2 (\hat{c}_1^{(1)})^2  \mathrm{e}^{\mathrm{i}2\omega t} + \mbox{c.c.})
    +  U^{(0)\prime} \hat{I}^2 (\hat{c}_0^{(2)} + \hat{c}_2^{(2)}  \mathrm{e}^{\mathrm{i}2\omega t} + \mbox{c.c.})\\
    &+ (\beta - 1/2)(R^{(0)})^2( 2\hat{I}^2 + \hat{I}^2\mathrm{e}^{\mathrm{i}2\omega t} + \mbox{c.c.}) \\
    &\pm R^{(0)\prime} \hat{I}^2 (\hat{c}_1^{(1)} + \mbox{c.c.})
    \pm R^{(0)\prime} \hat{I}^2 (\hat{c}_1^{(1)} \mathrm{e}^{\mathrm{i}2\omega t} + \mbox{c.c.})
    + O(\hat{I}^3).
\end{align}
Here we used that $E^{-1}(x) = 2x + 4(\beta-1/2)x^2+O(x^3)$ for $|x|\ll1$.
Collecting terms with factors of $1,\mathrm{e}^{\mathrm{i}\omega t},\mathrm{e}^{\mathrm{i}2\omega t}$, we can read off the Fourier coefficients 
\begin{align}
    \widehat{V}_0 &= U^{(0)} + \hat{I}^2 \left[U^{(0)\prime\prime}  |\hat{c}_1^{(1)}|^2 + U^{(0)\prime} \hat{c}_0^{(2)} + 2(\beta - 1/2)(R^{(0)})^2  \pm 2 R^{(0)\prime}  \mathrm{Re}\{\hat{c}_1^{(1)}\}\right] + O(\hat{I}^3), \\
    \widehat{V}_1 &= \hat{I}\left[ U^{(0)\prime} \hat{c}_1^{(1)}  \pm R^{(0)} \right] + O(\hat{I}^3),\\
    \widehat{V}_2 &= \hat{I}^2 \left[\frac{1}{2} U^{(0)\prime\prime}  (\hat{c}_1^{(1)})^2 + U^{(0)\prime}  \hat{c}_2^{(2)} + (\beta - 1/2)(R^{(0)})^2 \pm R^{(0)\prime}  \hat{c}_1^{(1)} \right] + O(\hat{I}^3),
\end{align}
and the impedances corresponding to the $\hat{I}^2$ terms are ($z_1^{(1)}$ is still (\ref{eq:z00_z11_derivation}))
\begin{align}
    z_0^{(2)} &= 2\mathrm{Re}\left[\frac{1}{2} U^{(0)\prime\prime}  |\hat{c}_1^{(1)}|^2 + \frac{1}{2} U^{(0)\prime} \hat{c}_0^{(2)} + (\beta - 1/2)(R^{(0)})^2  \pm R^{(0)\prime}  \hat{c}_1^{(1)} \right], \label{eq:z02_derivation}\\
    z_2^{(2)} &= \frac{1}{2} U^{(0)\prime\prime}  (\hat{c}_1^{(1)})^2 + U^{(0)\prime}  \hat{c}_2^{(2)} + (\beta - 1/2)(R^{(0)})^2 \pm R^{(0)\prime}  \hat{c}_1^{(1)}, \label{eq:z22_derivation}
\end{align}
where all terms are evaluated at $r=1$.

The impedances (\ref{eq:z00_z11_derivation}), (\ref{eq:z02_derivation}), (\ref{eq:z22_derivation}) hold for both electrodes $\pm$, and so correspond to (\ref{eq:z_11_pm}), (\ref{eq:z22_pm}), (\ref{eq:z02_pm}) in the paper, respectively.  

\subsection{Including double-layer capacitance}
\label{sec:Derivation-with-C}
Here we derive the impedances including double-layer capacitance, $Z_{1,\pm}^{(1)}$, $Z_{2,\pm}^{(2)}$, $Z_{0,\pm}^{(2)}$ in (\ref{eq:Z11_pm})-(\ref{eq:Z02_pm}), given the expressions when capacitance is neglected, $z_{1,\pm}^{(1)}$, $z_{2,\pm}^{(2)}$, $z_{0,\pm}^{(2)}$. To do so, observe that the double layer appears in parallel to the electrochemical reaction and dynamics considered in the previous section, which gave rise to the set of impedances $z_{1,\pm}^{(1)}$, $z_{2,\pm}^{(2)}$, $z_{0,\pm}^{(2)}$. The derivation here holds for a set of impedances $z$ due to any model dynamics, but we have verified the final formulae for this SPM by repeating the analysis of section \ref{sec:Derivation-no-C} with capacitance included\textemdash however that method involves significantly more algebra. Here we also drop the $\pm$ subscripts for brevity, as our analysis applies to both electrodes. 

Given the current $\pm I(t)$ entering electrode $\pm$ let $J_R$ be the current going into the reaction ($J_R = j/\xi$) and $J_C$ the current that charges the double layer, with $\pm I = J_R + J_C$. Equation (\ref{eq:capacitance_ODE_nondim}) is then
\begin{align}
    C\frac{\mathrm{d}V}{\mathrm{d}t} &= \pm J_C, \label{eq:J_C_equation}
\end{align}
where the $\pm$ corresponds to electrode $\pm$.
Considering $I(t) = \hat{I}\mathrm{e}^{\mathrm{i}\omega t} + \hat{I}\mathrm{e}^{-\mathrm{i}\omega t}$, then $J_R$, $J_C$ will both have fundamental frequency components but also higher harmonics. In particular, we have $J_R = \sum_{n=-\infty}^\infty \hat{J}_{R,n} \mathrm{e}^{\mathrm{i}n\omega t}$ and for $\hat{I}\ll1$ the harmonics behave as
\begin{align}
   \hat{J}_{R,0} &= O(\hat{I}^2), \label{eq:J_R_0_asym}\\
   \hat{J}_{R,1} &= \hat{J}_{R,1}^{(1)}\hat{I} + O(\hat{I}^3), \\
   \hat{J}_{R,2} &= \hat{J}_{R,2}^{(2)}\hat{I}^2 + O(\hat{I}^4),\label{eq:J_R_2_asym}
\end{align}
or in general $\hat{J}_{R,n} = O(\hat{I}^n), n\geq 1$ (note that $\hat{J}_{R,0}$ is not $O(1)$ since it must vanish when $\hat{I}=0$). 

The key to our approach is to notice that the voltage $V$ should be able to be expressed purely in terms of the current going into the reaction, $J_R$. This current is not yet known, but we may leverage the behaviour of its Fourier coefficients with respect to the small applied current amplitude $\hat{I}$. As $J_R$ is small in this limit, we can expand $V = U^{(0)} + \sum_{m=1}^\infty A^{(m)} J_R^m$, or in the frequency domain,
\begin{align}
    \widehat{V}_n &= U^{(0)}\delta_{n0} + A^{(1)} \hat{J}_{R,n}+ A^{(2)} \widehat{(J_R^2)}_n + \cdots, \label{eq:V_n_expansion_in_J_R}
\end{align}
where $\delta_{ij}$ is the Kronecker delta and $\widehat{(J_R^2)}_n$ is the $n$-th Fourier coefficient of $J_R^2$. By squaring the Fourier series for $J_R$, and using (\ref{eq:J_R_0_asym})-(\ref{eq:J_R_2_asym}), we have
\begin{align}
    \widehat{(J_R^2)}_n &= \begin{dcases}
            2|\hat{J}_{R,1}|^2 + O(\hat{I}^4), & n=0 \\
            \hat{J}_{R,1}^2 + O(\hat{I}^3), & n=2 \\
            O(\hat{I}^3) & n = 1,3,4,5,\ldots 
    \end{dcases}
\end{align}
Setting $n=1$ in (\ref{eq:V_n_expansion_in_J_R}) then gives
\begin{align}
    \widehat{V}_1 &= A^{(1)} \hat{J}_{R,1} + O(\hat{I}^3),
\end{align}
which is simply the linear response of the reaction at fundamental frequency ($\omega$) due to current component $\hat{J}_{R,1} \mathrm{e}^{\mathrm{i}\omega t}$, so can be written in terms of the usual linear impedance $z_1^{(1)}(\omega)$: 
\begin{align}
    \widehat{V}_1 &= z_1^{(1)}(\omega) \hat{J}_{R,1} + O(\hat{I}^3).
\end{align}
Together with $C\mathrm{i}\omega\widehat{V}_1 = \hat{J}_{C,1}$ (capacitance equation (\ref{eq:J_C_equation}) in frequency domain) and $\pm\hat{I} = \hat{J}_{R,1} + \hat{J}_{C,1}$, we can eliminate $\hat{J}_{R,1}$ and $\hat{J}_{C,1}$ to give
\begin{align}
    \widehat{V}_1 &= \underbrace{ \frac{z_1^{(1)}(\omega)}{1\pm z_1^{(1)}(\omega)\mathrm{i}\omega C} }_{\displaystyle{Z_1^{(1)}}} \hat{I} + O(\hat{I}^3), \label{eq:Z11_with_C_derivation}
\end{align}
where $Z_1^{(1)}$ can be read off as the coefficient of $\hat{I}$.

If we set $n=2$ in (\ref{eq:V_n_expansion_in_J_R}) instead, 
\begin{align}
    \widehat{V}_2 &= A^{(1)} \hat{J}_{R,2} + A^{(2)} \hat{J}_{R,1}^2+ O(\hat{I}^3),
\end{align}
where the first term is the linear response to the component $\hat{J}_{R,2} \mathrm{e}^{\mathrm{i}2\omega t}$ (frequency $2\omega$), and the second is the quadratic response to the component  $\hat{J}_{R,1} \mathrm{e}^{\mathrm{i}\omega t}$ (frequency $\omega$). Hence, they are given by the impedances $z_1^{(1)}(2\omega)$ and $z_2^{(2)}(\omega)$, respectively:
\begin{align}
    \widehat{V}_2 &= z_1^{(1)}(2\omega) \hat{J}_{R,2} + z_2^{(2)}(\omega) \hat{J}_{R,1}^2+ O(\hat{I}^3).
\end{align}
Together with $C\mathrm{i}2\omega\widehat{V}_2 = \hat{J}_{C,2}$ and $0 = \hat{J}_{R,2} + \hat{J}_{C,2}$ (the applied current has no $2\omega$ component), we can eliminate $\hat{J}_{R,2}$ and $\hat{J}_{C,2}$ to give
\begin{align}
    \widehat{V}_2 &= \frac{\hat{J}_{C,2}}{\mathrm{i}2\omega C}
    = \underbrace{ \frac{z_2^{(2)}(\omega)}{[1\pm \mathrm{i}\omega C z_1^{(1)}(\omega)]^2 [1\pm \mathrm{i}2\omega C z_1^{(1)}(2\omega)]} }_{\displaystyle{Z_2^{(2)}}} \hat{I}^2 + O(\hat{I}^4), \label{eq:Z22_with_C_derivation}
\end{align}
with $Z_2^{(2)}$ given by the coefficient of $\hat{I}^2$.

The case $n=0$ is similar to $n=2$, except that $\hat{J}_{R,0} = -\hat{J}_{C,0} = 0$ (from zero mode of (\ref{eq:J_C_equation})), and similar manipulations on (\ref{eq:V_n_expansion_in_J_R}) lead to
\begin{align}
    \widehat{V}_0 &= U^{(0)} + \underbrace{ \frac{z_0^{(2)}(\omega)}{|1\pm \mathrm{i}\omega C z_1^{(1)}(\omega)|^2} }_{\displaystyle{Z_0^{(2)}}} |\hat{I}|^2 + O(\hat{I}^4). \label{eq:Z02_with_C_derivation}
\end{align}
Reapplying the $\pm$ subscripts, then impedance expressions in (\ref{eq:Z11_with_C_derivation}), (\ref{eq:Z22_with_C_derivation}), (\ref{eq:Z02_with_C_derivation}) correspond to the single electrode expressions (\ref{eq:Z11_pm}), (\ref{eq:Z22_pm}), (\ref{eq:Z02_pm}) given in the paper.

\section{Transfer functions for nonlinear solid-state diffusion}
\subsection{Equations for $H_{2}$ and $H_{0}$}
\label{sec:H2_H0}
Solutions for $H_{2}(\omega)$ and $H_{0}(\omega)$, which appear
in the expressions for $Z_{2}^{(2)}$ and $Z_{0}^{(2)}$, require
solving the following one dimensional ODEs. Define $M_{n}(r,\omega)$, where $n$ is the number of the harmonic,
to be the solution of 
\begin{align}
\frac{1}{r^{2}}\frac{\mathrm{d}}{\mathrm{d}r}\left(r^{2}\frac{\mathrm{d}M_{n}}{\mathrm{d}r}\right)-\mathrm{i}\omega nM_{n} & =f_{n}(r),   &    0<r<1 &,   \label{eq:M_n_ODE}\\
\frac{\mathrm{d}M_{n}}{\mathrm{d}r} & =0, & r=0 &, \label{eq:M_n_BC_1}\\
\frac{\mathrm{d}M_{n}}{\mathrm{d}r} & =g_{n}, & r=1 &, \label{eq:M_n_BC_2}
\end{align}
for given $f_{n}(r)$, $g_{n}$. Then $H_{1}(\omega)$ corresponds
to the boundary value 
\begin{equation}
H_{1}=M_{1}|_{r=1},\qquad\text{where }f_{1}=0,\,g_{1}=-1
\end{equation}
The solution for $M_{1}$ is simply
\begin{equation}
M_{1}(r,\omega)=\frac{\sinh(\sqrt{\mathrm{i}\omega} r)}{r(\sinh(\sqrt{\mathrm{i}\omega})-\sqrt{\mathrm{i}\omega}\cosh(\sqrt{\mathrm{i}\omega}))},
\end{equation}
giving the solution (\ref{eq:H_1}) for $H_{1}(\omega)$. Then $H_{2}(\omega)$
corresponds to
\begin{equation}
H_{2}=M_{2}|_{r=1},\qquad\text{where }f_{2}=\left(\frac{\mathrm{d}M_{1}}{\mathrm{d}r}\right)^{2}+\mathrm{i}\omega M_{1}^{2},\,g_{2}=\left.\left(M_{1}\frac{\mathrm{d}M_{1}}{\mathrm{d}r}\right)\right|_{r=1}. \label{eq:M_2_problem}
\end{equation}
The ODE for $M_2(r,\omega)$ (and hence $H_2(\omega)$) was solved using a finite difference scheme with 100 grid points in $r$. When $\omega \geq 1000$, to reduce computation time, we instead used the high frequency approximation (\ref{eq:H_2_high_omega}) which was accurate to within 10$^{-5}$ in magnitude. 

Finally, $H_{0}(\omega)$ corresponds to
\begin{equation}
H_{0}=M_{0}|_{r=1},\qquad\text{where }f_{0}=\left|\frac{\mathrm{d}M_{1}}{\mathrm{d}r}\right|^{2},\, g_{0}=\mathrm{Re}\left.\left\{ M_{1}\overline{\frac{\mathrm{d}M_{1}}{\mathrm{d}r}}\right\} \right|_{r=1}.
\end{equation}
Note that when $n=0$, only derivatives of $M_0$ appear in (\ref{eq:M_n_ODE})-(\ref{eq:M_n_BC_2}) and another condition is needed. We may use that the spatially averaged concentration at every order (except leading) must vanish, which implies $\int_0^1 M_0 r^2 \mathrm{d}r=0$.

\subsection{High frequency limit of $H_{2}$}
\label{sec:H_2_high_omega}

Here we consider the high frequency limit $\omega \to \infty$ in the ODE problem (\ref{eq:M_n_ODE})-(\ref{eq:M_n_BC_2}), (\ref{eq:M_2_problem}), for $M_2(r,\omega)$, and hence $H_2(\omega)=M_2(1,\omega)$. The equation for $M_2$ is forced by $M_1$ and its derivatives, so we need the behaviour of $M_1$ as $\omega \to \infty$. For $r=O(1)$ (away from the boundary $r=1$), $M_1 \to 0$ exponentially, with the oscillatory behaviour confined to a boundary layer $1 - r = O(\omega^{-1/2})$. In the boundary layer, $X = \omega^{1/2}(r-1) = O(1)$ and
\begin{align}
    M_1 & \sim - \frac{ \mathrm{e}^{\sqrt{\mathrm{i}}X} }{\sqrt{\mathrm{i}\omega}},\qquad \mbox{as }\omega \to \infty.
\end{align}
Substituting this into $f_2$ and $g_2$ in (\ref{eq:M_2_problem}) gives $f_2\sim 2\mathrm{e}^{2\sqrt{\mathrm{i}}X}$ and $g_2 = 1/\sqrt{\mathrm{i} \omega}$. From inspection of (\ref{eq:M_n_ODE}), this implies the expansion $M_2 = \omega^{-1}(\overline{M}_2 (X) + \cdots)$, where $\overline{M}_2(X)=O(1)$ is the leading order solution in the boundary layer, satisfying
\begin{align}
    \frac{\mathrm{d}^2\overline{M}_2}{\mathrm{d}X^2}-2\mathrm{i}\overline{M}_2 & = 2\mathrm{e}^{2\sqrt{\mathrm{i}}X}, &  X &< 0,   \label{eq:tilde_M_2_ODE}\\
    \overline{M}_2 & \to 0, & X &\to -\infty, \label{eq:tilde_M_2_BC_1}\\
    \frac{\mathrm{d}\overline{M}_2}{\mathrm{d}X} & = \frac{1}{\sqrt{\mathrm{i}}}, & X &= 0, \label{eq:tilde_M_2_BC_2}
\end{align}
where (\ref{eq:tilde_M_2_BC_1}) is the matching condition with the outer solution. This problem is easily solved to give
\begin{align}
    M_2(r,\omega) & \sim \frac{1}{\omega} \overline{M}_2 = \frac{1}{\mathrm{i}\omega} \left[ \mathrm{e}^{2\sqrt{\mathrm{i}}X} - \frac{1}{\sqrt{2}}\mathrm{e}^{\sqrt{2\mathrm{i}}X}\right],
\end{align}
and hence
\begin{align}
    H_2(\omega) & \sim \frac{1}{\omega} \left. \overline{M}_2 \right|_{X=0} = \frac{1}{\mathrm{i}\omega}\left(1 - \frac{1}{\sqrt{2}} \right).
\end{align}

\section{Parameter set used for synthetic data}
\label{sec:Synthetic-param-groups}

The parameter set used for the generation of synthetic NLEIS data, both dimensional and nondimensional parameter groupings, is given in Tables \ref{tab:Marquis-params}, \ref{tab:Param-groups}.

\begin{table}[]
\centering
\footnotesize
\begin{tabular}{@{}llll@{}}
\toprule
\multirow{2}{*}{Parameter} & \multirow{2}{*}{Description {[}unit{]}}            & \multicolumn{2}{c}{Value}   \\ \cmidrule(l){3-4} 
                           &                                                    & negative (-) & positive (+) \\ \midrule 
$\mathcal{R}_\pm^*$        & Particle radius {[}m{]}                            & $1\times 10^{-5}$         & $1\times 10^{-5}$     \\
$c_{\mathrm{max},\pm}^*$   & Maximum Li concentration {[}mol m$^3${]}           & 24983                     & 51218                 \\
$\mathcal{D}_\pm^*$        & Diffusivity of Li in electrode {[}m$^2$s$^{-1}${]} & $3.9\times 10^{-15}$ (chosen)      & $10^{-14}$  (chosen)   \\
$m_\pm^*$                  & Reaction rate {[}A m$^{-2}$ (m$^3$/mol)$^{1.5}${]}   & $2\times 10^{-5}$         & $6\times 10^{-7}$     \\
$\beta_\pm$                & Cathodic charge transfer coefficient               & 0.45 (chosen)             & 0.55 (chosen)         \\
$\epsilon_\pm$             & Active material volume fraction                    & 0.6                       & 0.5                   \\
$L_\pm^*$                  & Electrode thickness {[}m{]}                        & $100\times 10^{-6}$       & $100\times 10^{-6}$   \\
$C_{\mathrm{dl},\pm}^*$    & Double-layer capacitance {[}F m$^{-2}${]}          & $10^{-2}$ \cite{Murbach2018} & $2.5\times 10^{-2}$ \cite{Murbach2018}  \\
$U_\pm^* (c_\pm^*)$        & Open circuit potential relative to Li/Li$^+$ [V]   & \cite{Marquis2019}      & \cite{Marquis2019}    \\ \midrule
$c_{e}^*$                  & Li$^+$ concentration in electrolyte {[}mol m$^3${]} & \multicolumn{2}{c}{$10^{3}$}                      \\
$\mathcal{A}^*$            & Current collector surface area {[}m$^2${]}         & \multicolumn{2}{c}{0.1}                           \\
$R_s^*$                    & Series resistance [$\Omega$]                        & \multicolumn{2}{c}{0.05 (chosen)}               \\
$F^*$                      & Faraday's constant [C mol$^{-1}$]                  & \multicolumn{2}{c}{96485}                         \\
$T^*$                      & Ambient temperature [K]                            & \multicolumn{2}{c}{298.15}                        \\ 
$R_g^*$                    & Universal gas constant [J mol$^{-1}$ K$^{-1}$]     & \multicolumn{2}{c}{8.314}                         \\
$\tau^*$                   & Reference timescale [s]                           & \multicolumn{2}{c}{1 (chosen)}                     \\
$J^*$                      & Reference current scale [A]                        & \multicolumn{2}{c}{1 (chosen)}                    \\
\bottomrule
\end{tabular}
\caption{Dimensional parameter set used for the generation of synthetic impedance data, values taken from \cite{Marquis2019} except where specified.}
\label{tab:Marquis-params}
\end{table}

\begin{table}[]
\centering
\footnotesize
\begin{tabular}{@{}lllll@{}}
\toprule
\multirow{2}{*}{}               & \multirow{2}{*}{Parameter groups} & \multirow{2}{*}{Units}    & \multicolumn{2}{c}{Value}                 \\ \cmidrule(l){4-5} 
                                &                                   &                           & negative (-)          & positive (+)      \\ \midrule 
\multirow{3}{*}{Dimensional}    & $\tau_{\mathrm{d},\pm}^*$         & s                         & $2.564\times 10^{4}$  & $1.000\times 10^{4}$    \\
                                & $R_{\mathrm{ct,typ},\pm}^*$       & $\Omega$                  & 0.969                 & 37.718                  \\
                                & $Q_{\mathrm{th},\pm}^*$           & A h                       & 4.018                 &  $6.864$             \\ \midrule
\multirow{5}{*}{Dimensionless}  & $\tau_{\mathrm{d},\pm}$           & -                         & $2.564\times 10^{4}$  & $1.000\times 10^{4}$       \\
                                & $\xi_\pm$                         & -                         & $2.305\times 10^{-5}$ & $1.349\times 10^{-5}$   \\
                                & $\chi_\pm$                        & -                         & 0.0249                & $0.969$               \\
                                & $C_\pm$                           & -                         & 0.180                 & $0.375$                          \\
                                & $c_\pm^{0\%}$                     & -                         & 0.8                   & 0.6                     \\
\bottomrule
\end{tabular}
\caption{Parameter groupings (dimensionless and dimensional) for the parameter set in Table \ref{tab:Marquis-params}, and used for the generation of synthetic impedance data.}
\label{tab:Param-groups}
\end{table}

\clearpage
\section{Further experimental information}
\label{sec:Further-Exp-Info}

\begin{figure}
\centering
\includegraphics[width=0.45\textwidth]{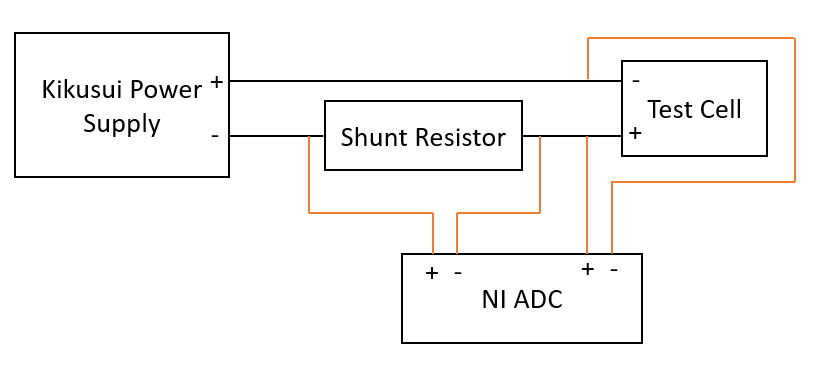}
\caption{Wiring diagram for NLEIS data collection experiments.}
\label{fig:wiring}
\end{figure}

\subsection{NLEIS Data collection procedure}

\begin{figure}
\centering
\includegraphics[width=0.50\textwidth]{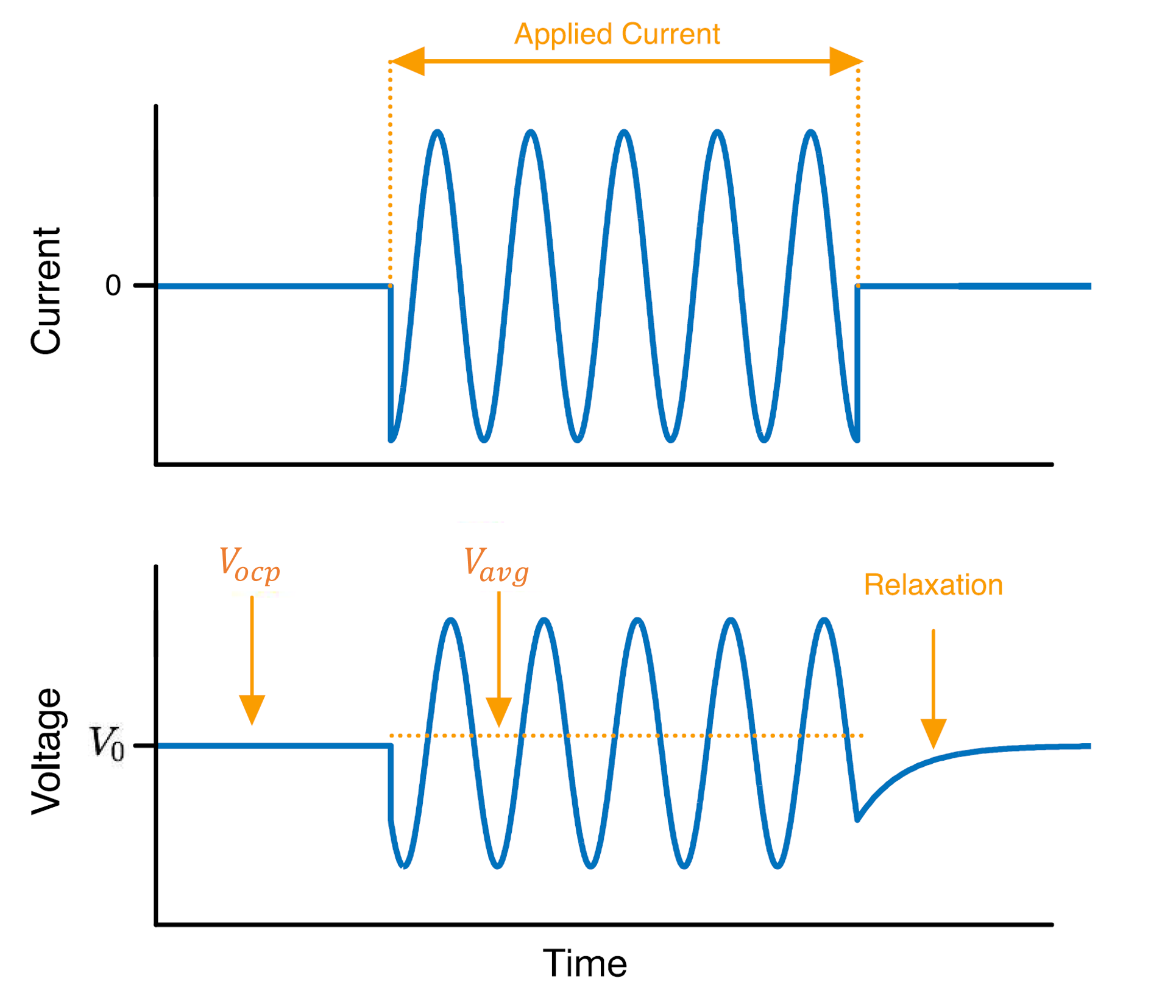}
\caption{Exemplary current and voltage NLEIS measurement instance include non-excitation periods at the beginning and end}
\label{fig:InitialData}
\end{figure}

NLEIS data, in the form of time-domain voltage and current sinusoids, was collected using the following set of procedures:
\begin{enumerate}
    \item The cell was placed in a thermal chamber (Vötsch VT 4002) at \SI{19}{\celsius} and initially left for 90 minutes to reach thermal equilibrium.
    \item Thermal limits were checked, as per Section \ref{section:thermal}.
    \item The following steps were executed at each DoD, starting at 10\% DoD and ending at 90\% DoD, at intervals of 10\%:
    \begin{enumerate}
        \item NLEIS data was measured
        \item The cell was discharged by 10\% DoD using constant \SI{500}{mA} current
        \item The cell was rested for 10 minutes.
    \end{enumerate}
\end{enumerate}
To ensure a complete dataset, data recording was started before current excitation began and ended after the current excitation was stopped. This left each recorded data instance as shown in Fig.\ \ref{fig:InitialData}.
The extra measurement information either side of the perturbation facilitated capturing the OCV of the cell before each test, the average voltage of the cell  under load, and the relaxation period after the load was removed.

\subsection{Thermal limits}\label{section:thermal}
A type-T nickel-copper-constantin thermocouple was attached to the surface of the pouch cell with tape and  connected to a National Instruments (NI) 9213 temperature input module. The thermocouple was used firstly as a safety check to stop the experiment in the event of overheating, and secondly to augment the current and voltage data with time dependant temperature data. 

Before each measurement, thermal limits were checked, and testing was stopped if these were exceeded. Specifically, the temperature before and after any measurement had to be less than \SI{25}{\celsius}, and the difference between the beginning and end temperatures had to be less than \SI{2}{\celsius}.

\subsection{Control and data acquisition}

The power supply/load (Kikusui PBZ60-6.7) was externally controlled from a  computer (Windows, Intel Core i9-9800HK \SI{32}{GB} RAM and \SI{1}{TB} RAID 5 storage). The data acquisition system (NI 9239 ADC module, 9213 temperature input module, and cDAQ-9188 back-plane) was also connected to the same computer, and a MATLAB script was written to control the system and measure data at \SI{24}{bits} and \SI{50}{kHz}, including careful consideration of memory issues due to the lengthy duration of the tests. 

\subsection{Data processing}

The measurements were stored in a database spanning 8 DoDs and 61 frequencies for each experimental campaign. For processing, first each dataset was trimmed to leave just sinusoids.
This was done by locating the first peak in the current signal following the rising edge, then iterating the following steps until all other peaks were found:
\begin{enumerate}
    \item The first current value within 99\% of the expected peak current was established and marked as the start of the clipped signal. Failing this, the search was attempted again with decreasing accuracy levels: 1\%, 5\%, 15\%, 25\%, or notified an error.
    \item The expected number of samples to the next peak was calculated by dividing the sampling frequency by the test excitation frequency. 
    \item If another peak was then found, the algorithm went back to step 2). Otherwise the last available peak was marked as the end of the clipped signal.
\end{enumerate}
The DC component of voltage was removed by taking the average voltage under load and  subtracting this from the measured voltages. To quantify the harmonics the voltage and current signals were multiplied by a Hann window function then Fourier transformed. 

\subsection{OCP Functions and electrode balancing}
\label{sec:OCPs-and-electrode-balancing}

Three-electrode data for the Kokam SLPB533459H4 cell from McTurk et al.\ \cite{McTurk2015} gives $U_\pm^*$ as a function of discharge capacity $Q^*$, which we converted to normalised discharge capacity $Q=Q^*/Q_{\mathrm{cap}}^*=\text{DoD}/100$ where $Q_{\mathrm{cap}}^* = $ \SI{749}{mAh}. Analytical expressions were  fitted to this data using the nonlinear least squares minimiser \texttt{lsqnonlin} in MATLAB, giving
\begin{align}
    U_{-}^* =& \,\,0.9577 + \frac{0.0575 Q -0.0533 Q^2}{1-0.9877 Q^2}- 0.0760 \tanh[25.6264(Q-0.9063)] \nonumber\\
     & + 0.0171 \tanh[18.2444(Q-0.6530)] - 0.0051 \tanh[7.8242(Q-0.5474)]\nonumber\\
     & + 0.9226 \tanh[9.3014(Q-1.0513)]+ 0.0127 \tanh[34.6668(Q-0.2353)], \label{eq:negative_OCP_fit}
\end{align}
and
\begin{align}
    U_{+}^* =& \,\,9.8219 -0.9241Q -5.3099 \tanh[8.7293(Q+0.3418)] \nonumber\\
     & -0.7503\tanh[6.2419(Q-0.7826)]+1.6105\tanh[3.8390(Q-0.8676)] \nonumber\\
     & -0.5861\tanh[6.7659(Q-0.9774)] -0.2770\tanh[0.0109(Q-0.5488)], \label{eq:positive_OCP_fit}    
\end{align}
with fit RSMEs of \SI{5.3}{mV} and \SI{6.4}{mV}, respectively. These are shown in the paper in Figs.\ \ref{fig:kokam_OCPs}($a$)-($b$).

The stoichiometry limits $c_\pm^{0\%}$ and $c_\pm^{100\%}$, corresponding to 0\% and 100\% DoD ($Q=0$ and 1), were chosen using knowledge of the electrode chemistries. In the negative electrode, which is graphite, they were chosen so that the staging between the different phases (indicated by the peaks in $\mathrm{d}U_-^*/\mathrm{d}Q$) occurs in the expected ranges. In particular, choosing the centre of the first and second visible peaks (Stages 2 and 3) to be at $c_-=0.5$ and $c_-=0.24$, results in $c_-^{0\%}=0.66$, $c_-^{100\%}=0.01$. The positive electrode consists of NMC, but the precise mixture of materials has not been characterised for these cells to our knowledge. Thus we chose stoichiometry limits $c_+^{0\%}=0.4$, $c_+^{100\%}=1$, typical for similar cells from the same manufacturer \cite{Ecker2015a,Ecker2015}. At these values, the capacity parameter values $\xi_\pm$ follow from (\ref{eq:ksi_stoichiometry_relation}), giving $\xi_- = 8.13\times 10^{-5}$, $\xi_+ = 7.41\times 10^{-5}$. The OCPs can then also be expressed as functions of $c_\pm$ directly by substituting $Q = (c_\pm - c_\pm^{0\%})(c_\pm^{100\%} - c_\pm^{0\%})$ into (\ref{eq:negative_OCP_fit})-(\ref{eq:positive_OCP_fit}).

Derivatives with respect to $c_\pm$ may be calculated using the chain rule,
\begin{align}
    U_\pm^{*\prime}(c_\pm) & = \frac{\mathrm{d}U_\pm^{*}}{\mathrm{d}c_\pm} = \frac{1}{c_\pm^{100\%} - c_\pm^{0\%}}\frac{\mathrm{d}U_\pm^{*}}{\mathrm{d}Q}. 
\end{align}
The second derivatives, however, are not captured sufficiently well by the expressions (\ref{eq:negative_OCP_fit})-(\ref{eq:positive_OCP_fit}), which is clear from looking at the slope of $\mathrm{d}U_+^*/\mathrm{d}Q$ in Fig.\ \ref{fig:kokam_OCPs}. Hence, in the results section, when calculation of the second derivatives was necessary, we did this directly from the data using second-order central differences.

\section{Further parameter estimation details}
\label{sec:Further-parameter-estimation}

\subsection{Maximum likelihood estimation (MLE)}
Assuming the voltage harmonic data is normally distributed around the model predictions, as in (\ref{eq:synthetic-data-noise}) of the paper, with residuals on the real and imaginary components of the $n$-th harmonic distributed as $\varepsilon_{n,r},\varepsilon_{n,i}\sim N(0,\sigma_n^{*2})$, the joint probability distribution of the observations $\widehat{V}_{1,j}^{*,\mathrm{data}},\widehat{V}_{2,j}^{*,\mathrm{data}}$ across the set of frequencies $\omega_j, j=1,2,\ldots,N_\omega$ is given by
\begin{align}
    p(\widehat{V}_{1,j}^{*,\mathrm{data}},\widehat{V}_{2,j}^{*,\mathrm{data}}|\omega_i, \bm{\theta}, \sigma_n^{*}) =& 
    \prod_{n=1}^2 \prod_{j=1}^{N_\omega}
    (\sqrt{2\pi\sigma_n^{*2}})^{-2} \exp\left( 
        -\frac{1}{2\sigma_n^{*2}}
        \left|\widehat{V}_{n}^{*}(\omega_j;\bm{\theta})-\widehat{V}_{n,j}^{*,\mathrm{data}} \right|^2
    \right)\label{eq:joint-prob}
\end{align}
which is also the likelihood $L(\bm{\theta},\sigma_1^{*},\sigma_2^{*})$ of the parameters $\bm{\theta},\sigma_1^{*},\sigma_2^{*}$ given the data. Maximising $L$ over $\bm{\theta},\sigma_1^{*},\sigma_2^{*}$ then gives a maximum likelihood estimator. Equivalently, we may minimise the negative log-likelihood, $ -\log L$, which is (substituting $\widehat{V}_n^{*} = (\hat{I}^*)^n Z_n^*$):
\begin{align}
    -\log L(\bm{\theta},\sigma_{1}^*,\sigma_2^*)
    =& \,\, 2N_\omega \log(2\pi \sigma_1^{*2}) + 2N_\omega \log(2\pi \sigma_2^{*2}) \nonumber \\
    & + \sum_{j=1}^{N_\omega}  \left[ \frac{|\hat{I}^*|^2}{2\sigma_1^{*2}}\left|Z_1^{*(1)}(\omega_j;\bm{\theta})-Z_{1,j}^{*,\mathrm{data}} \right|^2 + \frac{|\hat{I}^*|^4}{2\sigma_2^{*2}}\left|Z_2^{*(2)}(\omega_j;\bm{\theta})-Z_{2,j}^{*,\mathrm{data}} \right|^2 \right]. \label{eq:-logL}
\end{align}
Minimizing over each $\sigma_n^{*}$, by setting $\partial (-\log L)/\partial \sigma_n^* = 0$ and rearranging for $\sigma_n^*$, gives
\begin{align}
    \sigma_n^{*2} &= \frac{|\hat{I}^*|^{2n}}{2N_\omega} \sum_{j=1}^{N_\omega}\left|Z_n^{*(n)}(\omega_j;\bm{\theta})-Z_{n,j}^{*,\mathrm{data}} \right|^2,\qquad
    n=1,2,
\end{align}
which can be used to eliminate the $\sigma_n^{*}$ in (\ref{eq:-logL}), leaving
\begin{align}
    -\log L(\bm{\theta}) =& 
    \,\, 2N_\omega \left[ l_{1}(\bm{\theta}) + l_{2}(\bm{\theta}) 
     +  \log\left(\frac{\pi |\hat{I}^*|^2}{N_\omega}\right) + \log\left(\frac{\pi |\hat{I}^*|^4}{N_\omega}\right) +1\right].
\end{align}
where
\begin{align}
    l_{1}(\bm{\theta}) &= 
    \log\left(\sum_{j=1}^{N_\omega}\left|Z_1^{*(1)}(\omega_j;\bm{\theta})-Z_{1,j}^{*,\mathrm{data}} \right|^2\right),\\
    l_{2}(\bm{\theta}) &= 
    \log\left(\sum_{j=1}^{N_\omega}\left|Z_2^{*(2)}(\omega_j;\bm{\theta})-Z_{2,j}^{*,\mathrm{data}} \right|^2\right),
\end{align}
Therefore, the MLE problem for $\bm{\theta}$ is reduced to
\begin{align}
    \hat{\bm{\theta}}_{\mathrm{NLEIS}} &= \underset{\bm{\theta}}{\mathrm{argmin}}\,(l_1(\bm{\theta}) + l_2(\bm{\theta})).
\end{align}
with total log-likelihood $l_{12}(\bm{\theta}) = l_1 + l_2$. 

\subsection{Numerical optimization}
The minimization problems above were solved using nonlinear optimization routines in MATLAB employing bound constraints. When using synthetic data, a local gradient-based minimizer \texttt{fmincon} was used, initialised at the true parameter values (except for $\beta_\pm$ when using EIS data\textemdash see the Results section. This allowed efficient and robust convergence to the known global minimum. We employed the SQP algorithm with optimality, constraint, and step size tolerances of $10^{-7}$, $10^{-10}$, and $10^{-12}$. The objective function may have several local minima thus, when using experimental data where the true global minimum is unknown, we used a global minimizer \texttt{GlobalSearch} which starts the aforementioned local solver (\texttt{fmincon}) at multiple start points (which we set to 300) generated with a scatter search algorithm within the parameter bounds, identifying basins of attraction. In all cases, the chosen parameter bounds were $\chi_{\pm}, C_\pm \in [0, 10]$, $\beta_\pm\in [0, 1]$, $\tau_{\mathrm{d},\pm}\in [0, 10^7]$, unless otherwise specified. 

\section{Large separation of timescales}
\label{sec:large-separation-of-timescales}

\begin{figure}
    \centering
    \includegraphics[width=1.0\textwidth]{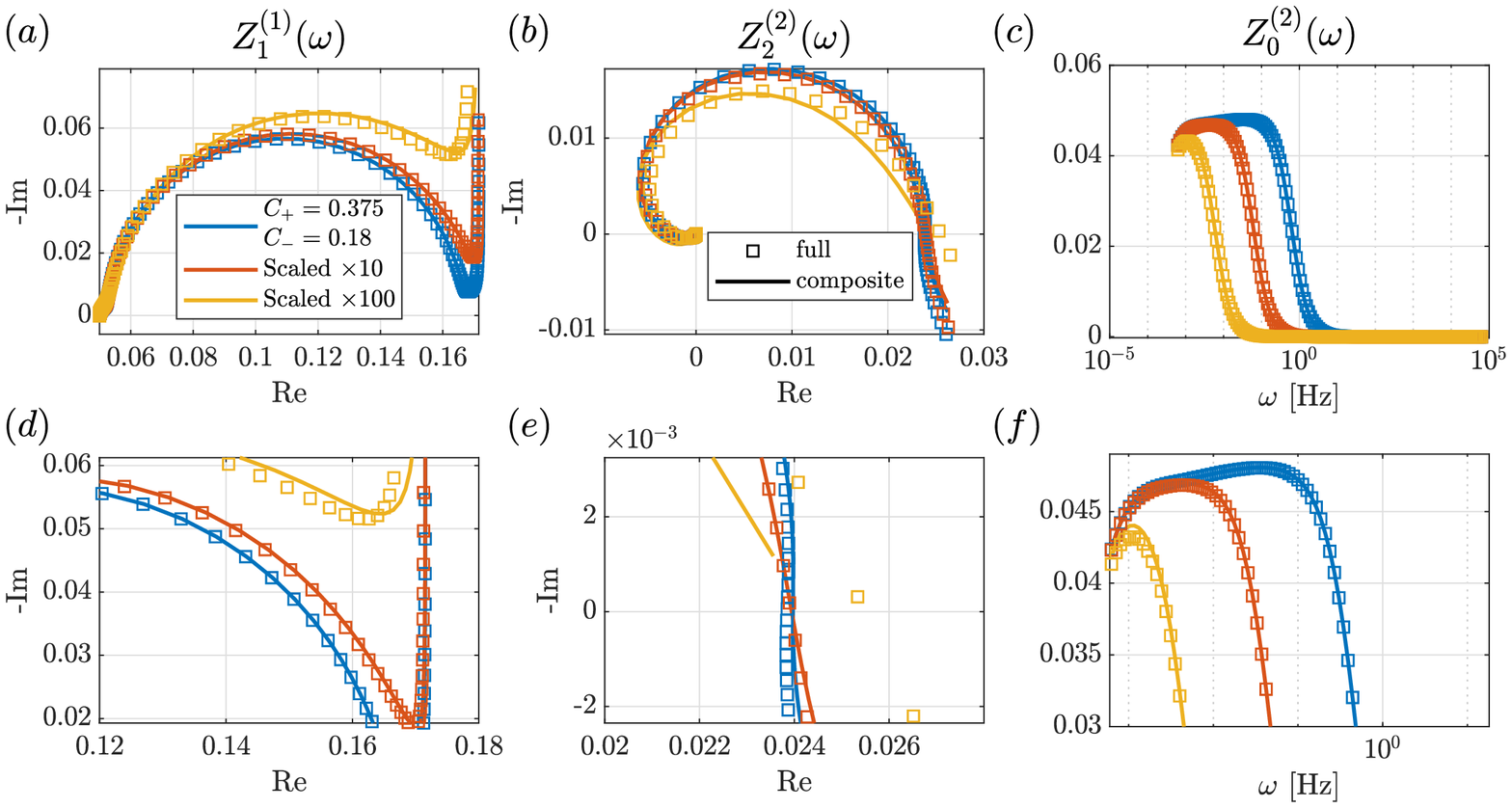}
    \caption{Comparison of impedance formulae for $Z_{1}^{(1)}$, $Z_{2}^{(2)}$, and $Z_{0}^{(2)}$. The full formula (squares) uses (\ref{eq:Z11_pm})-(\ref{eq:Z02_pm}), and composite formula uses (\ref{eq:Z11_pm_comp})-(\ref{eq:Z02_pm_comp}), for parameters in Table \ref{tab:Marquis-params} at $\mathrm{DoD}=50\%$. Bottom panels zoom into the ``overlap region" in the panels above. Frequency range $10^{-4}\leq \omega/2\pi \leq 10^{4}$ Hz. The different colours are for increasing values of the capacitances $C_{\pm}$, i.e.\ scaled by 10 (red) and 100 (yellow).} 
    \label{fig:composite-comparison}
\end{figure}

In this section, we compare the exact impedance expressions (\ref{eq:Z11_pm}), (\ref{eq:Z22_pm}) and (\ref{eq:Z02_pm}) to the simplified (composite) expressions (\ref{eq:Z11_pm_comp}), (\ref{eq:Z22_pm_comp}) and (\ref{eq:Z02_pm_comp}) which assume that the capacitive timescale $R_\pm^{(0)}C_\pm$ is much shorter than the diffusive one $\tau_{\mathrm{d},\pm}=1/D_\pm^{(0)}$. Fig.\ \ref{fig:composite-comparison} shows the full-cell impedances as calculated from both sets of expressions\textemdash for the parameters in Tables \ref{tab:Marquis-params}-\ref{tab:Param-groups}, and when the nondimensional capacitances $C_\pm$ are increased. The simplified formulae were derived assuming $C_\pm R_\pm^{(0)}/\tau_{\mathrm{d},\pm} \ll 1$, and for the base case here (blue lines), these ratios are $O(10^{-4})$, $O(10^{-6})$ for the positive and negative electrodes, respectively. The well-separated timescales assumption is clearly satisfied, and thus the composite expressions agree excellently with the full expression for all frequencies. Indeed, excellent agreement is observed even when $C_\pm$ values (and hence the capacitance timescales) are increased by up to 2 orders of magnitude. As $C_\pm$ values are increased, the composite expression should first breakdown for frequencies in the ``overlap region" (see (\ref{eq:omega_overlap})), where still only small discrepancies are observed. The mean relative errors (mean absolute error scaled by the maximum magnitude of the exact impedance), defined as
\begin{align}
    \mbox{MRE}_{n,m} &= \frac{\frac{1}{N_\omega}\sum_\omega |Z_n^{(m),\mathrm{comp}}(\omega) - Z_n^{(m)}(\omega)|}{\max_\omega |Z_n^{(m)}(\omega)|},
\end{align}
were found to be 0.03\%, 0.64\%, 0.04\% for $Z_1^{(1)}$, $Z_2^{(2)}$, $Z_0^{(2)}$ at the original values of $C_\pm$, and still only 0.39\%, 1.69\%, 0.45\% at the largest values of $C_\pm$.

\bibliographystyle{elsarticle-num} 
\bibliography{NLEIS-refs-no-URLs}